\documentclass[aps,prc,showpacs,showkeys,nofootinbib]{revtex4}
\usepackage[dvips]{graphicx}
\usepackage{dcolumn}% Align table columns on decimal point
\usepackage{amsthm}
\usepackage{bm}% bold math

\newtheorem{dfn}{Definition}

\begin{document}
%-----------------------------------------------------------------------------------------------------------------------

%-----------------------------------------------------------------------------------------------------------------------
\title{A fully quantum model of Big Bang}

% \author{S.~P.~Maydanyuk\thanks{\emph{E-mail:} maidan@kinr.kiev.ua}}

\author{S.~P.~Maydanyuk$^{1,}$\footnote[5]{\emph{E-mail:} maidan@kinr.kiev.ua},
A.~Del Popolo$^{2,3,}$\footnote[6]{\emph{E-mail:} adelpopolo@oact.inaf.it},
V.~S.~Olkhovsky$^{1,}$\footnote[7]{\emph{E-mail:} olkhovsk@kinr.kiev.ua}}
% E.~Recami$^{4,}$\footnote[8]{\emph{E-mail:} recami@mi.infn.it}}

\affiliation{$^{(1)}$Institute for Nuclear Research, National Academy of Sciences of Ukraine \\
$^{(2)}$Istituto di Astronomia dellпїЅUniversit`a di Catania, Viale A. Doria 6, I-95125 Catania, Italy \\
$^{(3)}$Dipartimento di Matematica, UniversitпїЅ Statale di Bergamo, via dei Caniana 2, 24127 Bergamo, Italy}
% $^{(4)}$Facolt\`{a} di Ingegneria, Universit\`{a} statale di Bergamo, Bergamo and INFN-Sezione di Milano, Milan, Italy}

% \affiliation{Institute for Nuclear Research, National Academy of Science of Ukraine, Kiev, 03680, Ukraine}

\date{\small\today}
\begin{abstract}
In the paper the closed Friedmann--Robertson--Walker model with quantization in the presence of the positive cosmological constant and radiation is studied. For analysis of tunneling probability for birth of an asymptotically deSitter, inflationary Universe as a function of the radiation energy a new definition of a ``free'' wave propagating inside strong fields is proposed.
On such a basis, tunneling boundary condition is corrected, penetrability and reflection concerning to the barrier are calculated in fully quantum stationary approach. For the first time non-zero interference between the incident and reflected waves has been taken into account which turns out to play important role inside cosmological potentials and could be explained by non-locality of barriers in quantum mechanics.
Inside whole region of energy of radiation the tunneling probability for the birth of the inflationary Universe is found to be close to its value obtained in semiclassical approach.
The reflection from the barrier is determined for the first time (which is essentially differs on 1 at the energy of radiation close to the barrier height).
The proposed method could be easily generalized on the cosmological models with the barriers of arbitrary shape, that has been demonstrated for the FRW-model with included Chaplygin gas.
Result is stable for variations of the studied barriers, accuracy are found to be 11--18 digits for all coefficients and energies below the barrier height.
\end{abstract}
%-----------------------------------------------------------------------------------------------------------------------

%-----------------------------------------------------------------------------------------------------------------------

% \textbf{Keywords:}

\maketitle
% *******************************************************************************************************************

% *******************************************************************************************************************
\section{Introduction
\label{sec.introduction}}

In order to understand what really happens in the formation of the Universe, many people came to the point of view that a quantum consideration of this process is necessary. After the publication of the first paper on the quantum description of Universe formation~\citep{DeWitt.1967,Wheeler.1968}, a lot of other papers appeared in this topic (for example, see Refs.~\citep{Vilenkin.1982.PLB,Hartle.1983.PRD,Linde.1984.LNC,Zeldovich.1984.LNC,Rubakov.1984.PLB,Vilenkin.1984.PRD,Vilenkin.1986.PRD,Atkatz.1984.PRD} and some discussions in Refs.~\citep{Vilenkin.1994.PRD,Rubakov.1999} with references therein).

Today, among all variety of models one can select two approaches which are the prevailing ones: these are the Feynman formalism of path integrals in multidimensional spacetime, developed by the Cambridge group and other researchers, called the \emph{``Hartle--Hawking method''} (for example, see Ref.~\citep{Hartle.1983.PRD}), and a method based on direct consideration of tunneling in 4-dimensional Euclidian spacetime, called the \emph{``Vilenkin method''}
(for example, see Refs.~\citep{Vilenkin.1982.PLB,Vilenkin.1984.PRD,Vilenkin.1986.PRD,Vilenkin.1994.PRD}). Here, according to Ref.~\citep{Vilenkin.1995}, in the quantum approach we have the following picture of the Universe creation: a closed Universe with a small size is formed from ``nothing'' (vacuum), where by the word ``nothing'' one refers to a quantum state without classical space and time. A wave function is used for a probabilistic description of the creation of the Universe and such a process is connected with transition of a wave through an effective barrier. Determination of penetrability of this barrier is a key point in the estimation of duration of the formation of the Universe, and the dynamics of its expansion in the first stage.

However, in the majority of these models, with the exception of some exactly solvable models, tunneling is mainly studied in details in the semiclassical approximation (see Refs.~\citep{Vilenkin.1994.PRD,Rubakov.1999}). An attractive side of such an approach is its simplicity in the construction of decreasing and increasing partial solutions for the wave function in the tunneling region, the outgoing wave function in the external region, and the possibility to define and to estimate in an enough simply way the penetrability of the barrier, which can be used for obtaining the duration of the nucleation of the Universe. The \emph{tunneling boundary condition} \citep{Vilenkin.1995,Vilenkin.1994.PRD} could seem to be the most natural and clear description, where the wave function should represent an outgoing wave only in the enough large value of the scale factor $a$. However, is really such a wave free in the asymptotic region? In order to draw attention on the increase of the modulus
of the potential with increasing value of the scale factor $a$ and increasing magnitude of the gradient of such a potential,
%, used with opposite sign and having a sense of force,
acting on this wave ``through the barrier'', then one come to a serious contradiction: \emph{the influence of the potential on this wave increases strongly with $a$!}
Now a new question has appeared: what should the wave represent in general in the cosmological problem?
This problem connects with another and more general one in quantum physics --- the real importance \emph{to define a ``free'' wave inside \underline{strong} fields}. To this aim we need a mathematical stable tool to study it. It is unclear whether a connection between exact solutions for the wave function at turning point and ``free'' wave defined in the asymptotic region is correct.

{\bf Note that the semiclassical formula of the penetrability of the barrier is constructed on the basis of wave which is defined concerning zero potential at infinity, i.e. this wave should be free outgoing in the asymptotic region. But in the cosmological problem we have opposite case, when the force acting on the wave increases up to infinity in the asymptotic region. At the same time, deformations of the shape of the potential outside the barrier cannot change the penetrability calculated in the framework of the semiclassical approach (up to the second order).}
An answer to such problem can be found in non-locality of definition of the penetrability in quantum mechanics, which is reduced to minimum in the semiclassical approach (i.~e. this is so called ``error'' of the cosmological semiclassical approach).

% {\bf In order to calculate penetrability without application of the semiclassical approximation, we should define such a wave as maximal accurately as a possible.}
The problem of the correct definition of the wave in cosmology is reinforced else more, if one wants to calculate the incident and reflected waves in the internal region. \emph{Even with the known exact solution for the wave function there is uncertainty in determination of these waves!} But, namely, the standard definition of the coefficients of penetrability and reflection is based on them. In particular, we have not found papers where the coefficient of reflection is defined and estimated in this problem (which differs essentially from unity at the energy of radiation close to the height of the barrier and, therefore, such a characteristics could be interesting from a physical point of view). Note that the semiclassical approximation put serious limits to the possibility of its definition at all \citep{Landau.v3.1989}.

Thus, in order to estimate probability of the formation of the Universe as accurately as possible, we need a fully quantum definition of the wave. Note that the non-semiclassical penetrability of the barrier in the cosmological problems has not been studied in detail and, therefore, a development of fully quantum methods for its estimation is a perspective task.

Researches in this direction exist~\citep{AcacioDeBarros.2007.PRD}, and in these papers was estimated the penetrability on the basis of tunneling of wave packet through the barrier. However, a stationary boundary condition has uncertainty that could lead to different results in calculations of the penetrability. The stationary approach could allow to clarify this issue. It is able to give stable solutions for the wave function (and results in Ref.~\citep{Maydanyuk.2008.EPJC} have confirmed this at zero energy of radiation), using the standard definition of the coefficients of the penetrability and reflection, is more accurate to their estimation.

Aims of this Chapter are:
(1) to define the wave in the quantum cosmological problem;
(2) to construct the fully quantum (non-semiclassical) stationary method of determination of the coefficients of penetrability of the barriers and reflection from them on the basis of such a definition of the wave;
(3) to estimate how much the semiclassical approach differs in the estimation of the penetrability from the fully quantum one.
In order to achieve this goal, we need to construct tools for calculation of partial solutions of the wave function. In order to resolve the questions pointed out above, we shall restrict ourselves to a simple cosmological model, where the potential has a barrier and internal above-barrier region.

\section{Cosmological model in the Friedmann--Robertson--Walker metric with radiation
\label{sec.model}}

\subsection{Dynamics of Universe in the Friedmann--Robertson--Walker metric
\label{sec.model.1}}

Let us consider a simple model of the homogeneous and isotropic Universe in \emph{Friedmann--Robertson--Walker (FRW) metric} (see Ref.~\citep{Weinberg.1975}, p.~438; also see
Refs.~\citep{Rubakov.RTN2005,Linde.2005,Trodden.TASI-2003,Brandenberger.1999}):
\begin{equation}
\begin{array}{cccccc}
  ds^{2} = - dt^{2} + a^{2}(t) \cdot \biggl(\displaystyle\frac{dr^{2}}{h(r)} + r^{2} (d\theta^{2} + \sin^{2}{\theta} \, d\phi^{2}) \biggr), &
  h(r) = 1-kr^{2},
\end{array}
\label{eq.model.1.1}
\end{equation}
where $t$ and $r$, $\theta$, $\phi$ are time and space spherical coordinates,
the signature of the metric is $(-,+,+,+)$ as in Ref.~\citep{Trodden.TASI-2003} (see~p.~4),
$a(t)$ is an unknown function of time and $k$ is a constant, the value of which equals $+1$, $0$ or $-1$, with appropriate choice of units for $r$. Further, we shall use the following system of units: $\hbar = c = 1$.
For $k = -1$, 0 the space is infinite (Universe of open type), and for $k=+1$ the space is finite (Universe of closed type). For $k=1$ one can describe the space as a sphere with radius $a(t)$ embedded in a 4-dimensional Euclidian space. The function $a(t)$ is referred to as the \emph{``radius of the Universe''} and is called the \emph{cosmic scale factor}. This function contains information of the dynamics of the expansion of the Universe, and therefore its determination is an actual task.

One can find the function $a(t)$ using the Einstein equations taking into account the cosmological constant $\Lambda$ in this metric (we use the signs according to the chosen signature, as in Ref.~\citep{Trodden.TASI-2003} p.~8; the Greek symbols $\mu$ and $\nu$ denote any of the four coordinates $t$, $r$, $\theta$ and $\phi$):
\begin{equation}
  R_{\mu\nu} - \displaystyle\frac{1}{2} \, g_{\mu\nu} \, R = 8\pi \: G \, T_{\mu\nu} + \Lambda,
\label{eq.model.1.2}
\end{equation}
where $R_{\mu\nu}$ is the Ricci tensor, $R$ is the scalar curvature, $T_{\mu\nu}$ is the energy-momentum tensor, and $G$ is Newton's constant.
From (\ref{eq.model.1.1}) we find the Ricci tensor $R_{\mu\nu}$ and the scalar curvature $R$:
\begin{equation}
\begin{array}{ll}
  \vspace{2mm}
  R_{tt} = -3 \displaystyle\frac{\ddot{a}}{a}, &
  \hspace{10mm}
  R_{rr} = \displaystyle\frac{a\ddot{a}}{h} +2\displaystyle\frac{\dot{a}^{2}}{h} -\displaystyle\frac{h^{\prime}}{hr} = \displaystyle\frac{2\dot{a}^{2} + a\ddot{a} + 2k}{1-kr^{2}}, \\
  R_{\phi\phi} = R_{\theta\theta} \, \sin^{2}{\theta}, &
  \hspace{10mm}
  R_{\theta\theta} = a\ddot{a}\,r^{2} + 2\dot{a}^{2}\,r^{2} - h - \displaystyle\frac{h^{\prime}r}{2} + 1 = 2\dot{a}^{2}\,r^{2} + a\ddot{a}^{2}\,r^{2} + 2kr^{2}
\end{array}
\label{eq.model.1.3}
\end{equation}
\begin{equation}
  R = g^{tt} R_{tt} + g^{rr} R_{rr} + g^{\theta\theta} R_{\theta\theta} + g^{\phi\phi} R_{\phi\phi} =
  \displaystyle\frac{6\dot{a}^{2} + 6a\ddot{a} + 6k}{a^{2}}.
\label{eq.model.1.4}
\end{equation}

The \emph{energy-momentum tensor} has the form (see~\citep{Trodden.TASI-2003}, p.~8):
$T_{\mu\nu} = (\rho + p) \: U_{\mu} U_{\nu} + p \: g_{\mu\nu}$,
where $\rho$ and $p$ are energy density and pressure.
Here, one needs to use the normalized vector of 4-velocity $U^{t} = 1$, $U^{r} = U^{\theta} = U^{\phi} = 0$.
Substituting the previously calculated components (\ref{eq.model.1.2}) of the Ricci tensor $R_{\mu\nu}$, the scalar curvature (\ref{eq.model.1.4}), the components of the energy-momentum tensor $T_{\mu\nu}$ and including the component $\rho_{\rm rad}(a)$, describing the radiation in the initial stage (equation of state for radiation: $p(a)=\rho_{\rm rad}(a)/3$), into the Einstein's equation (\ref{eq.model.1.2}) at $\mu=\nu=0$),
we obtain the \emph{Friedmann equation} with the cosmological constant (see p.~8 in Ref.~\citep{Trodden.TASI-2003}; p.~3 in Ref.~\citep{Brandenberger.1999}; p.~2 in Ref.~\citep{Vilenkin.1995}):
\begin{equation}
\begin{array}{ll}
  \dot{a}^{2} + k -
  \displaystyle\frac{8\pi\,G}{3}\,
  \Bigl\{\displaystyle\frac{\rho_{\rm rad}}{a^{2}(t)} + \rho_{\Lambda}\,a^{2}(t) \Bigr\} = 0, &
  \hspace{5mm}
  \rho_{\Lambda} = \displaystyle\frac{\Lambda}{8\pi\,G},
\end{array}
\label{eq.model.1.5}
\end{equation}
where $\dot{a}$ is derivative $a$ at time coordinate.
From here, we write a general expression for the energy density:
\begin{equation}
  \rho\,(a) = \rho_{\Lambda} + \displaystyle\frac{\rho_{\rm rad}}{a^{4}(t)}.
\label{eq.model.1.6}
\end{equation}
%-----------------------------------------------------------------------------------------------------------------------

%-----------------------------------------------------------------------------------------------------------------------
\subsection{Action, lagrangian and quantization
\label{sec.model.2}}

We define the action as in Ref.~(\citep{Vilenkin.1995}, see~(1), p.~2):
\begin{equation}
  S = \displaystyle\int \sqrt{-g}\: \biggl( \displaystyle\frac{R}{16\pi\,G} - \rho \biggr)\; dx^{4}.
\label{eq.model.2.1}
\end{equation}
Substituting the scalar curvature (\ref{eq.model.1.4}), then integrating item at $\ddot{a}$ by parts with respect to variable $t$, we obtain the \emph{lagrangian}
(see Ref.~\citep{Vilenkin.1995}, (11), p.~4):
\begin{equation}
  \mathcal{L}\,(a,\dot{a}) =
  \displaystyle\frac{3\,a}{8\pi\,G}\:
  \biggl(-\dot{a}^{2} + k - \displaystyle\frac{8\pi\,G}{3}\; a^{2}\,\rho(a) \biggr).
\label{eq.model.2.2}
\end{equation}
Considering the variables $a$ and $\dot{a}$ as generalized coordinate and velocity respectively, we find a generalized momentum conjugate to $a$:
\begin{equation}
  p_{a} = \displaystyle\frac{\partial\, \mathcal{L}\,(a,\dot{a})}{\partial \dot{a}} =
  - \,\displaystyle\frac{3}{4\pi\,G}\; a\,\dot{a}
\label{eq.model.2.3}
\end{equation}
and then hamiltonian:
\begin{equation}
\begin{array}{ccl}
  \vspace{2mm}
  h\,(a,p_{a}) & = & p\,\dot{a} - \mathcal{L}\,(a,\dot{a}) =
  -\:\displaystyle\frac{1}{a}\;
  \biggl\{
    \displaystyle\frac{2\pi\,G}{3}\: p_{a}^{2} +
    a^{2}\,\displaystyle\frac{3\,k}{8\pi\,G} -
    a^{4}\,\rho(a) \biggr\}.
\end{array}
\label{eq.model.2.4}
\end{equation}
The passage to the quantum description of the evolution of the Universe is obtained by the standard procedure of canonical quantization in the Dirac formalism for systems with constraints. In result, we obtain the \emph{Wheeler--De Witt (WDW) equation} (see Ref.~\citep{Vilenkin.1995}, (16)--(17), in p.~4, \citep{Wheeler.1968,DeWitt.1967,Rubakov.2002.PRD}),
which {\bf can be written as}
\begin{equation}
\begin{array}{ccl}
  \biggl\{ -\:\displaystyle\frac{\partial^{2}}{\partial a^{2}} + V\,(a) \biggr\}\; \varphi(a) =
  E_{\rm rad}\; \varphi(a), &
  V\, (a) =
    \biggl( \displaystyle\frac{3}{4\pi\,G} \biggr)^{2}\: k\,a^{2} -
    \displaystyle\frac{3\,\rho_{\Lambda}}{2\pi\,G}\; a^{4}, &
  E_{\rm rad} = \displaystyle\frac{3\,\rho_{\rm rad}}{2\pi\,G},
\end{array}
\label{eq.model.2.5}
\end{equation}
where $\varphi(a)$ is wave function of Universe.
This equation looks similar to the one-dimensional stationary Schr\"{o}dinger equation on a semiaxis (of the variable $a$) at energy $E_{\rm rad}$ with potential $V\,(a)$.
It is convenient to use system of units where $8\pi\,G \equiv M_{\rm p}^{-2} = 1$, and
to rewrite $V\,(a)$ in a generalized form as
\begin{equation}
  V(a) = A\,a^{2} - B\,a^{4}.
\label{eq.model.2.6}
\end{equation}
In particular, for the Universe of the closed type ($k=1$) we obtain $A = 36$, $B = 12\,\Lambda$ (this potential coincides with Ref.~\citep{AcacioDeBarros.2007.PRD}).

%-----------------------------------------------------------------------------------------------------------------------

%-----------------------------------------------------------------------------------------------------------------------
\subsection{Potential close to the turning points: non-zero energy case
\label{sec.model.3}}

In order to find the wave function we need to know the shape of the potential close to the turning points. Let us find the \emph{turning points} $a_{\rm tp,\,in}$ and $a_{\rm tp,\,out}$
concerning the potential (\ref{eq.model.2.6}) at energy $E_{\rm rad}$:
\begin{equation}
\begin{array}{cc}
\vspace{3mm}
  a_{\rm tp,\, in} =
    \sqrt{\displaystyle\frac{A}{2B}} \cdot
    \sqrt{1 - \sqrt{1 - \displaystyle\frac{4BE_{\rm rad}}{A^{2}}}},&
  a_{\rm tp,\, out} =
    \sqrt{\displaystyle\frac{A}{2B}} \cdot
    \sqrt{1 + \sqrt{1 - \displaystyle\frac{4BE_{\rm rad}}{A^{2}}}}.
\end{array}
\label{eq.model.3.1}
\end{equation}
Let us expand the potential $V(a)$ (\ref{eq.model.3.1}) in powers of $q_{\rm out}=a-a_{\rm tp}$ (where the point $a_{\rm tp,\, in}$ or $a_{\rm tp,\, out}$ is used as $a_{\rm tp}$. Expansion is calculated at these points), where (for small $q$) we restrict ourselves to the liner term:
\begin{equation}
  V(q) = V_{0} + V_{1}\,q,
\label{eq.model.3.2}
\end{equation}
where the coefficients $V_{0}$ and $V_{1}$ are:
\begin{equation}
\begin{array}{lcl}
\vspace{1mm}
  V_{0} & = &
    V(a=a_{\rm tp,\, in}) =
    V(a=a_{\rm tp,\, out}) =
    A\, a_{\rm tp}^{2} - B\, a_{\rm tp}^{4} = E_{\rm rad}, \\
\vspace{1mm}
  V_{1}^{\rm (out)} & = &
    -\: 2\, A \cdot
    \sqrt{\displaystyle\frac{A}{2B}\:
    \biggl(1 - \displaystyle\frac{4BE_{\rm rad}}{A^{2}}\biggr)\,
    \biggl(1 + \sqrt{1 - \displaystyle\frac{4BE_{\rm rad}}{A^{2}}}\biggr)}, \\
  V_{1}^{\rm (int)} & = &
    2\,A \cdot
    \sqrt{\displaystyle\frac{A}{2B}\:
    \biggl(1 - \displaystyle\frac{4BE_{\rm rad}}{A^{2}}\biggr)\,
    \biggl(1 - \sqrt{1 - \displaystyle\frac{4BE_{\rm rad}}{A^{2}}}\biggr)}.
\end{array}
\label{eq.model.3.3}
\end{equation}
Now eq.~(\ref{eq.model.3.3}) transforms into a new form at variable $q$ with potential $V(q)$:
\begin{equation}
  -\displaystyle\frac{d^{2}}{dq^{2}}\, \varphi(q) +
  V_{1}\, q\: \varphi(q) = 0.
\label{eq.model.3.4}
\end{equation}
% *******************************************************************************************************************

% *******************************************************************************************************************
\section{Tunneling boundary condition in cosmology
\label{sec.5}}

% \subsection{A fully quantum definition of ``free'' wave inside strong fields
% \label{sec.5}}

\subsection{A problem of definition of ``free'' wave in cosmology and correction of the boundary condition
\label{sec.5.1}}

Which boundary condition should be used to obtain a wave function that describes how the wave function leaves the barrier accurately? A little variation of the boundary condition leads to change of the fluxes concerning the barrier and, as result, it changes the coefficients of penetrability and reflection. So, a proper choice of the boundary condition is extremely important. However before, let us analyze how much the choice of the boundary condition
is natural in the asymptotic region.
\begin{itemize}
\item
In description of collisions and decay in nuclear and atomic physics potentials of interactions tend to zero asymptotically. {\bf So, in these calculations, the boundary conditions are imposed on the wave function at infinity.}
In cosmology we deal with another, different type of potential: its modulus increases with increasing of the scale factor $a$. The gradient of the potential
%used with opposite sign and having a sense of force acting on the wave,
also increases. Therefore, \emph{here there is nothing common to the free propagation of the wave in the asymptotic region}. Thus, a direct passage of the application of the boundary condition in the asymptotic region into cosmological problems looks questionable.

\item
The results in Ref.~\citep{Maydanyuk.2008.EPJC}, which show that when the scale factor $a$ increases the region containing solutions for the wave function enlarges  (and its two partial solutions), reinforce the seriousness of this problem. According to~Ref.~\citep{Maydanyuk.2008.EPJC}, the scale factor $a$ in the external region is larger, the  period of oscillations of each partial solution for the wave function is \underline{smaller}. One has to decrease the time--step and as a consequence increase the calculation time. {\bf This increases errors in computer calculations of the wave function close the barrier (if it is previously fixed by the boundary condition in the asymptotic region).}
From here a natural conclusion follows on the impossibility to use practically the boundary condition at infinity for calculation of the wave (in supposition if we know it maximally accurately in the asymptotic region), if we like to pass from the semiclassical approach to the fully quantum one. Another case exists in problems of decay in nuclear and atomic physics where calculations of the wave in the asymptotic region are the most stable and accurate.

\item
One can add a fact that it has not been known yet whether the Universe expands at extremely large value of the scale factor $a$. Just the contrary, it would like to clarify this from a solution of the problem, however imposing a condition that the Universe expands in the initial stage.
\end{itemize}

On such a basis, we shall introduce the following \underline{\textbf{definition of the boundary condition:}}
\begin{quote}
\emph{\bf The boundary condition should be imposed on the wave function at such value of the scale factor $a$, where the potential minimally acts on the wave, determined by this wave function.}
\end{quote}
The propagation of the wave defined in such a way is close to free one for the potential and at used value of the scale factor $a$ (we call such a wave conditionally ``free''). However, when we want to give a mathematical formulation of this definition we have to answer two questions:

\begin{enumerate}
\item
What should the free wave represent in a field of a cosmological potential of arbitrary shape? How could it be defined in a correct way close to an arbitrary selected point?
%the most correctly in enough small neighborhood of arbitrary selected coordinate?

\item
{\bf Where should the boundary condition be imposed? }
\end{enumerate}
%-----------------------------------------------------------------------------------------------------------------------

%-----------------------------------------------------------------------------------------------------------------------
To start with, let us consider the second question namely where we must impose the boundary condition on the wave function.
{\bf One can suppose that this coordinate could be (1) a turning point (where the potential coincides with energy of radiation), or (2) a point where a gradient from the potential (having a sense of \emph{force of interaction}) becomes zero, or (3) a point where the potential becomes zero.
But the clear condition of free propagation of the wave is the minimal influence of the potential on this wave.
So, we define these coordinate and force in the following way:
\begin{quote}
\emph{The point in which we impose the boundary condition is the coordinate where the force acting on the wave is minimal. The force is defined as minus the gradient of the potential.}
\end{quote}
}
It turns out that according to such a (local) definition the force is minimal at the external turning point $a_{\rm tp,\,out}$.
%in the external region .
Also, the force, acting on the wave incident on the barrier in the internal region and on the wave reflected from it, has a minimum at the internal turning point $a_{\rm tp,\,in}$. Thus, we have just obtain the internal and external turning points where we should impose the boundary conditions in order to determine the waves.

%-----------------------------------------------------------------------------------------------------------------------

%-----------------------------------------------------------------------------------------------------------------------
\subsection{Boundary condition at $a=0$: stationary approach versus non-stationary one
\label{sec.5.4}}

A choice of the proper boundary condition imposed on the wave function is directly connected with the question: could the wave function be defined at $a=0$, and which value should it be equal to at this point in such a case?
The wave function is constructed on the basis of its two partial solutions which should be linearly independent.
In particular, these two partial solutions can be real (not complex), without any decrease of accuracy in determination of the total wave function. \emph{For any desirable boundary condition imposed on the total wave function, such methods should work.}
In order to achieve the maximal linear independence between two partial solutions, we choose one solution to be increasing in the region of tunneling and another one to be decreasing in this tunneling region. For the increasing partial solution we use as starting point $a_{x}$ the internal turning point $a_{\rm tp,\, in}$ {\bf at $E_{\rm rad} \ne 0$ or zero point $a_{x}=0$ at $E_{\rm rad}=0$.} For the second decreasing partial solution the starting point $a_{x}$ is chosen as the external turning point $a_{\rm tp,\, out}$. Such a choice of starting points turns out to give us higher accuracy in calculations of the total wave function than starting calculations of both partial solutions from zero or from only one turning point.

% So, eq.~(\ref{eq.3.1.3.1}) and furthers in Subsection~\ref{sec.5.4.1} are defined relatively arbitrary non-zero point $a_{x}$ in a general case. These equations in Appendixes~\ref{sec.3.1} and \ref{sec.3.2} do not absolutely contradict to any desirable boundary condition, and they could be extended on the case $a_{x}=0$. \emph{In latter case the methods in Appendix~B should be working at any finite value of the total wave function at $a=0$.}

In order to obtain the total wave function, we need to connect two partial solutions using one boundary condition, which should be obtained from physical motivations.
{\bf According to analysis in Introduction and previous section,
% Sect.~4.1 and 4.2 reinforced by interference between the incident and reflected waves in Sect.~5 (which is invisible in the semiclassical approach),
it is natural not to define the wave function at zero (or at infinity), but to find outgoing wave at \underline{finite} value of $a$ in the external region, where this wave corresponds to observed Universe at present time. But, in practical calculations, we shall define such a wave at point where forces minimally act on it. This is an initial condition imposed on the outgoing wave in the external region\footnote{For example, on the basis of such a boundary condition for $\alpha$-decay problem we obtain the asymptotic region where the wave function is spherical outgoing wave.}. }

Let us analyze a question: which value has the wave function {\bf at $a=0$? }
In the paper the following ideas are used:
\begin{itemize}
%\vspace{-2mm}
\item
\emph{the wave function should be continuous in the whole spatial region of its definition},

\vspace{-2mm}
\item
\emph{we have outgoing non-zero flux in the asymptotic region defined on the basis of the total wave function},

\vspace{-2mm}
\item
\emph{we consider the case when this flux is constant}.
\end{itemize}
The non-zero outgoing flux defined at arbitrary point requires the wave function to be complex and non-zero. The condition of continuity of this flux in the whole region of definition of the wave function requires this wave function to be complex and non-zero in the entire region. If we include point $a=0$ into the studied region, then we should obtain the non-zero and complex wave function also at such point. If we use the above ideas, then we cannot obtain zero wave function at $a=0$.
One can use notions of nuclear physics, field in which the study of such questions and their possible solutions have longer history then in quantum cosmology. As example, one can consider elastic scattering of particles on nucleus (where we have zero radial wave function at $r=0$, and we have no divergences), and alpha decay of nucleus (where we cannot obtain zero wave function at $r=0$).
\emph{A possible divergence of the radial wave function at zero in nuclear decay problem could be explained by existence of source at a point which creates the outgoing flux in the asymptotic region (and is the source of this flux).}
Now the picture becomes clearer: any quantum decay could be connected with source at zero. This is why the vanishing of the total wave function at $a=0$, after introduction of the wall at this point (like in Ref.~\citep{AcacioDeBarros.2007.PRD}), is not obvious and is only one of the possibilities.

If we wanted to study physics at zero $a=0$, we should come to two cases:
\begin{itemize}
\item
If we include the zero point into the region of consideration, we shall use to quantum mechanics with included sources. In such a case, the condition of constant flux is broken. But a more general integral formula of non-stationary dependence of the fluxes on probability can include possible sources and put them into frameworks of the standard quantum mechanics also (see eq.~(19.5) in Ref.~\citep{Landau.v3.1989}, p.~80). Perhaps, black hole could be another example of quantum mechanics with sources and sinks.

\item
We can consider only quantum mechanics with constant fluxes and without sources. Then we should eliminate the zero point $a=0$ from the region of our consideration. In this way, the formalism proposed in this paper works and is able to calculate the penetrability and reflection coefficients without any lost of accuracy.
\end{itemize}
This could be a \underline{stationary} picture of interdependence between the wave function at zero and the outgoing flux in the asymptotic region. In order to study the non-stationary case, then we need initial conditions which should define also the  evolution of the Universe.
%from the problem of possible singularity of the wave function at zero and sources .
In such a case, after defining the initial state (through set of parameters) it is possible to connect zero value of wave packet at $a=0$ (i.~e. without singularity at such a point) with non-zero outgoing flux in the asymptotic region. In such direction different proposals have been made in frameworks of semiclassical models in order to describe inflation, to introduce time or to analyze dynamics of studied quantum system (for example, see \citep{Finelli.1998.PRD,Tronconi.2003.PRD}).

\section{Direct fully quantum method
\label{sec.6}}

\subsection{Wave function of Universe: calculations and analysis
\label{sec.6.1}}

% \subsubsection{Behavior of the wave function and its partial solutions
% \label{sec.6.1.1}}

The wave function is known to oscillate above the barrier and increase (or decrease) under the barrier without any oscillations. So, in order to provide an effective linear independence between two partial solutions for the wave function, we look for a first partial solution increasing in the region of tunneling and a second one decreasing in this tunneling region. To start with, we define each partial solution and its derivative at a selected starting point, and then we calculate them in the region close enough to this point using the \emph{method of beginning of the solution} presented in Subsection~\ref{sec.5.4.1}. Here, for the partial solution which increases in the barrier region, as starting point we use the internal turning point $a_{\rm tp,\, in}$ at non-zero energy $E_{\rm rad}$ or equals to zero $a=0$ at null energy $E_{\rm rad}$, and for the second partial solution, which decreases in the barrier region, we choose the starting point to be equal to the external turning point $a_{\rm tp,\, out}$. Then we calculate both partial solutions and their derivatives in the whole required range of $a$ using the \emph{method of continuation of the solution} presented in Subsection~\ref{sec.5.4.2}, which is improvement of the Numerov method with constant step. In this way, we obtain two partial solutions for the wave function and their derivatives in the whole studied region.

\begin{figure}[h]
\centerline{
\includegraphics[width=47mm]{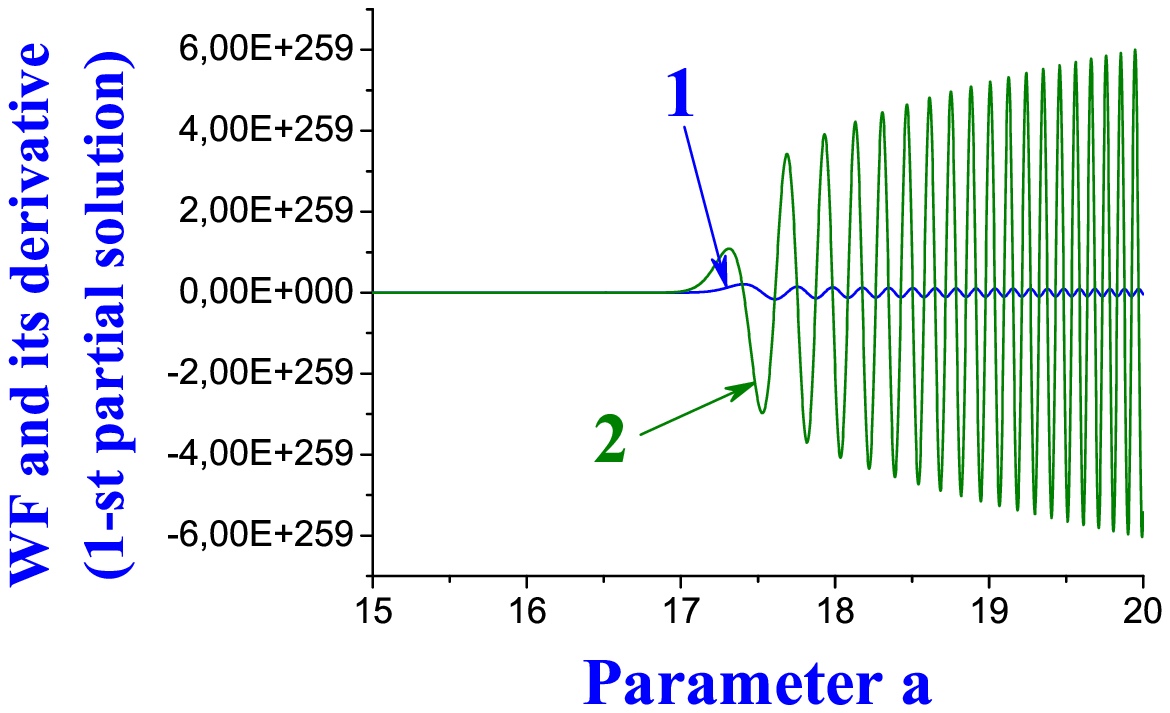}
\hspace{-3mm}\includegraphics[width=47mm]{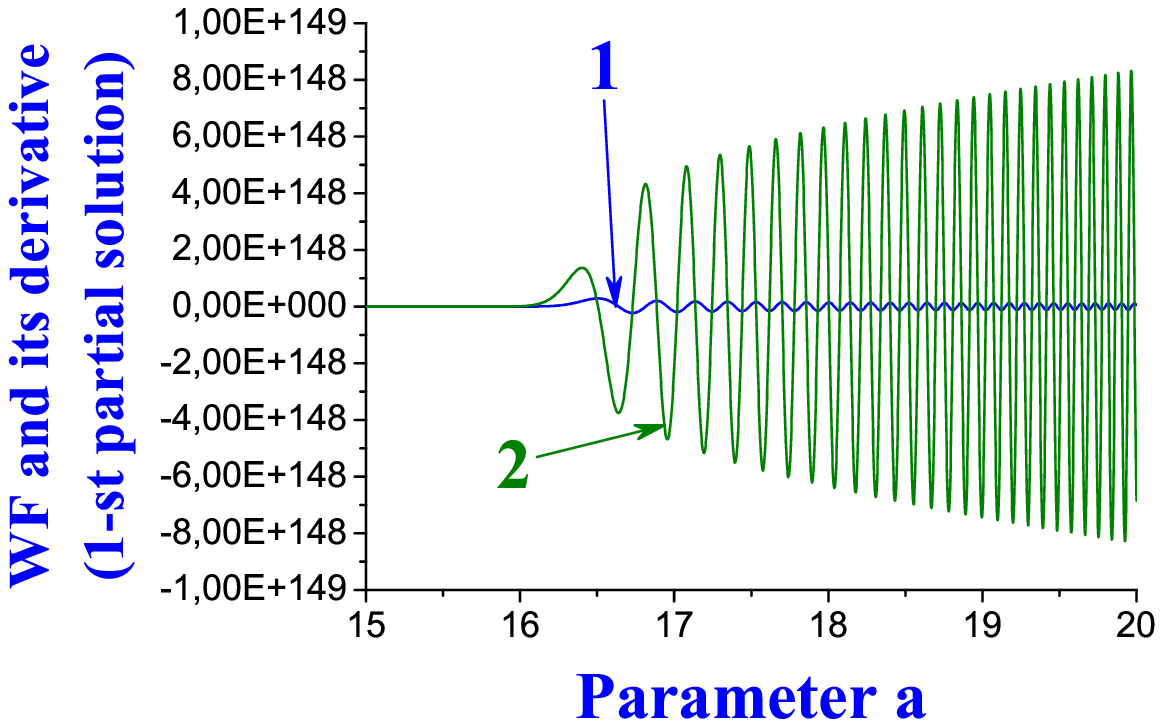}
\hspace{-3mm}\includegraphics[width=47mm]{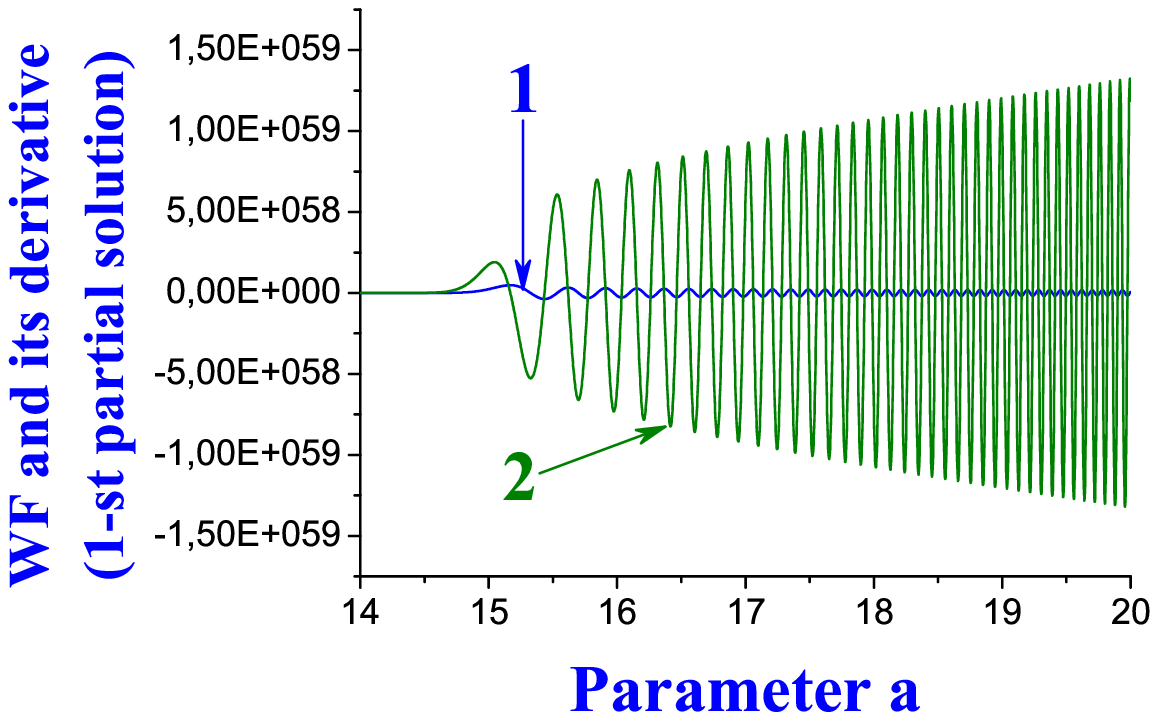}}
\caption{\small The first partial solution for the wave function and its derivative at different values of the energy of radiation $E_{\rm rad}$, increasing in the tunneling region. The blue plot represents the wave function; the green one, the derivative of this wave function):
(a) $E_{\rm rad}=10$; (b) $E_{\rm rad}=1000$; (c) $E_{\rm rad}=2000$
\label{fig.1}}
\end{figure}

\begin{figure}[h]
\centerline{
\includegraphics[width=47mm]{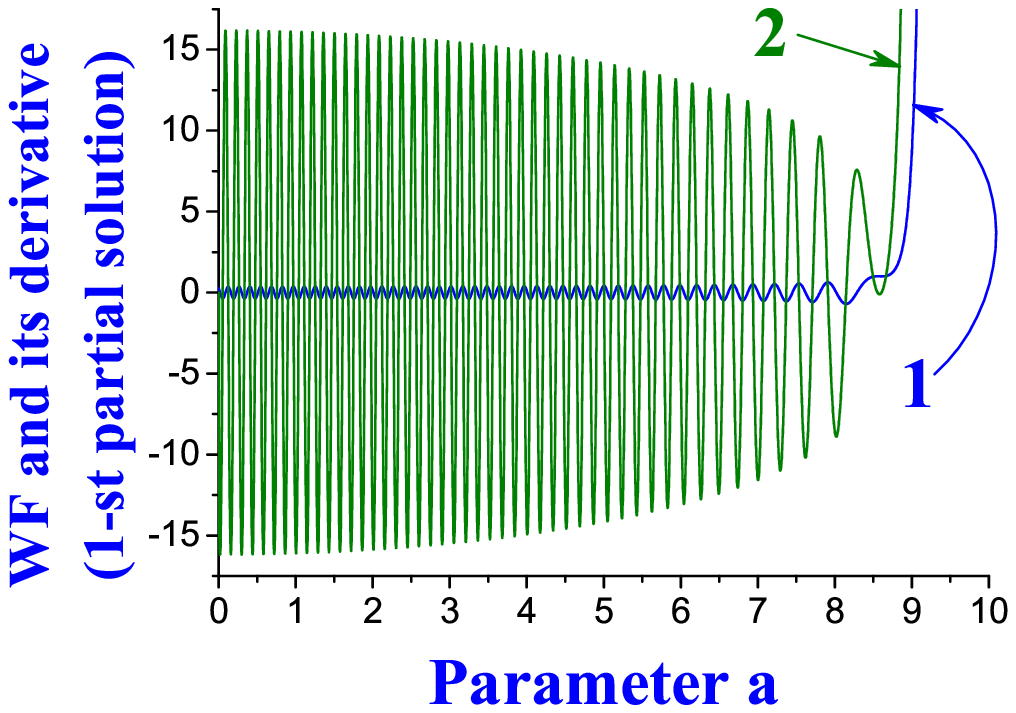}
\hspace{-3mm}\includegraphics[width=47mm]{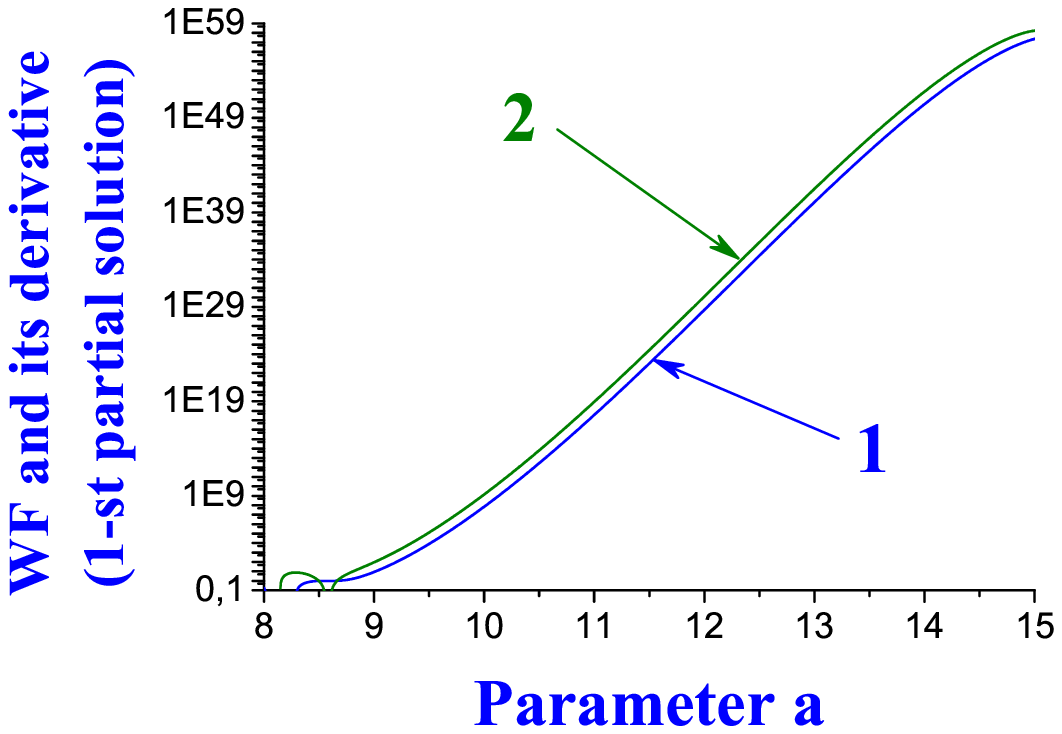}
\hspace{-3mm}\includegraphics[width=47mm]{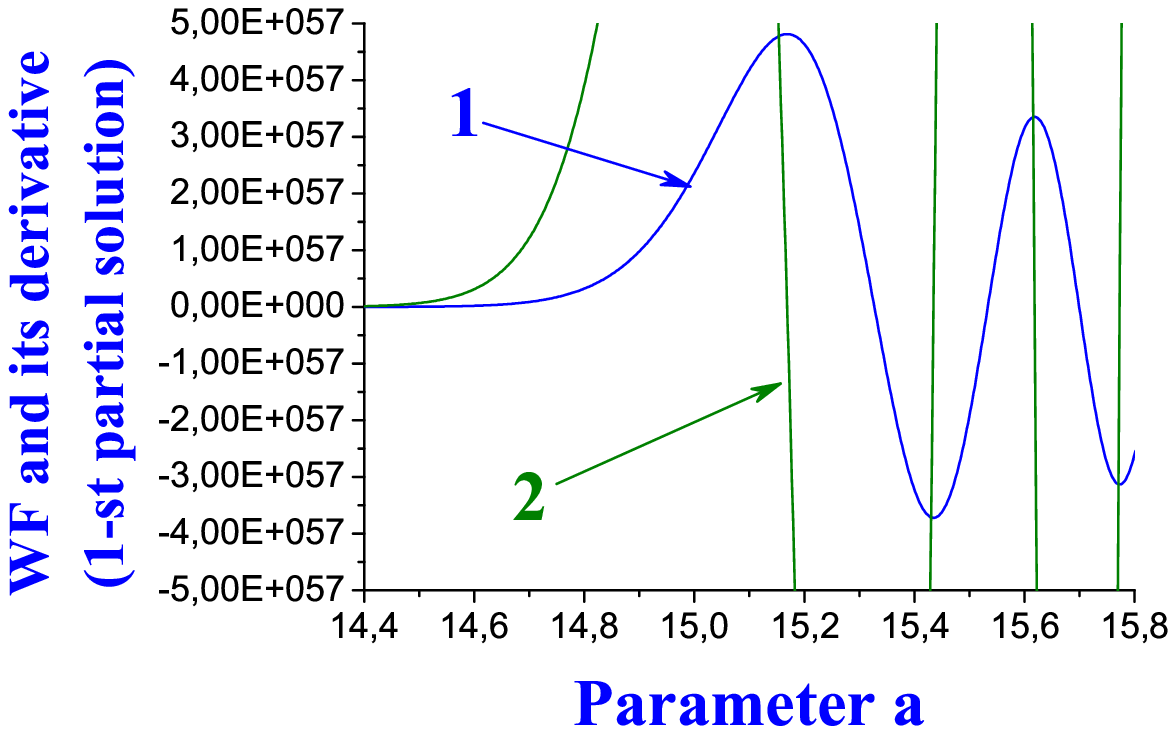}}
\caption{\small The first partial solution for the wave function and its derivative at the energy of radiation $E_{\rm rad}=2000$. The blue line represents the wave function; the green one, the derivative of this wave function)
\label{fig.2}}
\end{figure}

In order to clarify how the proposed approach gives convergent (stable) solutions, we compare our results with the paper of ~\citep{AcacioDeBarros.2007.PRD}.
%with the published results of calculations of the wave function.
Let us consider the behavior of the wave function. The first partial solution for the wave function and its derivative in my calculation are presented in Fig.~\ref{fig.1}, which increase in the tunneling region and have been obtained at different values of the energy of radiation $E_{\rm rad}$.
From these figures one can see that the wave function satisfies the rules satisfied by the wave function inside the sub-barrier and in above-barrier regions \citep{Zakhariev.1990.PEPAN}. Starting from very small $a$, the wave function has oscillations and its maxima increase monotonously with increasing of $a$. This corresponds to the behavior of the wave function in the internal region before the barrier (this becomes more obvious after essential increasing of scale, see left panel in Fig.~\ref{fig.2}). Moreover, for larger values of $a$, the wave function increases monotonously without any oscillation, that points out on transition into the tunneling region (one can see this in a logarithmic presentation of the wave function, see central panel in Fig.~\ref{fig.2}). A boundary of such a transformation in behavior of the wave function must be the point of penetration of the wave into the barrier, i.~e. the internal turning point  $a_{\rm tp,\,in}$. Further, with increasing of $a$ the oscillations are appeared in the wave function, which could be possible inside the above barrier region only (in the right panel of Fig.~\ref{fig.2} one can see that such a transition is extremely smooth that characterizes the accuracy of the method positively).
{\bf The boundary of such a transformation in the behavior of the wave function should be the external turning point $a_{\rm tp,\,out}$. Like Ref.~\citep{Maydanyuk.2008.EPJC}, but at arbitrary non-zero energy $E_{\rm rad}$ we obtain monotonous increasing of maximums of the derivative of the wave function and smooth decreasing of this wave function in the external region. One can see that the derivative is larger than the wave function. At large values of $a$ we obtain the smooth continuous solutions
up to $a=100$ (in Ref.~\citep{AcacioDeBarros.2007.PRD} the maximal presented limit is $a=30$). }

\begin{figure}[h]
\centerline{
\includegraphics[width=47mm]{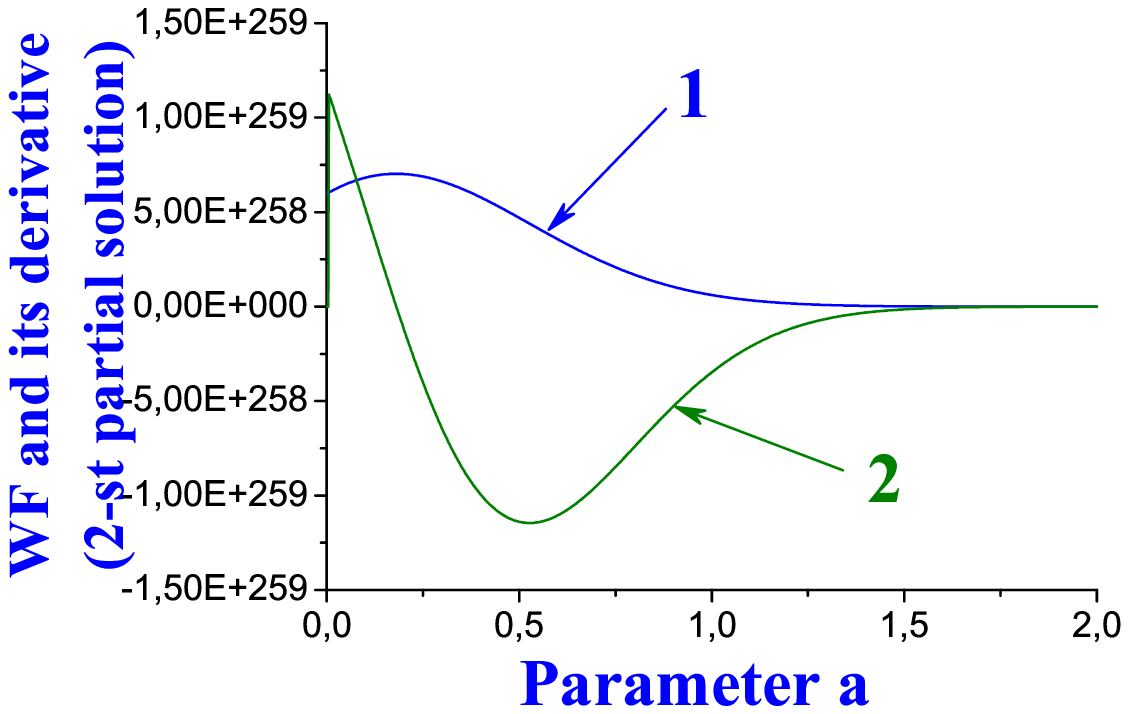}
\hspace{-3mm}\includegraphics[width=47mm]{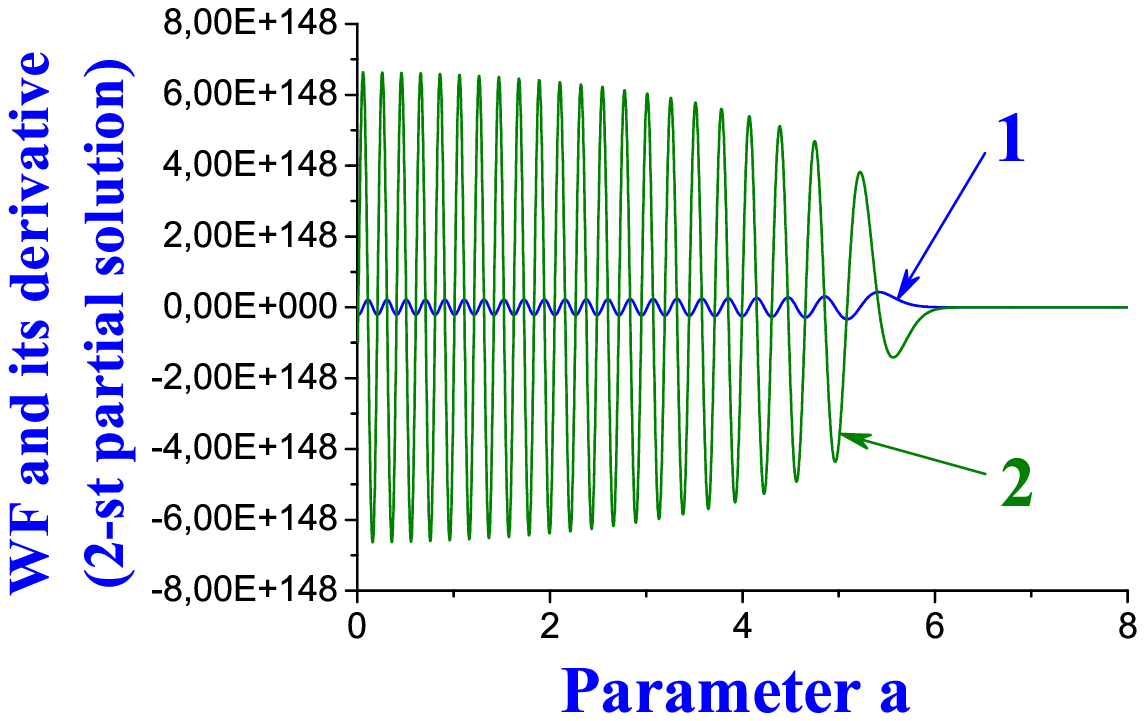}
\hspace{-3mm}\includegraphics[width=47mm]{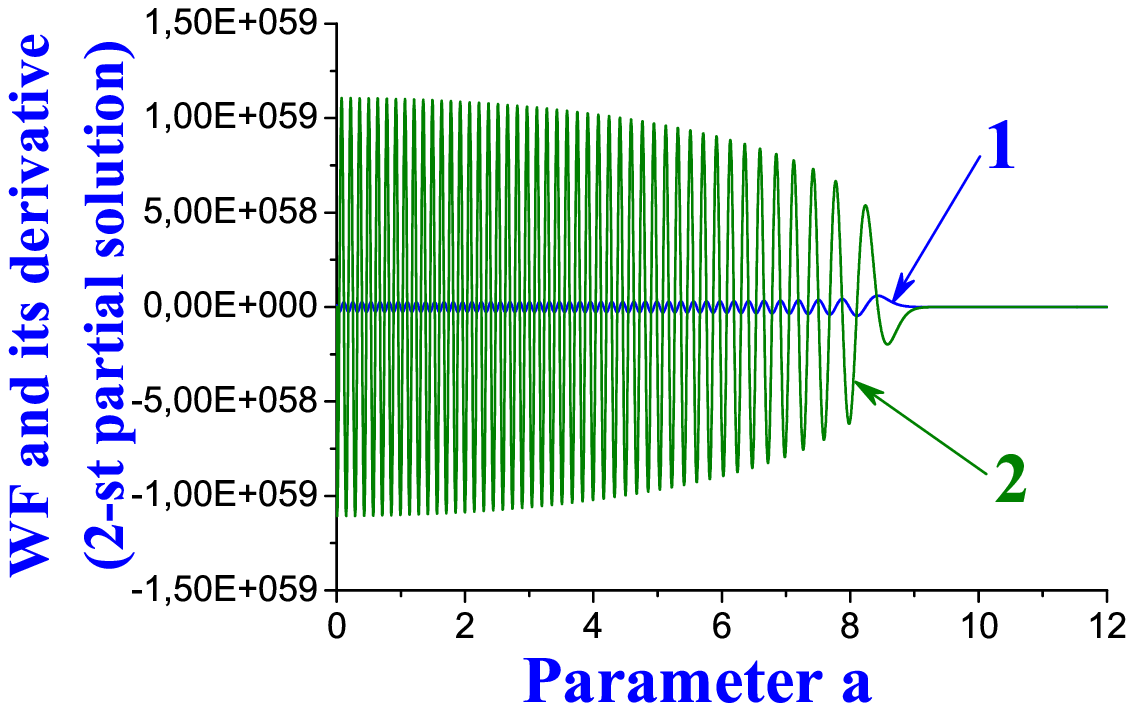}}
\caption{\small The second partial solution for the wave function and its derivative at different values of the energy of radiation $E_{\rm rad}$, decreasing in the tunneling region (curve 1, blue, is for the wave function; curve 2, green, for the derivative of this wave function):
(a) $E_{\rm rad}=10$; (b) $E_{\rm rad}=1000$; (c) $E_{\rm rad}=2000$
\label{fig.3}}
\end{figure}

In Fig.~\ref{fig.3}, it is presented the second partial solution of the wave function  and its derivative at different values of the energy of radiation $E_{\rm rad}$
%, which decrease in the region of tunneling.
%
According to the analysis, this solution close to the turning points, in the tunneling region, in the sub-barrier and above-barrier regions looks like the first partial solution, but with the difference that now the maxima of the wave function and their derivatives are larger essentially in the external region in a comparison with the internal region, and amplitudes in the tunneling region decrease monotonously.

Comparing the previous pictures of the wave function with the results of Ref.~\citep{AcacioDeBarros.2007.PRD}, one can see that the wave function, in this approach, is essentially more continuous, has no divergencies and its behavior is everywhere clear. From here we conclude that \emph{
the developed method for the determination of the wave function and its derivative at arbitrary energy of radiation is essentially more quick, more stable and accurate in  comparison with the non-stationary quantum approach in Ref.~\citep{AcacioDeBarros.2007.PRD}}.
Note that:
\begin{itemize}

\item
With increasing $a$, the period of the oscillations, both for the wave function and its derivative, decreases uniformly in the external region and increases uniformly in the internal region (this result was partially obtained earlier in Ref.~\citep{Maydanyuk.2008.EPJC} at $E_{\rm rad}=0$).

\item
At larger distance from the barrier (i.~e. for increasing values of $a$, in the external region, and at decreasing value of $a$, in the internal region) it becomes \underline{more difficult} to get the convergent continuous solutions for the wave function and its derivative (this result was partially obtained earlier in Ref.~\citep{Maydanyuk.2008.EPJC} at $E_{\rm rad}=0$).

\item
\emph{A number of oscillations of the wave function in the internal region increases with increasing of the energy of radiation $E_{\rm rad}$} (this is a new result).
\end{itemize}
%-----------------------------------------------------------------------------------------------------------------------

% -----------------------------------------------------------------------------------------------------------------------
\subsection{Definition of the wave minimally interacting with the potential
\label{sec.6.2}}

Now we shall be looking for a form of the wave function in the external region, which describes accurately the wave, whose propagation is the closest to the ``free'' one in the external region at the turning point $a_{\rm tp,\, out}$ and is directed outside. Let us return back to eq.~(\ref{eq.model.3.4}) where the variable $q = a - a_{\rm tp,\, out}$ has been introduced.
Changing this variable to
\begin{equation}
  \xi = \bigl|V_{1}^{\rm (out)}\bigr|^{1/3} q,
\label{eq.6.2.1}
\end{equation}
this equation is transformed into
\begin{equation}
  \displaystyle\frac{d^{2}}{d\xi^{2}}\, \varphi(\xi) + \xi \: \varphi(\xi) = 0.
\label{eq.6.2.2}
\end{equation}

From quantum mechanics we know two linearly independent exact solutions for the function $\varphi(\xi)$ in this equation --- these are the \emph{Airy functions} ${\rm Ai}\,(\xi)$ and ${\rm Bi}\,(\xi)$ (see Ref.~\citep{Abramowitz.1964}, p.~264--272, 291--294). Expansions of these functions into power series at small $\xi$, their asymptotic expansions at large $|\xi|$, their representations through Bessel functions, zeroes and their asymptotic expansions are known. We have some integrals of these functions, and also the form of the Airy functions in the semiclassical approximation (which can be applied at large $|\xi|$).
In some problems of the analysis of finite solutions $\varphi(\xi)$ in the whole range of $\xi$ it is convenient to use the integral representations of the Airy functions (see eq.~(10.4.32) in Ref.~\citep{Abramowitz.1964}, p.~265. In eq.~(10.4.1) we took into account the sign and $a=1/3$):
\begin{equation}
\begin{array}{ccl}
  {\rm Ai} \, (\pm\xi) & = &
  \displaystyle\frac{1}{\pi} \displaystyle\int\limits_{0}^{+\infty}
  \cos{\biggl(\displaystyle\frac{u^{3}}{3} \mp \xi u \biggr)} \; du, \\

  {\rm Bi}\, (\pm\xi) & = &
  \displaystyle\frac{1}{\pi} \displaystyle\int\limits_{0}^{+\infty}
  \biggl[
    \exp{\biggl(-\displaystyle\frac{u^{3}}{3} \mp\xi u \biggr)} +
    \sin{\biggl(\displaystyle\frac{u^{3}}{3} \mp\xi u \biggr)}
  \biggr] \; du.
\end{array}
\label{eq.6.2.3}
\end{equation}
Furthermore, we shall be interested in the solution
%of the function
$\varphi(\xi)$ which describes the \emph{outgoing wave} in the range of $a$ close to the $a_{tp}$ point. However, it is not clear what the wave represents in general near the point $a_{tp}$,
%in the potential studied,
and which linear combination of the ${\rm Ai}\,(\xi)$ and ${\rm Bi}\,(\xi)$ functions defines it in the most accurate way.

The clearest and most natural understanding of the outgoing wave is given by the semiclassical consideration of the tunneling process. However, at the given potential the semiclassical approach allows us to define the outgoing wave in the asymptotic region only (while we can join solutions in the proximity of $a_{tp}$ by the Airy functions). But it is not clear whether the wave, defined in the asymptotic region, remains outgoing near the $a_{tp}$. During the whole path of propagation outside the barrier the wave interacts with the potential, and this must inevitably lead to a deformation of its shape (like to appearance of a phase shift in the scattering of a wave by a radial potential caused by interaction in scattering theory).
{\bf Does the cosmological potentials deform the wave more than the potentials used for description of nuclear collisions in scattering theory?}
Moreover, for the given potential there is a problem in obtaining the convergence in the calculation of the partial solutions for the wave function in the asymptotic region. According to our calculations, a small change of the range of the definition of the wave in the asymptotic region leads to a significant increase of errors, which requires one to increase the accuracy of the calculations.
Therefore, we shall be looking for a way of defining the outgoing wave not in the asymptotic region, but in the closest vicinity of the point of escape, $a_{tp}$. In a search of solutions close to the point $a_{tp}$, i.~e. at small enough $|\xi|$, the validity of the semiclassical method breaks down as $|\xi|$ approaches zero. Therefore, we shall not use the semiclassical approach in this paper.

Assuming the potential $V(a)$ to have an arbitrary form, we define the wave at the point $a_{tp}$ in the following way.

\vspace{7mm}
\begin{dfn}[strict definition of the wave]
\label{def.wave.strict}
The wave is a linear combination of two partial solutions of the wave function such that the change of the modulus $\rho$ of this wave function is approximately constant under variation of $a$:
\begin{equation}
  \displaystyle\frac{d^{2}}{da^{2}}\, \rho(a) \biggl|_{a=a_{tp}} \to 0.
\label{eq.6.2.4}
\end{equation}
\end{dfn}
According to this definition, the real and imaginary parts of the total wave function have the closest behaviors under the same variation of $a$, and the difference between possible maximums and minimums of the modulus of the total wave function is the smallest. For some types of potentials (in particular, for a rectangular barrier) it is more convenient to define the wave less strongly.

\vspace{7mm}
\noindent
\underline{\textbf{Definition 2 (weak definition of wave):}}
\begin{quote}
\emph{The wave is a linear combination of two partial solutions of wave function such that the modulus $\rho$ changes minimally under variation of $a$:}
\begin{equation}
  \displaystyle\frac{d}{da}\, \rho(a) \biggl|_{a=a_{tp}} \to 0.
\label{eq.6.2.5}
\end{equation}
\end{quote}
According to this definition, the change of the wave function caused by variation of $a$ is characterized mainly by its phase (which can characterize the interaction between the wave and the potential).

Subject to this requirement, we shall look for a solution
%for the function $\varphi(\xi)$
in the following form:
\begin{equation}
  \varphi\, (\xi) = T \cdot \Psi^{(+)}(\xi),
\label{eq.6.2.6}
\end{equation}
where
\begin{equation}
\begin{array}{ccl}
  \Psi^{(\pm)} (\xi) & = &
    \displaystyle\int\limits_{0}^{u_{\rm max}}
    \exp{\pm\,i\,\Bigl(-\displaystyle\frac{u^{3}}{3} + f(\xi)\,u \Bigr)} \; du.
\end{array}
\label{eq.6.2.7}
\end{equation}
where $T$ is an unknown normalization factor, $f(\xi)$ is an unknown continuous function satisfying $f(\xi) \to {\rm const}$ at $\xi \to 0$, and $u_{\rm max}$ is the unknown upper limit of integration. In such a solution, the real part of the function $f(\xi)$ gives a contribution to the phase of the integrand function, while the imaginary part of $f(\xi)$ deforms its modulus.

Let us find the first and second derivatives of the function $\Psi(\xi)$ (a prime denotes a derivative with respect to $\xi$):
\begin{equation}
\begin{array}{ccl}
  \displaystyle\frac{d}{d\xi}\, \Psi^{(\pm)} (\xi) & = &
    \pm\, we \displaystyle\int\limits_{0}^{u_{\rm max}} f^{\prime} u \;
    \exp{\pm i\, \Bigl(-\displaystyle\frac{u^{3}}{3} + f(\xi) u \Bigr)} \; du, \\
  \displaystyle\frac{d^{2}}{d\xi^{2}}\, \Psi^{(\pm)} (\xi) & = &
    \displaystyle\int\limits_{0}^{u_{\rm max}}
    \Bigl(\pm\, if^{\prime\prime} u - (f^{\prime})^{2} u^{2} \Bigr) \:
    \exp{\pm i\, \Bigl(-\displaystyle\frac{u^{3}}{3} + f(\xi) u \Bigr)} \; du.
\end{array}
\label{eq.6.2.8}
\end{equation}
From this we obtain:
\begin{equation}
\begin{array}{c}
  \displaystyle\frac{d^{2}}{d\xi^{2}}\, \Psi^{(\pm)} (\xi) + \xi\: \Psi^{(\pm)} (\xi) =
  \displaystyle\int\limits_{0}^{u_{\rm max}}
    \Bigl(\pm\, if^{\prime\prime} u - (f^{\prime})^{2} u^{2} + \xi \Bigr) \:
    \exp{\pm i\, \Bigl(-\displaystyle\frac{u^{3}}{3} + f(\xi) u \Bigr)} \; du.
\end{array}
\label{eq.6.2.9}
\end{equation}

Considering the solutions at small enough values of $|\xi|$, we represent $f(\xi)$ in the form of a power series:
\begin{equation}
  f(\xi) = \sum\limits_{n=0}^{+\infty} f_{n}\, \xi^{n},
\label{eq.6.2.10}
\end{equation}
where $f_{n}$ are constant coefficients. The first and second derivatives of $f(\xi)$ are
\begin{equation}
\begin{array}{l}
  f^{\prime}(\xi) = \displaystyle\frac{d}{d\xi}\, f(\xi) =
    \sum\limits_{n=1}^{+\infty} n f_{n} \; \xi^{n-1} = \sum\limits_{n=0}^{+\infty} (n+1) \: f_{n+1} \; \xi^{n}, \\
  f^{\prime\prime}(\xi) = \displaystyle\frac{d^{2}}{d\xi^{2}}\, f(\xi) =
    \sum\limits_{n=0}^{+\infty} (n+1)\,(n+2) \: f_{n+2} \; \xi^{n}.
\end{array}
\label{eq.6.2.11}
\end{equation}
Substituting these solutions into eq.~(\ref{eq.6.2.8}), we obtain
\begin{equation}
\begin{array}{c}
  \displaystyle\frac{d^{2}}{d\xi^{2}}\, \Psi^{(\pm)} (\xi) + \xi\: \Psi^{(\pm)} (\xi) =
  \displaystyle\int\limits_{0}^{u_{\rm max}}
    \Biggl\{
      \Bigl(\pm\,2iu \: f_{2} - u^{2} \: f_{1}^{2} \Bigr) + \\
    +  \Bigl(\pm\,6iu \: f_{3} - 4 u^{2} \: f_{1}f_{2} + 1 \Bigr) \: \xi +
    + \sum\limits_{n=2}^{+\infty} \Bigl[\pm\,iu \: (n+1)(n+2) \: f_{n+2} - \\
    - u^{2} \sum\limits_{m=0}^{n} (n-m+1)(m+1) \: f_{n-m+1}f_{m+1} \Bigr] \: \xi^{n}
    \Biggr\} \:
    \exp{\pm\,i \Bigl(-\displaystyle\frac{u^{3}}{3} + fu \Bigr)} \; du.
\end{array}
\label{eq.6.2.12}
\end{equation}

Considering this expression at small $|\xi|$, we use the following approximation:
\begin{equation}
\begin{array}{ccc}
  \exp{\pm\,i \Bigl(-\displaystyle\frac{u^{3}}{3} + fu \Bigr)} & \to &
  \exp{\pm\,i \Bigl(-\displaystyle\frac{u^{3}}{3} + f_{0}u \Bigr)}.
\end{array}
\label{eq.6.2.13}
\end{equation}
Then from eq.~(\ref{eq.6.2.2}) we obtain the following condition for the unknown $f_{n}$:
\begin{equation}
\begin{array}{c}
  \displaystyle\int\limits_{0}^{u_{\rm max}}
    \Bigl(\pm\,2iu \: f_{2} - u^{2} \: f_{1}^{2} \Bigr)
    \exp{\pm i\, \Bigl(-\displaystyle\frac{u^{3}}{3} + f_{0}u \Bigr)} \; du \; + \\

  + \;
    \xi \cdot \displaystyle\int\limits_{0}^{u_{\rm max}}
    \Bigl(\pm\,6iu \: f_{3} - 4 u^{2} \: f_{1}f_{2} + 1 \Bigr) \:
    \exp{\pm i\, \Bigl(-\displaystyle\frac{u^{3}}{3} + f_{0}u \Bigr)} \; du \; + \\

  + \;
    \sum\limits_{n=2}^{+\infty}  \xi^{n} \cdot
    \displaystyle\int\limits_{0}^{u_{\rm max}}
    \Bigl[\pm\, iu \: (n+1)(n+2) \: f_{n+2} -
    u^{2} \sum\limits_{m=0}^{n} (n-m+1)(m+1) \: f_{n-m+1}f_{m+1} \Bigr] \: \times \\
  \times \;
    \exp{\pm i\, \Bigl(-\displaystyle\frac{u^{3}}{3} + f_{0}u \Bigr)} \; du \; = 0.
\end{array}
\label{eq.6.2.14}
\end{equation}
Requiring that this condition is satisfied for different $\xi$ and with different powers $n$, we obtain the following system:
\begin{equation}
\begin{array}{cc}
\xi^{0}: &
  \displaystyle\int\limits_{0}^{u_{\rm max}}
    \Bigl(\pm\,2iu \: f_{2} - u^{2} \: f_{1}^{2} \Bigr) \:
    \exp{\pm i\, \Bigl(-\displaystyle\frac{u^{3}}{3} + f_{0}u \Bigr)} \; du = 0, \\

\xi^{1}: &
  \displaystyle\int\limits_{0}^{u_{\rm max}}
      \Bigl(\pm\, 6iu \: f_{3} - 4 u^{2} \: f_{1}f_{2} + 1 \Bigr) \:
    \exp{\pm i\, \Bigl(-\displaystyle\frac{u^{3}}{3} + f_{0}u \Bigr)} \; du = 0, \\

\xi^{n}: &
  \displaystyle\int\limits_{0}^{u_{\rm max}}
    \Bigl[\pm\, iu \: (n+1)(n+2) \: f_{n+2} -
    u^{2} \sum\limits_{m=0}^{n} (n-m+1)(m+1) \: f_{n-m+1}f_{m+1} \Bigr] \: \times \\
    & \times \;
    \exp{\pm i\, \Bigl(-\displaystyle\frac{u^{3}}{3} + f_{0}u \Bigr)} \; du = 0.
\end{array}
\label{eq.6.2.15}
\end{equation}

Assuming that the coefficients $f_{0}$ and $f_{1}$ are known, we find the following solutions for the unknown $f_{2}$, $f_{3}$ and $f_{n}$:
\begin{equation}
\begin{array}{ccc}
  f_{2}^{(\pm)} = \pm\;\displaystyle\frac{f_{1}^{2}}{2i} \cdot \displaystyle\frac{J_{2}^{(\pm)}}{J_{1}^{(\pm)}}, &
  f_{3}^{(\pm)} =
    \pm\;\displaystyle\frac{4 f_{1}f_{2}^{(\pm)}\, J_{2}^{(\pm)} - J_{0}^{(\pm)}} {6i\, J_{1}^{(\pm)}},  \\
\end{array}
\label{eq.6.2.16}
\end{equation}
\begin{equation}
\begin{array}{ccc}
  f_{n+2}^{(\pm)} =
    \displaystyle\frac{\sum\limits_{m=0}^{n} (n-m+1)(m+1) \: f_{n-m+1}^{(\pm)}\,f_{m+1}^{(\pm)}}
      {i \: (n+1)(n+2)} \cdot
    \displaystyle\frac{J_{2}^{(\pm)}} {J_{1}^{(\pm)}},
\end{array}
\label{eq.6.2.16a}
\end{equation}
where the following notations for the integrals have been introduced:
\begin{equation}
\begin{array}{ccc}
  J_{0}^{(\pm)} =
    \displaystyle\int\limits_{0}^{u_{\rm max}}
    \exp{\pm i\, \Bigl(-\displaystyle\frac{u^{3}}{3} + f_{0}u \Bigr)} \; du, &
  J_{1}^{(\pm)} =
    \displaystyle\int\limits_{0}^{u_{\rm max}}
    u \: \exp{\pm i\, \Bigl(-\displaystyle\frac{u^{3}}{3} + f_{0}u \Bigr)} \; du,
\end{array}
\label{eq.6.2.17}
\end{equation}
\begin{equation}
\begin{array}{ccc}
  J_{2}^{(\pm)} =
    \displaystyle\int\limits_{0}^{u_{\rm max}}
    u^{2} \: e^{\pm i\, \Bigl(-\displaystyle\frac{u^{3}}{3} + f_{0}u \Bigr)} \; du.
\end{array}
\label{eq.6.2.17a}
\end{equation}
Thus, we see that the solution (\ref{eq.6.2.6}) taking into account eq.~(\ref{eq.6.2.7}) for the function $\varphi\,(\xi)$ has arbitrariness in the choice of the unknown coefficients $f_{0}$, $f_{1}$ and the upper limit of integration, $u_{\rm max}$. However, the solutions found, eqs.~(\ref{eq.6.2.16}), define the function $f(\xi)$ so as to ensure that the equality (\ref{eq.6.2.6}) is \underline{exactly} satisfied in the region of $a$ close to the escape point $a_{tp}$.
This proves that \emph{the function $\varphi\,(\xi)$ in the form (\ref{eq.6.2.6}), taking into account eq.~(\ref{eq.6.2.7}) for an arbitrary choice of $f_{0}$, $f_{1}$ and $u_{\rm max}$ is the \underline{exact} solution of the Schr\"{o}dinger equation near the escape point $a_{tp}$}. In order to write the solution $\Psi(\xi)$ in terms of the well-known Airy functions, ${\rm Ai}\,(\xi)$ and ${\rm Bi}\,(\xi)$,
we choose
\begin{equation}
\begin{array}{cc}
  f_{0} = 0, &
  f_{1} = 1.
\end{array}
\label{eq.6.2.18}
\end{equation}
For such a choice of the coefficients $f_{0}$ and $f_{1}$, the integrand function in the solution (\ref{eq.6.2.7}) (up to $\xi^{2}$) has a constant modulus and a varying phase. Therefore, one can expect that the solution (\ref{eq.6.2.6}) at the turning point $a_{tp}$ describes the wave accurately.

%-----------------------------------------------------------------------------------------------------------------------

%-----------------------------------------------------------------------------------------------------------------------
\subsection{Total wave function
\label{sec.5.3}}

\begin{figure}[h]
\centerline{
\includegraphics[width=47mm]{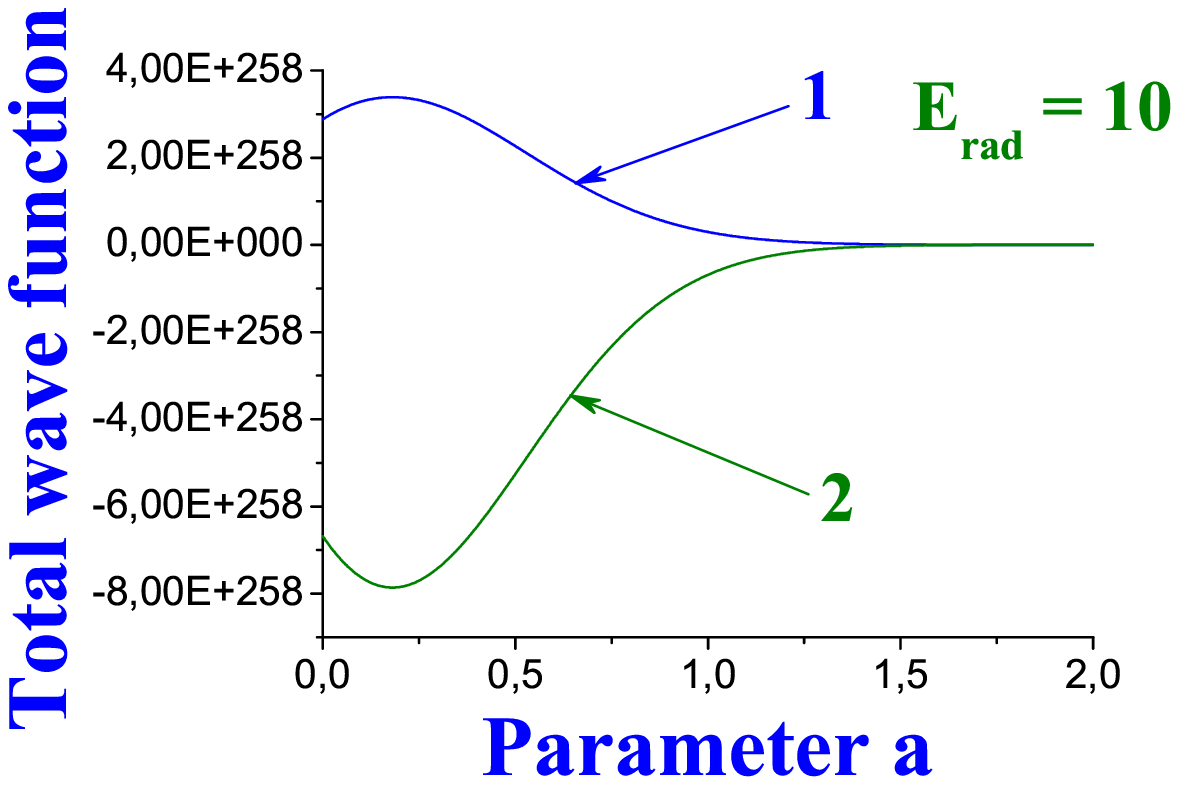}
\hspace{-3mm}\includegraphics[width=47mm]{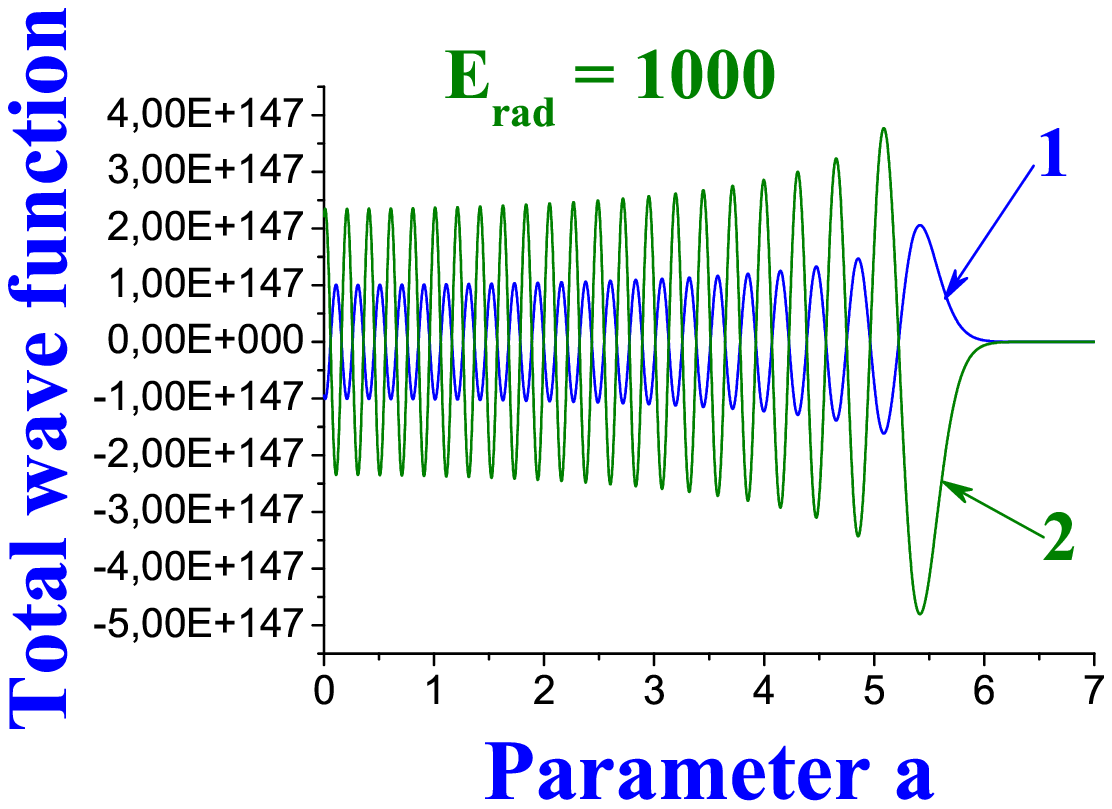}
\hspace{-3mm}\includegraphics[width=47mm]{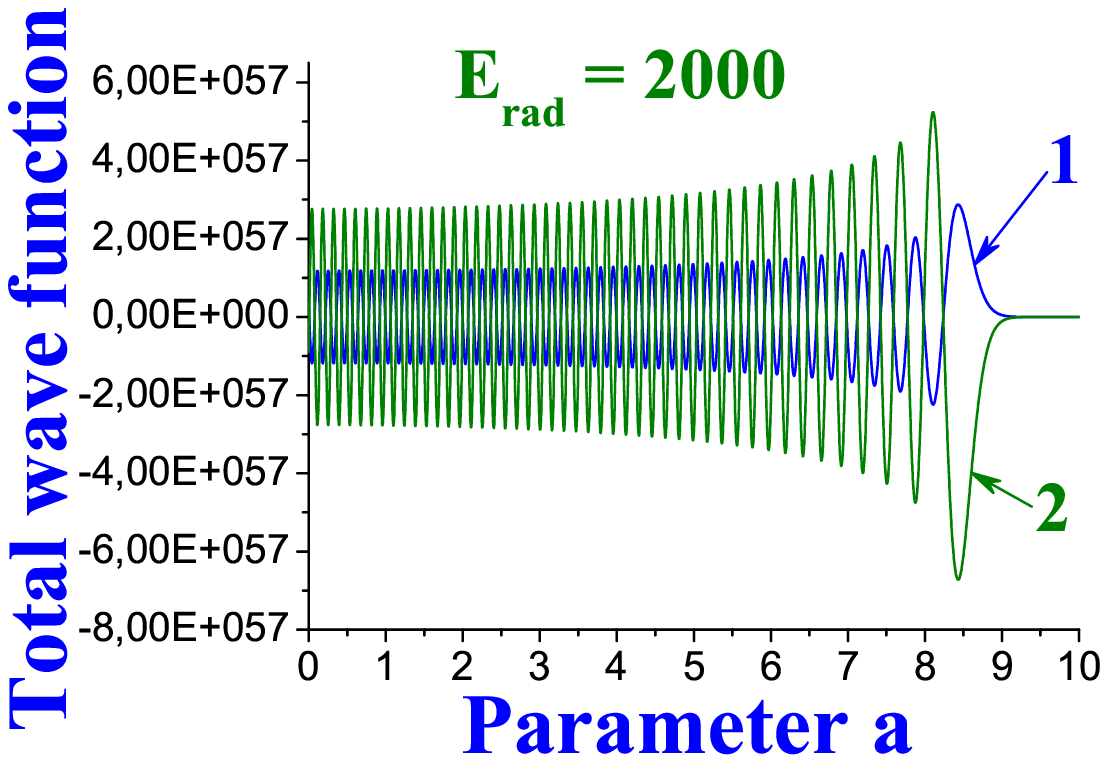}}
\caption{\small The wave function at selected values of the energy of radiation $E_{\rm rad}$ (the blue line, represents the real part of the wave function; the green line the imaginary part of the wave function):
(a) $E_{\rm rad}=10$; (b) $E_{\rm rad}=1000$; (c) $E_{\rm rad}=2000$
\label{fig.4}}
\end{figure}

Having obtained two linearly independent partial solutions $\varphi_{1}(a)$ and $\varphi_{2}(a)$,
we can write the general solution (a prime is for the derivative with respect to $a$) as:
\begin{equation}
  \varphi\,(a) = T \cdot \bigl(C_{1}\, \varphi_{1}(a) + C_{2}\,\varphi_{2}(a) \bigr),
\label{eq.6.3.1}
\end{equation}
\begin{equation}
\begin{array}{cc}
  C_{1} = \displaystyle\frac{\Psi\varphi_{2}^{\prime} - \Psi^{\prime}\varphi_{2}}
          {\varphi_{1}\varphi_{2}^{\prime} - \varphi_{1}^{\prime}\varphi_{2}} \bigg|_{a=a_{\rm tp,\, out}}, &
  C_{2} = \displaystyle\frac{\Psi^{\prime}\varphi_{1} - \Psi\varphi_{1}^{\prime}}
          {\varphi_{1}\varphi_{2}^{\prime} - \varphi_{1}^{\prime}\varphi_{2}} \bigg|_{a=a_{\rm tp,\, out}},
\end{array}
\label{eq.6.3.3}
\end{equation}
where $T$ is a normalization factor, $C_{1}$ and $C_{2}$ are complex constants found from the boundary condition introduced above: \emph{the $\varphi\,(a)$ function should represent an outgoing wave at turning point $a_{\rm tp,\, out}$}.
% So, the $\varphi\,(a)$ function and its derivative should equal to eq.~(\ref{eq.6.2.15}) at point $a_{\rm tp,\, out}$:
%
% \begin{equation}
% \begin{array}{cc}
%   \varphi\,(a_{\rm tp,\, out}) = T\, \Psi^{(+)} (\xi=0), &
%   \hspace{5mm}
%   \displaystyle\frac{d \varphi(a_{\rm tp,\, out})}{da} =
%   T\, |V_{1}|^{1/3} \cdot \displaystyle\frac{d \Psi (\xi=0)}{d\xi}.
% \end{array}
% \label{eq.6.3.2}
% \end{equation}
%
Fig.~\ref{fig.4} plots the total wave function calculated in this way for the potential (\ref{eq.model.2.6}) with parameters $A=36$, $B=12\,\Lambda$ at $\Lambda=0.01$
at different values of the energy of radiation $E_{\rm rad}$.
One can see that the number of oscillations of the wave function in the internal region increases with increasing of the energy of radiation. Another interesting property are \emph{the larger maxima of the wave function in the internal region at smaller distances to the barrier for arbitrary energy} (result found for the first time).

\begin{figure}[h]
\centerline{
\includegraphics[width=47mm]{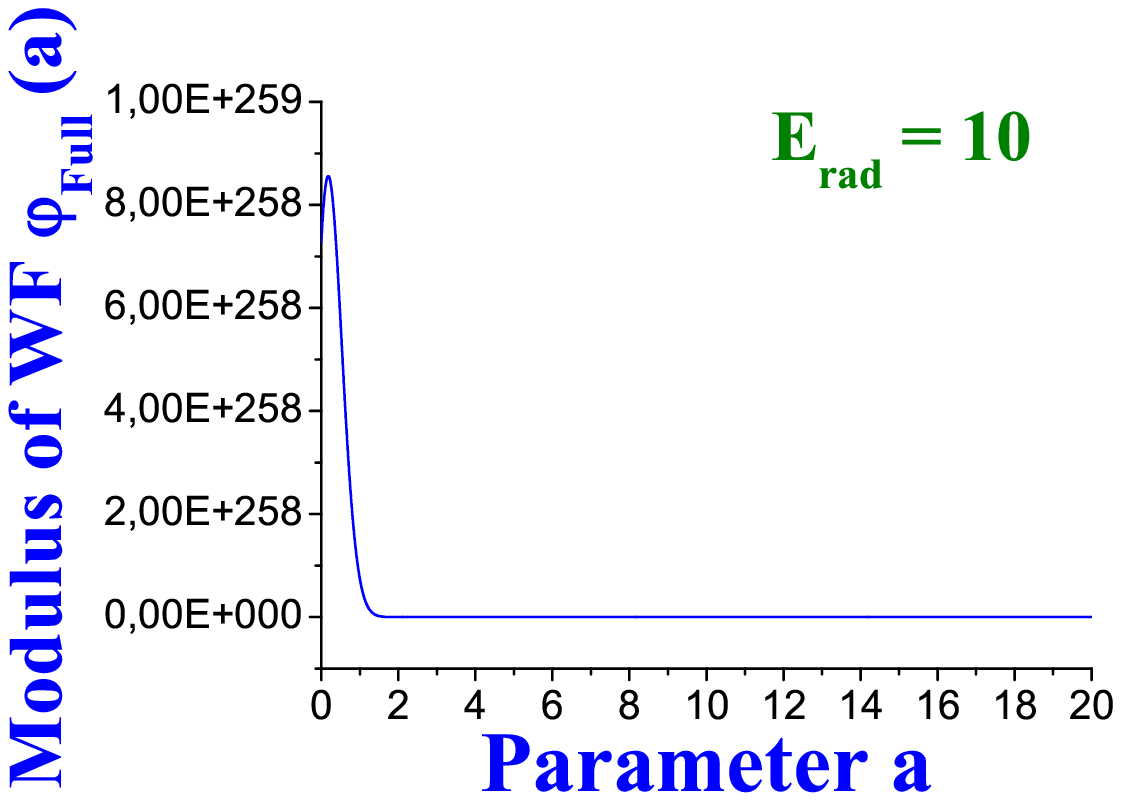}
\hspace{-3mm}\includegraphics[width=47mm]{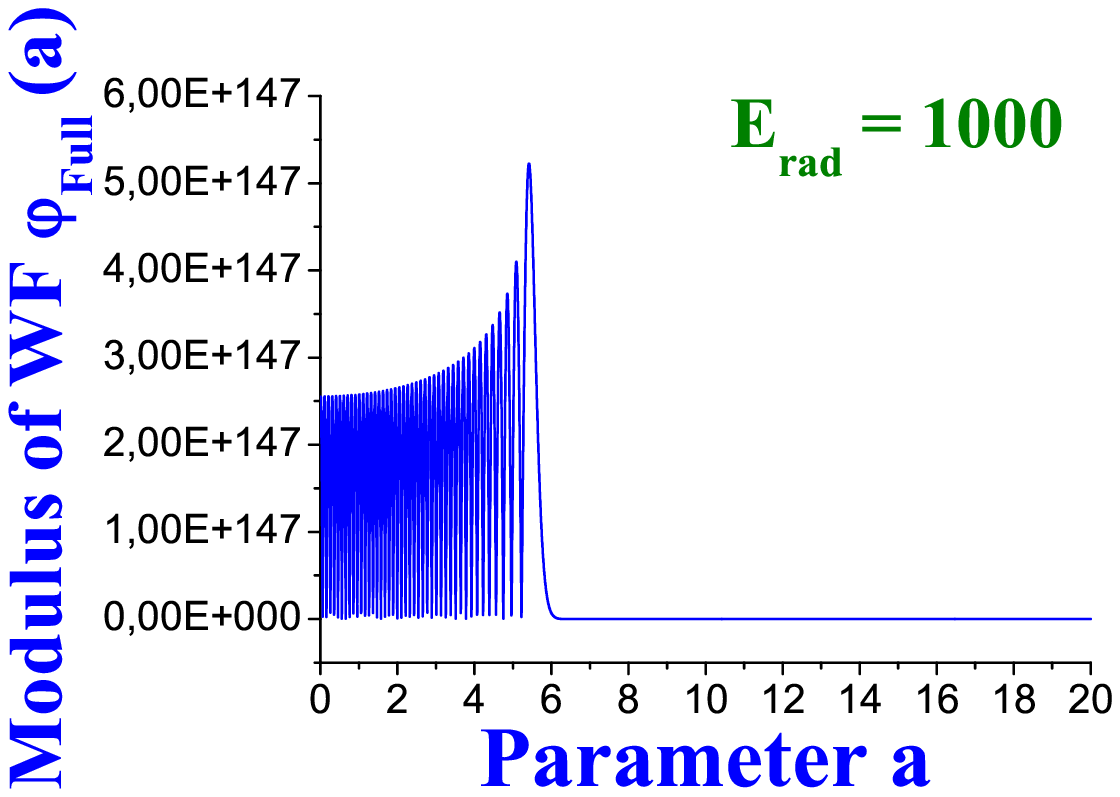}
\hspace{-3mm}\includegraphics[width=47mm]{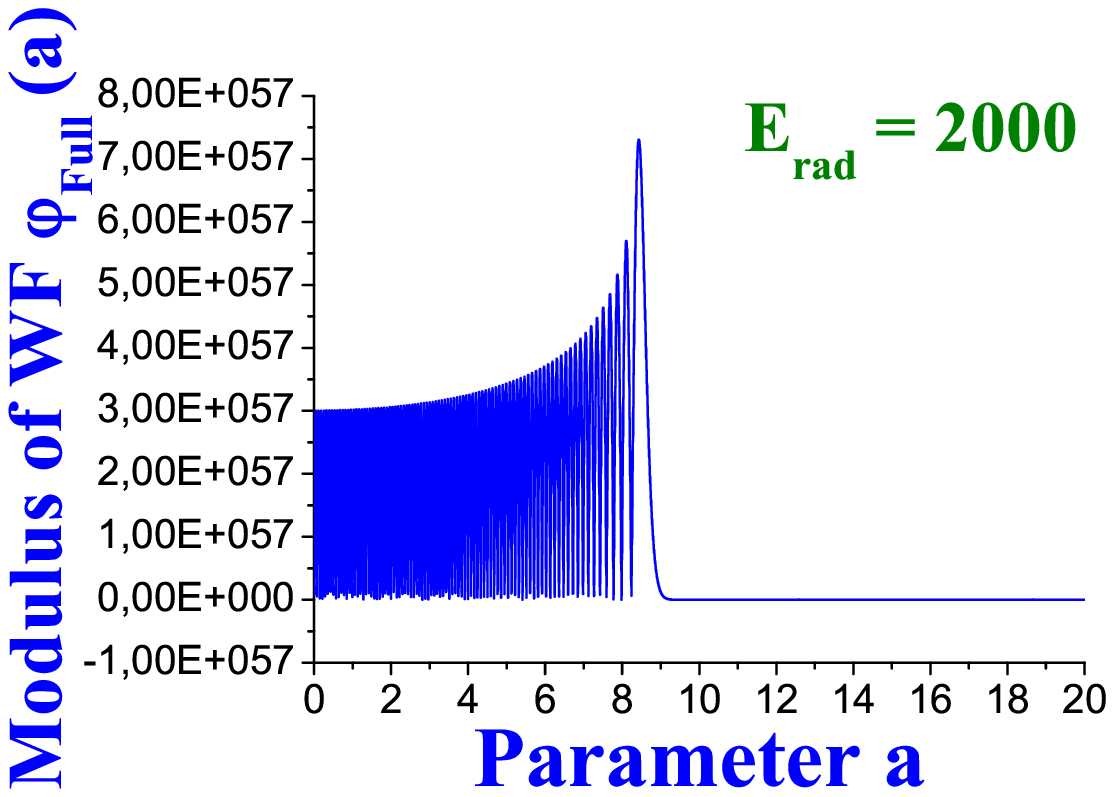}}
\caption{\small The behavior of the modulus of the wave function at the selected energies of radiation $E_{\rm rad}$:
(a) $E_{\rm rad}=10$; (b) $E_{\rm rad}=1000$; (c) $E_{\rm rad}=2000$.
\label{fig.5}}
\end{figure}

\begin{figure}[h]
\centerline{
\includegraphics[width=47mm]{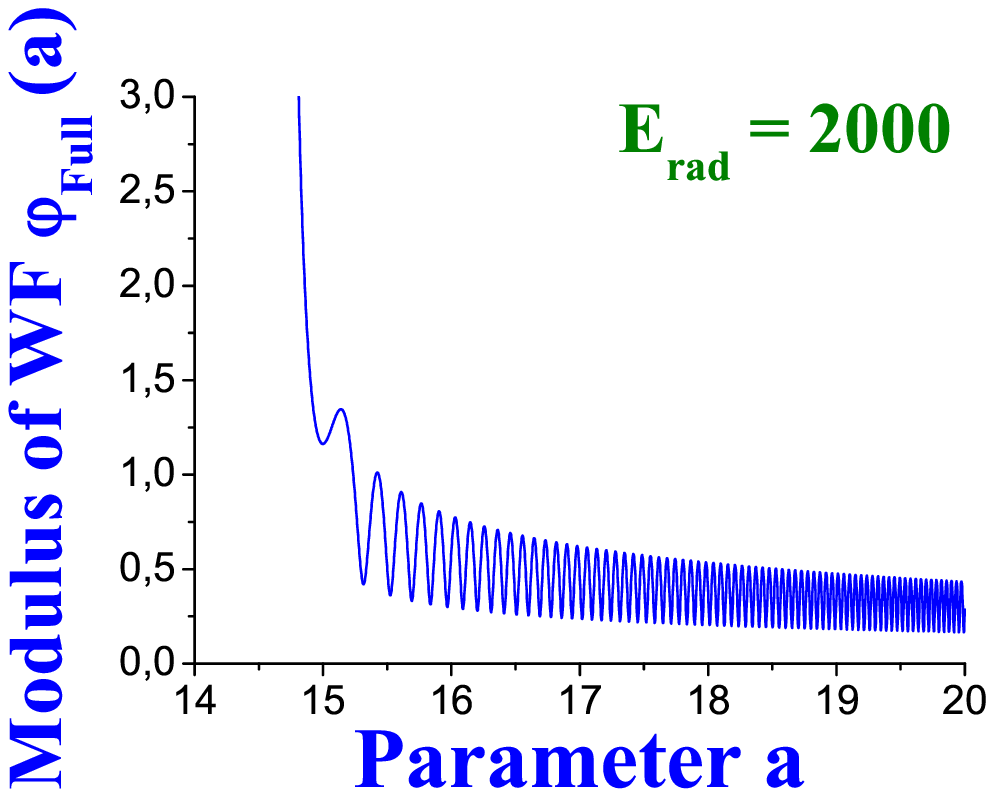}
\hspace{-3mm}\includegraphics[width=47mm]{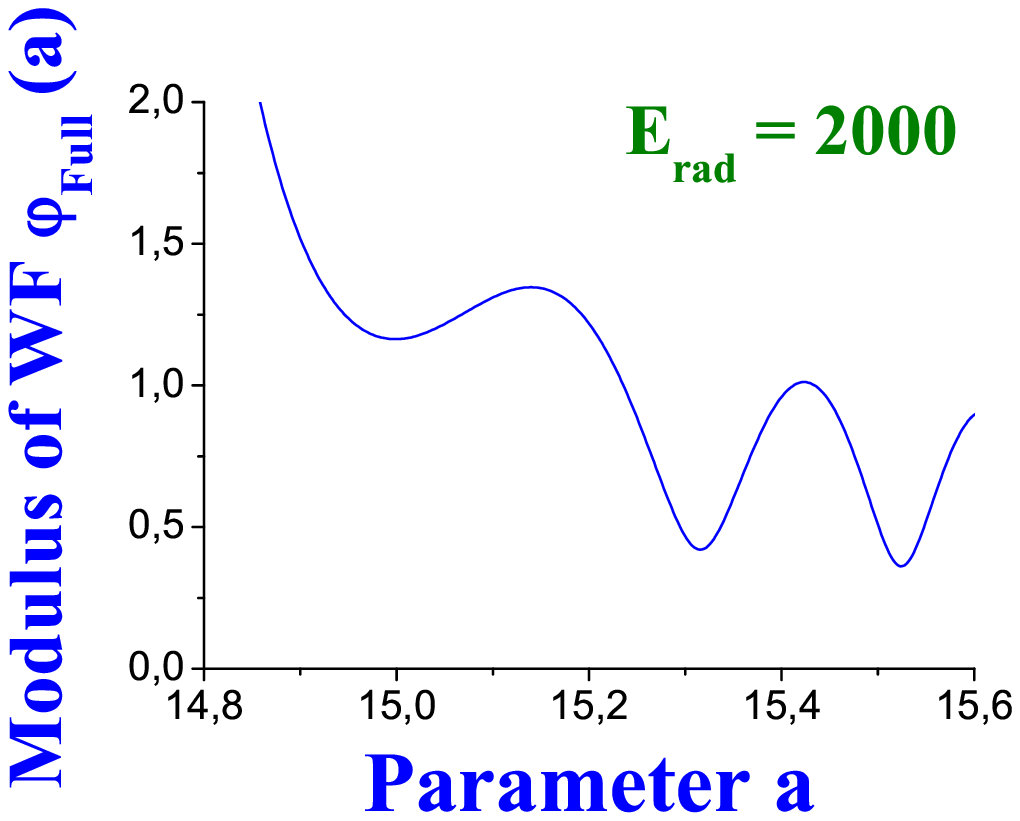}
\hspace{-3mm}\includegraphics[width=47mm]{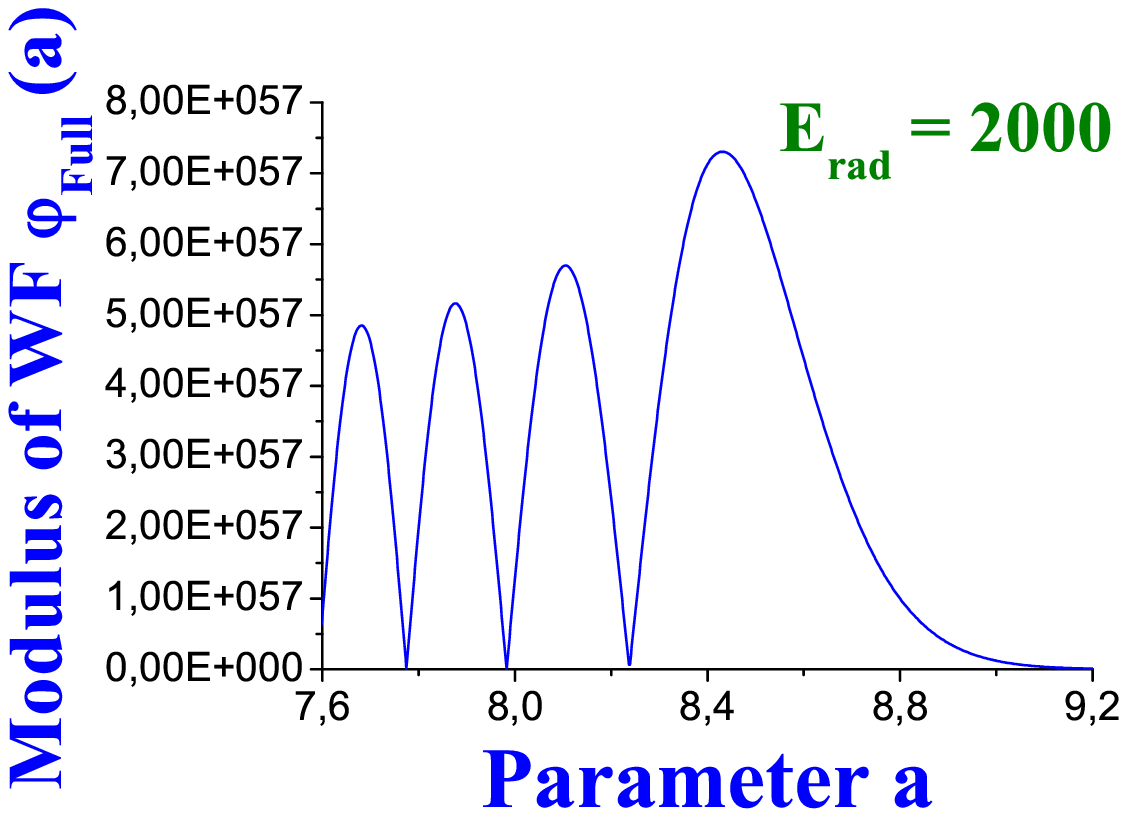}}
\caption{\small The behavior of the modulus of the total wave function at the energy of radiation $E_{\rm rad}=2000$, close to the turning points (for $a_{\rm tp,\,in}=8.58$, $a_{\rm tp,\,out}=15.04$, see also Table 1):
(a) the modulus decreases monotonously in the tunneling region, with increasing of $a$. It shows maxima and holes connected with the oscillations of the wave function in the external region, but the modulus is not equal to zero (thispoints out the existence of a \underline{non-zero} flux);
(b) when $a$ increases, the modulus reaches a minimum close to the external turning point $a_{\rm tp,\,out}$  (this demonstrates the practical fulfillment of the definition for the wave at such a point);
(c) transition close to $a_{\rm tp,\,in}$ is shown, where at increasing of $a$ the modulus with maximums and holes is transformed rapidly into a monotonously decreasing function without maximums and holes. This is connected with transition to the region of tunneling.
\label{fig.6}}
\end{figure}

In Fig.~\ref{fig.5} it has been shown how the modulus of this wave function changes at selected values of the energy of radiation.
From these figures it becomes clear why the coefficient of penetrability of the barrier is extremely small (up to the energy $E_{\rm rad}=2000$). In order to estimate, how effective is the boundary condition introduced above in building up the wave on the basis of the total wave function close to the external turning point $a_{\rm tp,\,out}$, it is useful to see how the modulus of this wave function changes close to this point. In Fig.~\ref{fig.6} we plot the modulus of the found wave function close to the turning points at the energy of radiation $E_{\rm rad}=2000$ is shown.
Here, one can see that the modulus at $a_{\rm tp,\,out}$ is practically constant (see left panel in Fig.~\ref{fig.6}). It is interesting to note that the modulus of the wave function, previously defined, does not change close to the internal turning point $a_{\rm tp,\,in}$, and is close to maximum (see right panel in Fig.~\ref{fig.6}).

%-----------------------------------------------------------------------------------------------------------------------

%-----------------------------------------------------------------------------------------------------------------------
\subsection{Calculations of the wave function of Universe
\label{sec.5.4}}

\subsubsection{Method of calculations of the wave function close to an arbitrary selected point $a_{x}$
% Method of beginning of the solution??????????????????????
\label{sec.5.4.1}}

Here, we look  for the regular partial solution of the wave function close to an arbitrary selected point $a_{x}$.
Let us write the wave function in the form:
\begin{equation}
\begin{array}{ll}
  \varphi(a) = c_{2} \sum\limits_{n=0}^{+\infty} b_{n} \: (a-a_{x})^{n}
  = c_{2} \sum\limits_{n=0}^{+\infty} b_{n} \: \bar{a}^{n}, &
  \bar{a} = a-a_{x}
\end{array}
\label{eq.5.4.1.1}
\end{equation}
and rewrite the potential through the variable $\bar{a}$:
\begin{equation}
  V(a) = C_{0} + C_{1}\,\bar{a} + C_{2}\,\bar{a}^{2} + C_{3}\,\bar{a}^{3} + C_{4}\,\bar{a}^{4},
\label{eq.5.4.1.4}
\end{equation}
where
\begin{equation}
\begin{array}{ccl}
  C_{0} & = & A\, a_{x}^{2} - B \,a_{x}^{4}, \\
  C_{1} & = & 2a_{x}(A-B\,a_{x}^{2}) - 2B\,a_{x}^{3} = 2A\,a_{x} - 4B\,a_{x}^{3}, \\
  C_{2} & = & A - B\,a_{x}^{2} - 4B\, a_{x}^{2} - B\,a_{x}^{2} = A - 6B\,a_{x}^{2}, \\
  C_{3} & = & -2B\,a_{x} - 2B\,a_{x} = -4B\,a_{x}, \\
  C_{4} & = & - B.
\end{array}
\label{eq.5.4.1.5}
\end{equation}
Substituting the wave function (\ref{eq.5.4.1.1}), its second derivative and the potential (\ref{eq.5.4.1.4}) into Schr\"{o}dinger equation,
we obtain recurrent relations for unknown $b_{n}$:
\begin{equation}
\begin{array}{ccccrl}
  b_{2} = \displaystyle\frac{(C_{0}-E)\,b_{0}}{2}, &
  b_{3} = \displaystyle\frac{(C_{0}-E)\,b_{1} + C_{1}\,b_{0}}{6}, &
  b_{4} = \displaystyle\frac{(C_{0}-E)\,b_{2} + C_{1}\,b_{1} + C_{2}\,b_{0}}{12}, &
\end{array}
\label{eq.5.4.1.8}
\end{equation}
\begin{equation}
\begin{array}{ccccrl}
  b_{5} = \displaystyle\frac{(C_{0}-E)\,b_{3} + C_{1}\,b_{2} + C_{2}\,b_{1} + C_{3}\,b_{0}}{20},
\end{array}
\label{eq.5.4.1.9}
\end{equation}
\begin{equation}
\begin{array}{ccccrl}
  b_{n+2} =
  \displaystyle\frac{(C_{0}-E)\,b_{n}+ C_{1}\,b_{n-1}+ C_{2}\,b_{n-2}+ C_{3}\,b_{n-3}+ C_{4}\,b_{n-4}}
  {(n+1)\,(n+2)} & \mbox{at } n \ge 4.
\end{array}
\label{eq.5.4.1.10}
\end{equation}
Given the values of $b_{0}$ and $b_{1}$ and using eqs.~(\ref{eq.5.4.1.8})--(\ref{eq.5.4.1.10}) one can calculate all $b_{n}$ needed. At limit $E_{\rm rad} \to 0$ and at $a_{x} = 0$ all found solutions for $b_{i}$ transform into the corresponding solutions (40), early obtained in \citep{Maydanyuk.2008.EPJC} at $E_{\rm rad}=0$.
Using $c_{2}=1$, from eqs.~(\ref{eq.5.4.1.1}) we find:
\begin{equation}
\begin{array}{cc}
  b_{0} = \varphi\,(a_{x}), & b_{1} = \varphi^{\prime}(a_{x}).
\end{array}
\label{eq.5.4.1.11}
\end{equation}
So, on the basis of the coefficients $b_{0}$ and $b_{1}$ one can obtain the values of the wave function and its derivative at point $a_{x}$.
{\bf Imposing two different boundary conditions via $b_{0}$ and $b_{1}$, we obtain two linear independent partial solutions $\varphi_{1}(a)$ and $\varphi_{2}(a)$ for the wave function. }
Using the internal turning point $a_{\rm tp,\,in}$ as the starting point, we calculate the first partial solution which increases in the barrier region (we choose: $b_{0} = 0.1$, $b_{1} = 1$), and using the external turning point $a_{\rm tp,\,out}$ as the starting point, we calculate the second partial solution which decreases in the barrier region (we choose: $b_{0} = 1$, $b_{1} = -0.1$). Such a choice provides effectively a linear independence between two partial solutions.

%-----------------------------------------------------------------------------------------------------------------------

%-----------------------------------------------------------------------------------------------------------------------
\subsubsection{Method of continuation of the solution
\label{sec.5.4.2}}

Let us rewrite equation~(\ref{eq.6.2.2})
in such a form\footnote{Here, we used the algorithm of \citep{Zaichenko_Kashuba.2001} }:
\begin{equation}
  \varphi^{\prime\prime}\,(a) = f\,(a)\: \varphi\,(a).
\label{eq.5.4.2.1}
\end{equation}
Let $\bigl\{a_{n} \bigr\}$ be a set of equidistant points $a_{n} = a_{0} + nh$. Denoting the values of the wave function $\varphi\,(a)$ at points $a_{n}$ as $\varphi_{n}$, we have constructed an algorithm of the ninth order to
determine $\varphi_{n+1}$ and $\varphi_{n}^{\prime}$ when $\varphi_{n}$ and $\varphi_{n-1}$ are known:
\begin{equation}
\begin{array}{ccl}
  \vspace{1mm}
  \varphi_{n+1} & = &
    \varphi_{n-1}\:\displaystyle\frac{g_{11} + g_{01}}{g_{01} - g_{11}} +
    \varphi_{n}\:\displaystyle\frac{g_{01}\,g_{10} - g_{00}\,g_{11}}{g_{01} - g_{11}} + O\,(h^{9}), \\

  \varphi_{n}^{\prime} & = &
    \varphi_{n-1}\:\displaystyle\frac{2}{g_{01} - g_{11}} +
    \varphi_{n}\:\displaystyle\frac{g_{10} - g_{00}}{g_{01} - g_{11}} + O\,(h^{9}),
\end{array}
\label{eq.5.4.2.2}
\end{equation}
where
\begin{equation}
\begin{array}{ccl}
\vspace{1mm}
  g_{00} & = &
    2  + h^{2}\,f_{n} +
    \displaystyle\frac{2}{4!}\: h^{4}\, \bigl(f_{n}^{\prime\prime} + f_{n}^{2}\bigr) +
    \displaystyle\frac{2}{6!}\: h^{6}\,
      \Bigl(f_{n}^{(4)} + 4\,\bigl(f_{n}^{\prime}\bigr)^{2} + 7\,f_{n}\,f_{n}^{\prime\prime} + f_{n}^{3}\Bigr) + \\
\vspace{3mm}
    & + &
    \displaystyle\frac{2}{8!}\: h^{8}\,
      \Bigl(
        f_{n}^{(6)} + 16\,f_{n}\,f_{n}^{(4)} + 26\,f_{n}^{\prime}\,f_{n}^{(3)} +
        15\,\bigl(f_{n}^{\prime\prime}\bigr)^{2} +
        22\,f_{n}^{2}\,f_{n}^{\prime\prime} +
        28\,f_{n}\,\bigl(f_{n}^{\prime}\bigr)^{2} + f_{n}^{4}
      \Bigr), \\

\vspace{2mm}
  g_{01} & = &
    \displaystyle\frac{2}{4!}\: h^{4}\, 2\,f_{n}^{\prime} +
    \displaystyle\frac{2}{6!}\: h^{6}\,
      \Bigl(4\,f_{n}^{(3)} + 6\,f_{n}\,f_{n}^{\prime} \Bigr) +
    \displaystyle\frac{2}{8!}\: h^{8}\,
      \Bigl(
        6\,f_{n}^{(5)} + 24\,f_{n}\,f_{n}^{(3)} + 48\,f_{n}^{\prime}f_{n}^{\prime\prime} +
        12\,f_{n}^{2}\,f_{n}^{\prime}
      \Bigr), \\

\vspace{2mm}
  g_{10} & = &
    \displaystyle\frac{2}{3!}\: h^{3}\,f_{n}^{\prime} +
    \displaystyle\frac{2}{5!}\: h^{5}\,\bigl(f_{n}^{(3)} + 4\,f_{n}\,f_{n}^{\prime}\bigr) +
    \displaystyle\frac{2}{7!}\: h^{7}\,
      \Bigl( f_{n}^{(5)} + 11\,f_{n}\,f_{n}^{(3)} +
             15\,f_{n}^{\prime}f_{n}^{\prime\prime} + 9\,f_{n}^{2}\,f_{n}^{\prime} \Bigr), \\

  g_{11} & = &
    2\,h +
    \displaystyle\frac{2}{3!}\:h^{3}\,f_{n} +
    \displaystyle\frac{2}{5!}\:h^{5}\, \bigl(3\,f_{n}^{\prime\prime} + f_{n}^{2}\bigr) +
    \displaystyle\frac{2}{7!}\: h^{7}\,
      \Bigl(5\,f_{n}^{(4)} + 13\,f_{n}\,f_{n}^{\prime\prime} + 10\,\bigl(f_{n}^{\prime}\bigr)^{2} + f_{n}^{3} \Bigr).
\end{array}
\label{eq.5.4.2.3}
\end{equation}
A local error of these formulas at point $a_{n}$ equals to:
\begin{equation}
  \delta_{n} =
    \displaystyle\frac{1}{10!}\: h^{10}\, f_{n}^{\prime}\, \varphi_{n}^{(7)}.
\label{eq.5.4.2.4}
\end{equation}
%-----------------------------------------------------------------------------------------------------------------------

% -----------------------------------------------------------------------------------------------------------------------
\subsection{The penetrability and reflection in the fully quantum approach
\label{sec.5.5}}

Let us analyze whether a known wave function in the whole region of its definition allows us to determine uniquely the coefficients of penetrability and reflection.

\subsubsection{Problem of interference between the incident and reflected waves
% An ambiguity in determination of the penetrability and reflection
\label{sec.6.2.1}}

Rewriting the wave function $\varphi_{\rm total}$ in the internal region through a summation of incident $\varphi_{\rm inc}$ wave
and reflected $\varphi_{\rm ref}$ wave:
\begin{equation}
  \varphi_{\rm total} =
  \varphi_{\rm inc} + \varphi_{\rm ref},
\label{eq.4.1.1}
\end{equation}
we consider the total flux:
\begin{equation}
\begin{array}{ccccc}
  j\, (\varphi_{\rm total}) & = &
  i\,
  \biggl[
    \Bigl( \varphi_{\rm inc} + \varphi_{\rm ref} \Bigr)
    \nabla \Bigl( \varphi_{\rm inc}^{*} + \varphi_{\rm ref}^{*} \Bigr) -
    \mbox{h.~c.}
  \Bigr) \biggr] & = &
  j_{\rm inc} + j_{\rm ref} + j_{\rm mixed},
\end{array}
\label{eq.4.1.2}
\end{equation}
where
\begin{equation}
\begin{array}{ccl}
  j_{\rm inc} =
  i\, \Bigl(\varphi_{\rm inc} \nabla \varphi_{\rm inc}^{*} - \mbox{h.~c.}\Bigr), &

  j_{\rm ref} =
  i\, \Bigl(\varphi_{\rm ref} \nabla \varphi_{\rm ref}^{*} - \mbox{h.~c.}\Bigr), \\

  j_{\rm mixed} =
  i\,
  \Bigl(
    \varphi_{\rm inc} \nabla \varphi_{\rm ref}^{*} +
    \varphi_{\rm ref} \nabla \varphi_{\rm inc}^{*} - \mbox{h.~c.}
  \Bigr). &
\end{array}
\label{eq.4.1.3}
\end{equation}
The $j_{\rm mixed}$ component describes interference between the incident and reflected waves in the internal region (let us call it \emph{mixed component of the total flux} or simply \emph{flux of mixing}).
From the constancy of the total flux $j_{\rm total}$ we find the flux $j_{\rm tr}$ for the wave
transmitted through the barrier, and:
\begin{equation}
\begin{array}{cc}
  j_{\rm inc} = j_{\rm tr} - j_{\rm ref} - j_{\rm mixed}, &
  j_{\rm tr} = j_{\rm total} = {\rm const}.
\end{array}
\label{eq.4.1.5}
\end{equation}
Now one can see that \emph{the mixed flux introduces ambiguity in the determination of the penetrability and reflection for the same known wave function.}
%-----------------------------------------------------------------------------------------------------------------------

% -----------------------------------------------------------------------------------------------------------------------
\subsection{Determination of the penetrability, reflection and interference coefficients % ???????????????????
% Non-locality in determination of the penetrability and reflection ???????????????????
\label{sec.4.2}}

In quantum mechanics the coefficients of penetrability and reflection are defined considering the potential as a whole, including asymptotic regions.
{\bf However, in the radial calculation of quantum decay such a consideration depends on how the incident and reflected waves are defined inside finite internal region from the left of the barrier. The question is: does the location of such a region influence the penetrability and reflection? }
In order to obtain these coefficients, we shall include into definitions coordinates where the fluxes
are defined (denote them as $x_{\rm left}$ and $x_{\rm right}$):
\begin{equation}
\begin{array}{ccc}
  T(x_{\rm left}, x_{\rm right}) = \displaystyle\frac{j_{\rm tr}(x_{\rm right})}{j_{\rm inc}(x_{\rm left})}, &
  R(x_{\rm left}) = \displaystyle\frac{j_{\rm ref}(x_{\rm left})}{j_{\rm inc}(x_{\rm left})}, &
  M(x_{\rm left}) = \displaystyle\frac{j_{\rm mixed}(x_{\rm left})}{j_{\rm inc}(x_{\rm left})}.
\end{array}
\label{eq.4.2.1}
\end{equation}
So, the $T$ and $R$ coefficients determine the probability of transmission (or tunneling) and reflection of the wave relatively the region of the potential with arbitrary selected boundaries $x_{\rm left}$, $x_{\rm right}$.
When $x_{\rm right}$ tends to the asymptotic limit, the coefficient defined before should transform into standard ones.
From eqs.~(\ref{eq.4.1.5}) and (\ref{eq.4.2.1}) we obtain
($j_{\rm tr}$ and $j_{\rm ref}$ are directed in opposite directions,
$j_{\rm inc}$ and $j_{\rm tr}$  --- in the same directions):
\begin{equation}
  |T| + |R| - M = 1.
\label{eq.4.2.3}
\end{equation}
\emph{Now we see that the condition
$|T| + |R| = 1$
has sense in quantum mechanics only if there is no interference between incident and reflected waves},
and for this is enough that:
\begin{equation}
  j_{\rm mixed} = 0.
\label{eq.4.2.5}
\end{equation}
A new question appears: \emph{does this condition allow to separate the total wave function into the incident and reflected components in a unique way?}
It turns out that the choice of the incident and reflected waves has essential influence on the barrier penetrability, and different forms of the incident $\varphi_{\rm inc}$and reflected $\varphi_{\rm ref}$ waves can give zero flux $j_{\rm mix}$.
Going from the rectangular internal well to the fully quantum treatment of the problem would become more complicated.
% Such a situation is typical in quantum cosmology.
% {\bf
% \emph{Whole importance of accurate definition of the wave in the quantum cosmological problem becomes clear from here for construction of the total wave function on the basis of its two partial solutions, and for the separation of the known wave function into the incident and reflected waves in the internal region from the left of the barrier.}
% }
%-----------------------------------------------------------------------------------------------------------------------

% -----------------------------------------------------------------------------------------------------------------------
\subsection{Wave incident on the barrier and wave reflected from it in the internal region
\label{sec.5.4}}

One can define the incident wave to be proportional to the function $\Psi^{(+)}$ and
the reflected wave to be proportional to the function $\Psi^{(-)}$:
\begin{equation}
\begin{array}{cclccl}
  \varphi_{\rm total}\,(a) = \varphi_{\rm inc}\,(a) + \varphi_{\rm ref}\,(a), &
  \varphi_{\rm inc}\,(a) = we \cdot \Psi^{(+)}\,(a), &
  \varphi_{\rm ref}\,(a) = R \cdot \Psi^{(-)}\,(a),
\end{array}
\label{eq.6.4.1}
\end{equation}
where $I$ and $R$ are new constants found from continuity condition of the total wave function $\varphi_{\rm total}$ and its derivative
at the internal turning point $a_{\rm tp,\, int}$:
\begin{equation}
\begin{array}{cclccl}
  we & = &
    \displaystyle\frac
      {\varphi_{\rm total}\,\Psi^{(-),\prime} - \varphi_{\rm total}^{\prime}\,\Psi^{(-)}}
      {\Psi^{(+)}\,\Psi^{(-),\prime} - \Psi^{(+),\prime}\,\Psi^{(-)}}
      \bigg|_{a=a_{\rm tp,\,int}}, &
  R & = &
    \displaystyle\frac
      {\varphi_{\rm total}^{\prime}\,\Psi^{(+)} - \varphi_{\rm total}\,\Psi^{(+),\prime}}
      {\Psi^{(+)}\,\Psi^{(-),\prime} - \Psi^{(+),\prime}\,\Psi^{(-)}}
      \bigg|_{a=a_{\rm tp,\,int}}.
\end{array}
\label{eq.6.4.2}
\end{equation}
On the basis of these solutions we obtain at the internal turning point $a_{\rm tp,\, int}$ the flux incident on the barrier, the flux reflected from it and the flux of mixing.
The flux transmitted through the barrier was calculated at the external turning point $a_{\rm tp,\, ext}$.
%-----------------------------------------------------------------------------------------------------------------------

%-----------------------------------------------------------------------------------------------------------------------
\subsection{Penetrability and reflection: fully quantum approach versus semiclassical one
\label{sec.6.5}}

Now we shall estimate through the method described above the coefficients of penetrability and reflection for the potential barrier with parameters $A=36$, $B=12\,\Lambda$, $\Lambda=0.01$ at different values of the energy of radiation $E_{\rm rad}$. We shall compare the coefficient of penetrability obtained with the values given by the semiclassical method.
In the semiclassical approach we shall consider two definitions of this coefficient:
\begin{equation}
\begin{array}{cc}
  P_{\rm penetrability}^{\rm WKB, (1)} = \displaystyle\frac{1}{\theta^{2}}, &
  P_{\rm penetrability}^{\rm WKB, (2)} = \displaystyle\frac{4}{\Bigl(2\theta + 1/(2\theta)^{2}\Bigr)^{2}},
\end{array}
\label{eq.6.5.1}
\end{equation}
where
\begin{equation}
  \theta =
%    \exp \displaystyle\int\limits_{a_{\rm tp}^{\rm (int)}}^{a_{\rm tp}^{\rm (ext)}} |p|\; da =
    \exp \displaystyle\int\limits_{a_{\rm tp}^{\rm (int)}}^{a_{\rm tp}^{\rm (ext)}} \bigl|V(a)-E\bigr|\; da.
\label{eq.6.2}
\end{equation}
One can estimate also \emph{the duration of the formation of the Universe},
using by definition (15) in Ref.~\citep{AcacioDeBarros.2007.PRD}:
\begin{equation}
  \tau = 2\, a_{\rm tp,\, int}\: \displaystyle\frac{1}{\rm P_{penetrability}}.
\label{eq.6.3}
\end{equation}

The results are presented in Tabl.~\ref{table.1}.
\begin{table}
\hspace{-20mm}
\begin{center}
\begin{tabular}{|c|c|c|c|c|c|c|} \hline
  Energy
  & \multicolumn{2}{|c|}{Penetrability $P_{\rm penetrability}$}
  & \multicolumn{2}{|c|}{Time $\tau$} & \multicolumn{2}{|c|}{Turning points} \\ \cline{2-7}
  $E_{\rm rad}$
  & Direct method & Method WKB & Direct method & Method WKB & $a_{\rm tp,\, in}$ & $a_{\rm tp,\, out}$ \\ \hline
   1.0 & $8.7126 \times 10^{-521}$ & $2.0888 \times 10^{-521}$ & $3.8260 \times 10^{+519}$ & $1.5958 \times 10^{+520}$ & 0.16 &  17.31 \\
   2.0 & $2.4225 \times 10^{-520}$ & $5.5173 \times 10^{-521}$ & $1.9460 \times 10^{+519}$ & $8.5448 \times 10^{+519}$ & 0.23 &  17.31 \\
   3.0 & $6.2857 \times 10^{-520}$ & $1.3972 \times 10^{-520}$ & $9.1863 \times 10^{+518}$ & $4.1326 \times 10^{+519}$ & 0.28 &  17.31 \\
   4.0 & $1.5800 \times 10^{-519}$ & $3.4428 \times 10^{-520}$ & $4.2201 \times 10^{+518}$ & $1.9367 \times 10^{+519}$ & 0.33 &  17.31 \\
   5.0 & $3.8444 \times 10^{-519}$ & $8.2935 \times 10^{-520}$ & $1.9392 \times 10^{+518}$ & $8.9892 \times 10^{+518}$ &  0.37 &  17.31 \\
   6.0 & $9.2441 \times 10^{-519}$ & $1.9701 \times 10^{-519}$ & $8.8350 \times 10^{+517}$ & $4.1455 \times 10^{+518}$ &  0.40 &  17.31 \\
   7.0 & $2.1678 \times 10^{-518}$ & $4.5987 \times 10^{-519}$ & $4.0694 \times 10^{+517}$ & $1.9183 \times 10^{+518}$ &  0.44 &  17.31 \\
   8.0 & $5.0192 \times 10^{-518}$ & $1.0621 \times 10^{-518}$ & $1.8790 \times 10^{+517}$ & $8.8797 \times 10^{+517}$ &  0.47 &  17.31 \\
   9.0 & $1.1604 \times 10^{-517}$ & $2.4316 \times 10^{-518}$ & $8.6212 \times 10^{+516}$ & $4.1140 \times 10^{+517}$ &  0.50 &  17.31 \\
  10.0 & $2.6279 \times 10^{-517}$ & $5.5016 \times 10^{-518}$ & $4.0128 \times 10^{+516}$ & $1.9168 \times 10^{+517}$ &  0.52 &  17.31 \\

100.0 & $1.6165 \times 10^{-490}$ & $3.1959 \times 10^{-491}$ & $2.0717 \times 10^{+490}$ & $ 1.0478 \times 10^{+491}$ & 1.67 & 17.23 \\
200.0 & $ 8.5909 \times 10^{-465}$ & $1.6936 \times 10^{-465}$ & $5.5397 \times 10^{+464}$ & $ 2.8100 \times 10^{+465}$ & 2.37 & 17.15 \\
300.0 & $ 6.8543 \times 10^{-441}$ & $1.3419 \times 10^{-441}$ & $8.5461 \times 10^{+440}$ & $ 4.3653 \times 10^{+441}$ & 2.92 & 17.07 \\
400.0 & $ 3.6688 \times 10^{-418}$ & $7.1642 \times 10^{-419}$ & $1.8531 \times 10^{+418}$ & $ 9.4900 \times 10^{+418}$ & 3.39 & 16.98 \\
500.0 & $ 2.6805 \times 10^{-396}$ & $5.2521 \times 10^{-397}$ & $2.8508 \times 10^{+396}$ & $ 1.4550 \times 10^{+397}$ & 3.82 & 16.89 \\
600.0 & $ 4.1386 \times 10^{-375}$ & $8.0511 \times 10^{-376}$ & $2.0338 \times 10^{+375}$ & $ 1.0454 \times 10^{+376}$ & 4.20 & 16.80 \\
700.0 & $ 1.7314 \times 10^{-354}$ & $3.3810 \times 10^{-355}$ & $5.2806 \times 10^{+354}$ & $ 2.7043 \times 10^{+355}$ & 4.57 & 16.70 \\
800.0 & $ 2.4308 \times 10^{-334}$ & $4.7497 \times 10^{-335}$ & $4.0448 \times 10^{+334}$ & $ 2.0701 \times 10^{+335}$ & 4.91 & 16.60 \\
900.0 & $ 1.3213 \times 10^{-314}$ & $2.5761 \times 10^{-315}$ & $7.9408 \times 10^{+314}$ & $ 4.0730 \times 10^{+315}$ & 5.24 & 16.50 \\
1000.0 & $ 3.0920 \times 10^{-295}$ & $6.0272 \times 10^{-296}$ & $3.5999 \times 10^{+295}$ & $ 1.8468 \times 10^{+296}$ & 5.56 & 16.40 \\
1100.0 & $ 3.4274 \times 10^{-276}$ & $6.6576 \times 10^{-277}$ & $3.4289 \times 10^{+276}$ & $ 1.7652 \times 10^{+277}$ & 5.87 & 16.29 \\
1200.0 & $ 1.9147 \times 10^{-257}$ & $3.7259 \times 10^{-258}$ & $6.4553 \times 10^{+257}$ & $ 3.3174 \times 10^{+258}$ & 6.18 & 16.18 \\
1300.0 & $ 5.8026 \times 10^{-239}$ & $1.1253 \times 10^{-239}$ & $2.2333 \times 10^{+239}$ & $ 1.1516 \times 10^{+240}$ & 6.47 & 16.06 \\
1400.0 & $ 9.9042 \times 10^{-221}$ & $1.9252 \times 10^{-221}$ & $1.3683 \times 10^{+221}$ & $ 7.0393 \times 10^{+221}$ & 6.77 & 15.93 \\
1500.0 & $ 1.0126 \times 10^{-202}$ & $1.9551 \times 10^{-203}$ & $1.3965 \times 10^{+203}$ & $ 7.2333 \times 10^{+203}$ & 7.07 & 15.81 \\
1600.0 & $ 6.2741 \times 10^{-185}$ & $1.2155 \times 10^{-185}$ & $2.3480 \times 10^{+185}$ & $ 1.2119 \times 10^{+186}$ & 7.36 & 15.67 \\
1700.0 & $ 2.4923 \times 10^{-167}$ & $4.8143 \times 10^{-168}$ & $6.1488 \times 10^{+167}$ & $ 3.1831 \times 10^{+168}$ & 7.66 & 15.53 \\
1800.0 & $ 6.4255 \times 10^{-150}$ & $1.2437 \times 10^{-150}$ & $2.4783 \times 10^{+150}$ & $ 1.2803 \times 10^{+151}$ & 7.96 & 15.38 \\
1900.0 & $ 1.1189 \times 10^{-132}$ & $2.1580 \times 10^{-133}$ & $1.4776 \times 10^{+133}$ & $ 7.6619 \times 10^{+133}$ & 8.26 & 15.22 \\
2000.0 & $ 1.3288 \times 10^{-115}$ & $2.5653 \times 10^{-116}$ & $1.2914 \times 10^{+116}$ & $ 6.6895 \times 10^{+116}$ & 8.58 & 15.04 \\
2100.0 & $ 1.1105 \times 10^{-98}$  & $2.1357 \times 10^{-99}$  & $1.6036 \times 10^{+99}$ & $ 8.3382 \times 10^{+99}$ & 8.90 & 14.85 \\
2200.0 & $6.6054 \times 10^{-82}$ & $1.2690 \times 10^{-82}$ & $2.7988 \times 10^{+82}$ & $ 1.4567 \times 10^{+83}$ & 9.24 & 14.64 \\
2300.0 & $2.8693 \times 10^{-65}$ & $5.4647 \times 10^{-66}$ & $6.6952 \times 10^{+65}$ & $ 3.5154 \times 10^{+66}$ & 9.60 & 14.41 \\
2400.0 & $9.1077 \times 10^{-49}$ & $1.7297 \times 10^{-49}$ & $2.1959 \times 10^{+49}$ & $ 1.1562 \times 10^{+50}$ & 10.00 &  14.14 \\
2500.0 & $2.1702 \times 10^{-32}$ & $4.0896 \times 10^{-33}$ & $9.6290 \times 10^{+32}$ & $5.1098 \times 10^{+33}$ & 10.44 & 13.81 \\
2600.0 & $3.9788 \times 10^{-16}$ & $7.3137 \times 10^{-17}$ & $5.5322 \times 10^{+16}$ & $3.0096 \times 10^{+17}$ & 11.00 & 13.37 \\

2610.0 & $1.6663 \times 10^{-14}$ & $3.0428 \times 10^{-15}$ & $1.3290 \times 10^{+15}$ & $7.2780 \times 10^{+15}$ & 11.07 & 13.31 \\
2620.0 & $6.9240 \times 10^{-13}$ & $1.2606 \times 10^{-13}$ & $3.2187 \times 10^{+13}$ & $1.7678 \times 10^{+14}$ & 11.14 & 13.25 \\
2630.0 & $2.8842 \times 10^{-11}$ & $5.2116 \times 10^{-12}$ & $7.7789 \times 10^{+11}$ & $4.3050 \times 10^{+12}$ & 11.21 & 13.19 \\
2640.0 & $1.2002 \times 10^{-9}$ & $2.1495 \times 10^{-10}$ & $1.8825 \times 10^{+10}$ & $1.0511 \times 10^{+11}$ & 11.29 & 13.12 \\
2650.0 & $4.9881 \times 10^{-8}$ & $8.8401 \times 10^{-9}$ & $4.5642 \times 10^{+8}$ & $2.5754 \times 10^{+9}$ & 11.38 & 13.05 \\
2660.0 & $2.0738 \times 10^{-6}$ & $3.6263 \times 10^{-7}$ & $1.1068 \times 10^{+7}$ & $6.3303 \times 10^{+7}$ & 11.47 & 12.97 \\
2670.0 & $8.7110 \times 10^{-5}$ & $1.4836 \times 10^{-5}$ & $2.6596 \times 10^{+5}$ & $1.5615 \times 10^{+6}$ & 11.58 & 12.87 \\
2680.0 & $3.6953 \times 10^{-3}$ & $6.0519 \times 10^{-4}$ & $6.3369 \times 10^{+3}$ & $3.8693 \times 10^{+4}$ & 11.70 & 12.76 \\
2690.0 & $1.5521 \times 10^{-1}$ & $2.4634 \times 10^{-2}$ & $1.5293 \times 10^{+2}$ & $9.3602 \times 10^{+2}$ & 11.86 & 12.61 \\
\hline
\end{tabular}
\end{center}
\caption{\small The penetrability $P_{\rm penetrability}$ of the barrier and the duration $\tau$ of the formation of the Universe defined by eq.~(\ref{eq.6.3}) in the fully quantum and semiclassical approaches
\label{table.1}}
\end{table}
In calculations the coefficients of penetrability, reflection and mixing are defined by eqs.~(\ref{eq.4.2.1}), the fluxes by eqs.~(\ref{eq.4.1.3}) (calculated $P_{\rm penetrability}^{\rm WKB, (2)}$ coincide with $P_{\rm penetrability}^{\rm WKB, (1)}$ up to the first 7 digits for energies in range $0 \le E_{\rm rad} \le 2500$).

From this table one can see that inside the entire range of energy, the fully quantum approach gives value for the coefficient of penetrability enough close to its value obtained by the semiclassical approach. This differs essentially from results in the non-stationary approach~\citep{AcacioDeBarros.2007.PRD}. This difference could be explained by difference in a choice of the boundary condition, which is used in construction of the stationary solution of the wave function.

% An advantage of the fully quantum method is possibility to calculate the coefficient of reflection. However, the calculations show that this coefficient inside the whole region of the energy used in Tabl.~1 equals to 1 practically. But this coefficient differs visibly from 1 only at the energy of radiation enough close to the height of the barrier, that is explained by essential decreasing of a role of the barrier in penetration of the wave through it. In order to estimate an accuracy of the found coefficients, we obtain property (\ref{eq.4.2.3}) up to the first 11 digits (see Tabl.~2 in Appendix) inside whole region of the energy of radiation. The coefficient of mixing is less than $10^{-19}$ (that is connected with computer error). So, \emph{there is no practical interference between the incident and reflected waves defined by such a way close to the internal turning point that points out their very accurate determination practically.}
% Now it becomes clear that the approach proposed in Ref.~\citep{AcacioDeBarros.2007.PRD} and the semiclassical methods do not give such an accuracy in the determination of the coefficients
% of penetrability and reflection.
%-----------------------------------------------------------------------------------------------------------------------

%-----------------------------------------------------------------------------------------------------------------------
\subsection{The penetrability in the FRW-model with the Chaplygin gas
\label{sec.6.6}}

{\bf In order to connect universe with dust and its accelerating stage, in Ref.~\citep{Kamenshchik.2001.PLB} a new scenario with the \emph{Chaplygin gas} was proposed.} A quantum FRW-model with the Chaplygin gas has been constructed on the basis of equation of state instead of $p\,(a)=\rho_{\rm rad}(a)/3$ (where $p\,(a)$ is pressure)
by the following (see also Refs.~\citep{Bilic.2002.PLB,Bento.2002.PRD}):
\begin{equation}
  p_{\rm Ch} = -\displaystyle\frac{A}{\rho_{\rm Ch}^{\alpha}},
\label{eq.7.1.1}
\end{equation}
where $A$ is positive constant and $0< \alpha \le 1$. In particular, for the standard Chaplygin gas we have $\alpha=1$.
Solution of equation of state~(\ref{eq.7.1.1}) gives the following dependence of density on the scale factor:
\begin{equation}
  \rho_{\rm Ch}(a) = \biggl( A + \displaystyle\frac{B}{a^{3\,(1+\alpha)}} \biggr)^{1/(1+\alpha)},
\label{eq.7.1.2}
\end{equation}
where $B$ is a new constant of integration. Using parameter $\alpha$, this model describes transition between stage, when Universe is filled with dust-like matter, and its accelerating expanding stage (through scenario of Chaplygin gas applied to cosmology, for details, see Refs.~\citep{Kamenshchik.2001.PLB,Bouhmadi-Lopez.2005.PRD,Bouhmadi-Lopez.2008.IJMPD}, also historical paper \citep{Chaplygin.1904}).

Let us combine expression for density which includes previous forms of matter and the Chaplygin gas in addition. At limit $\alpha \to 0$ eq.~(\ref{eq.7.1.2}) transforms into the $\rho_{\rm dust}$ component plus the $\rho_{\Lambda}$ component.
From such limit we find
\begin{equation}
\begin{array}{cc}
  A = \rho_{\Lambda}, &
  B = \rho_{\rm dust}
\end{array}
\label{eq.7.1.3}
\end{equation}
and obtain the following generalized density:
\begin{equation}
  \rho\,(a) =
    \biggl( \rho_{\Lambda} + \displaystyle\frac{\rho_{\rm dust}}{a^{3\,(1+\alpha)}} \biggr)^{1/(1+\alpha)} +
    \displaystyle\frac{\rho_{\rm rad}}{a^{4}(t)}.
\label{eq.7.1.4}
\end{equation}
Now we have:
\begin{equation}
  \dot{a}^{2} + k -
  \displaystyle\frac{8\pi\,G}{3}\,
    \Biggl\{
      a^{2}\:\biggl( \rho_{\Lambda} + \displaystyle\frac{\rho_{\rm dust}}{a^{3\,(1+\alpha)}} \biggr)^{1/(1+\alpha)} +
      \displaystyle\frac{\rho_{\rm rad}}{a^{2}(t)}
    \Biggr\}= 0.
\label{eq.7.1.5}
\end{equation}
After quantization we obtain the Wheeler-De Witt equation
\begin{equation}
\begin{array}{cc}
  \biggl\{ -\:\displaystyle\frac{\partial^{2}}{\partial a^{2}} + V_{\rm Ch}\,(a) \biggr\}\; \varphi(a) =
  E_{\rm rad}\; \varphi(a), &
  E_{\rm rad} = \displaystyle\frac{3\,\rho_{\rm rad}}{2\pi\,G},
\end{array}
\label{eq.7.2.1}
\end{equation}
where
\begin{equation}
\begin{array}{ccl}
  V_{\rm Ch}\,(a) & = &
    \biggl( \displaystyle\frac{3}{4\pi\,G} \biggr)^{2}\: k\,a^{2} -
    \displaystyle\frac{3}{2\pi\,G}\:
     a^{4}\, \biggl( \rho_{\Lambda} + \displaystyle\frac{\rho_{\rm dust}}{a^{3\,(1+\alpha)}} \biggr)^{1/(1+\alpha)}.
\end{array}
\label{eq.7.2.2}
\end{equation}
For the Universe of closed type (at $k=1$) at $8\pi\,G \equiv M_{\rm p}^{-2} = 1$ we have
(see eqs.~(6)--(7) in Ref.~\citep{Bouhmadi-Lopez.2005.PRD}):
\begin{equation}
\begin{array}{cc}
  V_{\rm Ch}\,(a) =
    36\,a^{2} -
    12\,a^{4}\,\Bigl(\Lambda + \displaystyle\frac{\rho_{\rm dust}}{a^{3\,(1+\alpha)}} \Bigr)^{1/(1+\alpha)}, &
    E_{\rm rad} = 12\, \rho_{\rm rad}.
\end{array}
\label{eq.7.2.3}
\end{equation}
\begin{figure}[h]
\centering{\includegraphics[width=50mm]{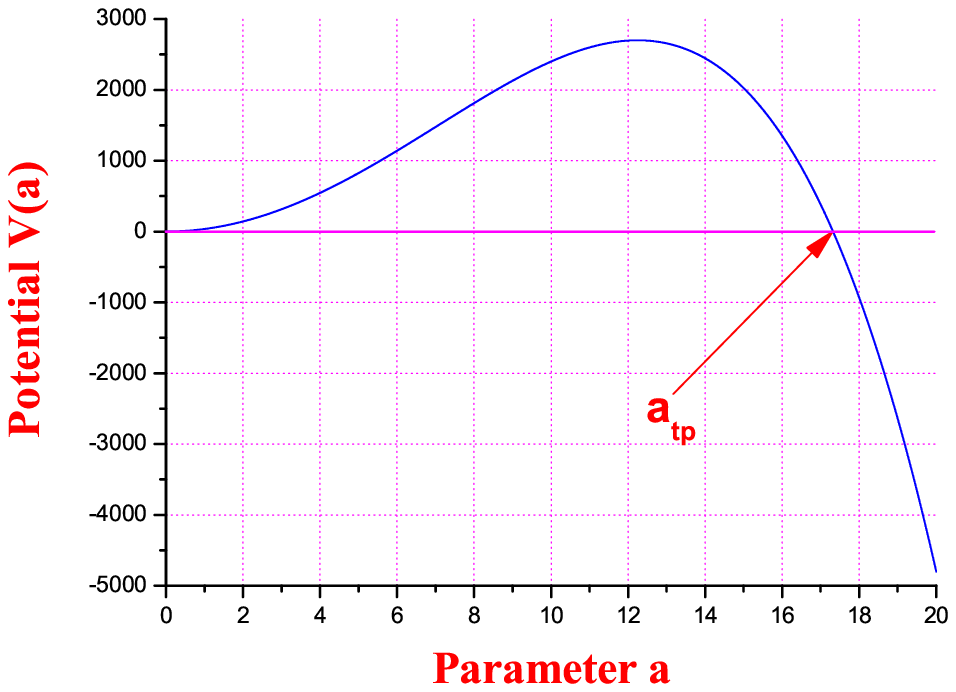}
\includegraphics[width=50mm]{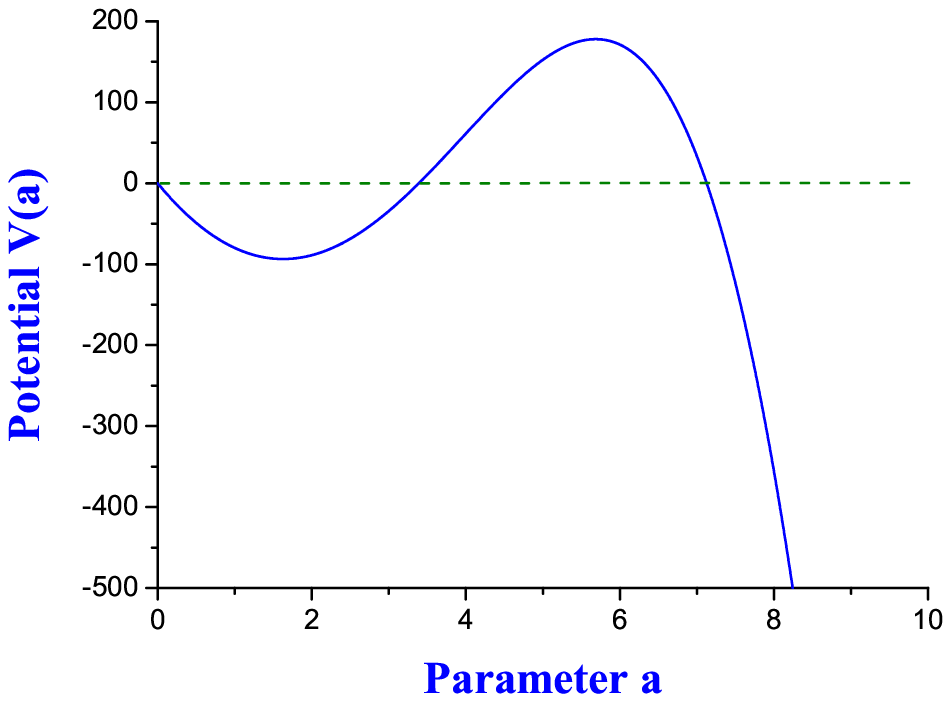}}
\caption{Cosmological potentials with and without Chaplygin gas:
Left panel is for potential $V(a) = 36\,a^{2} - 12\,\Lambda\,a^{4}$ with parameter $\Lambda=0.01$ (turning point $a_{tp} = 17.320508$ at zero energy $E_{\rm rad}=0$),
Right panel is for potential (\ref{eq.7.2.3}) with parameters $\Lambda=0.01$, $\rho_{\rm dust}=30$, $\alpha=0.5$
(minimum of the hole is -93.579 and its coordinate is 1.6262,
maximum of the barrier is 177.99 and its coordinate is 5.6866).
\label{fig.7}}
\end{figure}

% Assuming turning points to be determined numerically,
Let us expand the potential (\ref{eq.7.2.3}) close to arbitrary selected point $\bar{a}$ by powers of $q=a-\bar{a}$
and restrict ourselves to linear terms:
\begin{equation}
  V_{\rm Ch}\,(q) = V_{0} + V_{1}q.
\label{eq.7.3.1}
\end{equation}
For coefficients $V_{0}$ and $V_{1}$ we find:
\begin{equation}
\begin{array}{ccl}
  V_{0} & = & V_{\rm Ch}\,(a=\bar{a}), \\
  V_{1} & = &
    \displaystyle\frac{d V_{\rm Ch}\,(a)}{da} \bigg|_{a=\bar{a}} =
    72\,a +
    12\,a^{3}\,
    \Bigl\{ -4\,\Lambda - \displaystyle\frac{\rho_{\rm dust}}{a^{3\,(1+\alpha)}} \Bigr\} \cdot
    \Bigl( \Lambda + \displaystyle\frac{\rho_{\rm dust}}{a^{3\,(1+\alpha)}} \Bigr)^{-\alpha/(1+\alpha)}
\end{array}
\label{eq.7.3.3}
\end{equation}
and eq.~(\ref{eq.7.2.1}) has the form:
\begin{equation}
  -\displaystyle\frac{d^{2}}{dq^{2}}\, \varphi(q) + (V_{0} - E_{\rm rad} + V_{1}\, q)\: \varphi(q) = 0.
\label{eq.7.3.4}
\end{equation}
After the change of variable
\begin{equation}
\begin{array}{cc}
  \zeta = |V_{1}|^{1/3}\, q, &
  \displaystyle\frac{d^{2}}{dq^{2}} =
  \Bigl(\displaystyle\frac{d\zeta}{dq} \Bigr)^{2}\, \displaystyle\frac{d^{2}}{d\zeta^{2}} =
  |V_{1}|^{2/3}\; \displaystyle\frac{d^{2}}{d\zeta^{2}}
\end{array}
\label{eq.7.3.6}
\end{equation}
eq.~(\ref{eq.7.3.4}) becomes:
\begin{equation}
  \displaystyle\frac{d^{2}}{d\zeta^{2}}\, \varphi(\zeta) +
  \biggl\{\displaystyle\frac{E_{\rm rad} - V_{0}} {|V_{1}|^{2/3}} -
          \displaystyle\frac{V_{1}} {|V_{1}|}\: \zeta\biggr\}\: \varphi(\zeta) = 0.
\label{eq.7.3.7}
\end{equation}
After the new change
\begin{equation}
  \xi = \displaystyle\frac{E_{\rm rad} - V_{0}} {|V_{1}|^{2/3}} - \displaystyle\frac{V_{1}} {|V_{1}|}\: \zeta
\label{eq.7.3.8}
\end{equation}
we have
\begin{equation}
  \displaystyle\frac{d^{2}}{d\xi^{2}}\, \varphi(\xi) + \xi\, \varphi(\xi) = 0.
\label{eq.7.3.9}
\end{equation}
From eqs.~(\ref{eq.7.3.6}) and (\ref{eq.7.3.8}) we have:
\begin{equation}
  \xi = \displaystyle\frac{E_{\rm rad} - V_{0}} {|V_{1}|^{2/3}} - \displaystyle\frac{V_{1}} {|V_{1}|^{2/3}}\: q.
\label{eq.7.3.10}
\end{equation}

Using such corrections after inclusion of the density component of the Chaplygin gas, we have calculated the wave function and on its basis the coefficients of penetrability, reflection and mixing by the formalism presented above.
Now following the method of Sec.~\ref{sec.5.1}, we have defined the incident and reflected waves relatively to a new boundary which is located in the minimum of the hole in the internal region. Results are presented in Tabl.~3. One can see that penetrability changes up to 100 times, in such a coordinate, in dependence on the location of the boundary  or in the internal turning point (for the same barrier shape and energy $E_{\rm rad}$)! This confirms that the coordinate where incident and reflected waves are defined has essential influence on estimation of the coefficients of penetrability and reflection. This result shows that the method proposed in the present paper has physical sense. In the next Tabl.~4, we demonstrate the fulfillment of the property (\ref{eq.4.2.3}) inside the entire energy range, which is calculated on the basis of the coefficients of penetrability, reflection and mixing obtained before.
% One can see that accuracy is ????the first 10--12 digits????.
% Of course, every semiclassical calculation is not able to give such accuracy for penetrability for the studied cosmological barriers.

% *******************************************************************************************************************

% *******************************************************************************************************************
\section{Multiple internal reflections fully quantum method
\label{sec.7}}

\subsection{Passage to non-stationary WDW equation: motivations
\label{sec.7.1}}

Tunneling is a pure quantum phenomenon characterized by the fact that a particle crosses through a classically-forbidden region of the barrier. By such a reason, the process of incidence of the particle on the barrier and its further tunneling and reflection are connected by unite cause-effect relation.
So, the dynamical consideration of the tunneling process through cosmological barriers is a natural one.
%not only motivated, but also natural.
The rejection of the dynamical consideration of tunneling from quantum cosmology {\bf limits the possible connection between initial stage, when the wave is incident on the barrier, and next propagation of this wave. }
This leads to uncertainties in determination of penetrability and rates. According to quantum mechanics, a particle is a quantum object having properties both particle and wave. In the classically forbidden regions the wave properties of the studied object are evident. So, the wave description of tunneling is natural.

So, we define a non-stationary generalization of WDW equation as
\begin{equation}
\begin{array}{l}
  \biggl(
    \displaystyle\frac{\partial^{2}}{\partial a^{2}} -
    V_{\rm eff}\,(a)
  \biggr)
  \Psi(a, \tau) =
  - i\,\displaystyle\frac {\partial}{\partial \tau}\, \Psi(a, \tau),
\end{array}
\label{eq.8.1.1}
\end{equation}
where $\tau$ is a new variable describing dynamics of evolution of the wave function being analog of time.
According to quantum mechanics, the penetrability and reflection are stationary characteristics, and such characteristics, obtained in the following, are independent on the parameter $\tau$. Note that all these characteristics are solutions of stationary WDW equation, while non-stationary consideration of multiple packets moving along barrier gives clear understanding of the process.

\begin{figure}[h]
\centerline{\includegraphics[width=55mm]{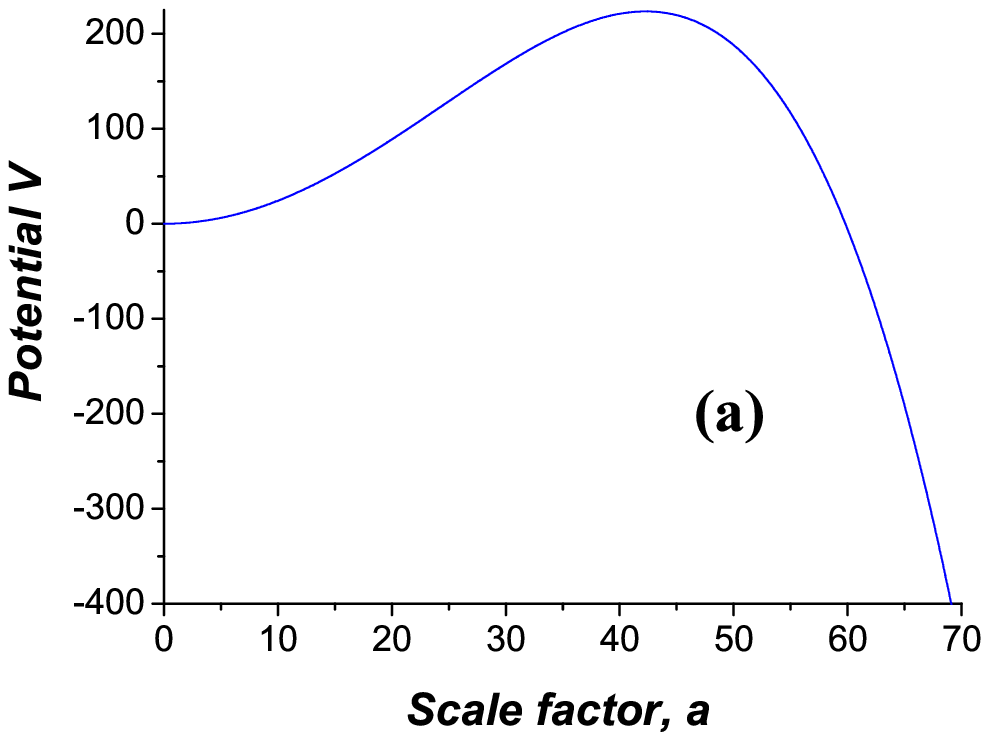}
\includegraphics[width=55mm]{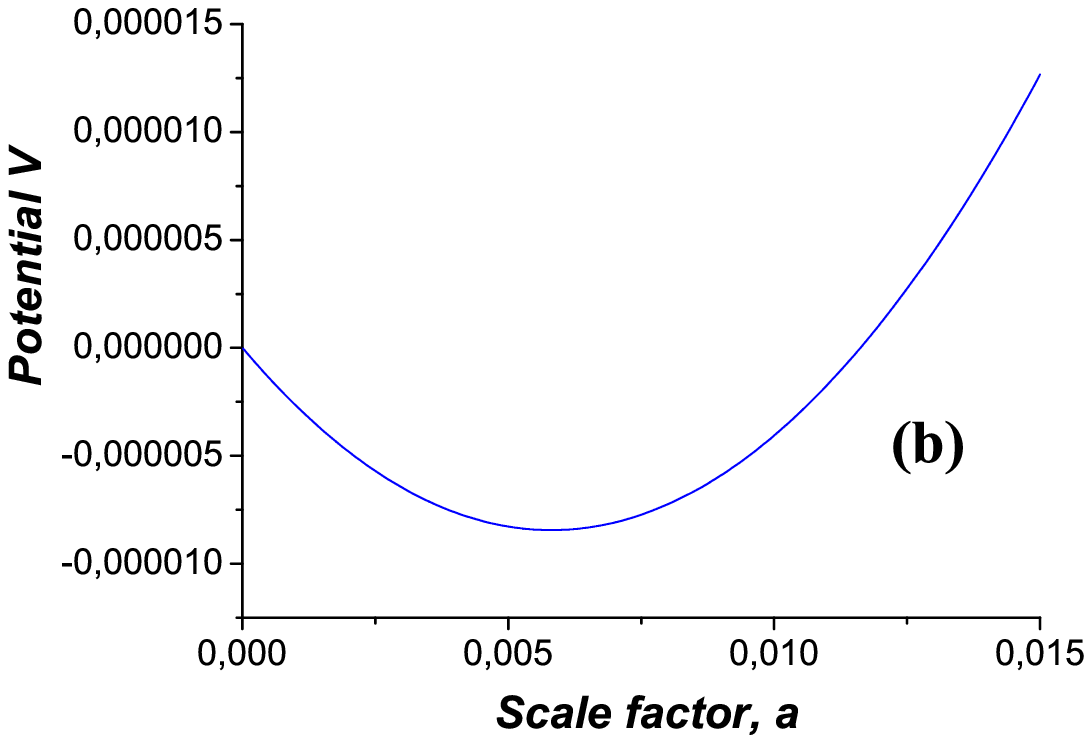}}
\vspace{-1mm}
\caption{\small
Behavior of the potential (\ref{eq.8.1.3}) after change $a_{\rm new} = a_{\rm old}\, \sqrt{12}$ at $\bar{A}=0.001$ and $\bar{B}=0.001$ (choice of parameters see in fig.~1, tables~I and II in \citep{Monerat.2007.PRD}):
(a) shape of the barrier ($V_{\rm max} = 223.52$ at $a=42.322$);
(b) there is a little internal well close to zero ($V_{\rm min} = -8.44$ at $a=0.00581$)
\label{fig.model_Monerat.1}}
\end{figure}

In order to give a basis to readers to estimate ability of the approach developed in this paper, let us consider results in \citep{Monerat.2007.PRD} (see eq.~(19)).
Here was studied the non-stationary WDW equation
\begin{equation}
  \biggl(
    \displaystyle\frac{1}{12}\,
    \displaystyle\frac{\partial^{2}}{\partial a^{2}} -
    V_{\rm eff}\,(a)
  \biggr)
  \Psi(a, \tau) =
  - i\,\displaystyle\frac {\partial}{\partial \tau}\, \Psi(a, \tau)
\label{eq.8.1.2}
\end{equation}
with the potential for the closed FRW model with the included generalized Chaplygin gas.
\begin{equation}
\begin{array}{cc}
\vspace{3mm}
  V_{\rm eff} (a) =
    3\,a^{2} -
    \displaystyle\frac{a^{4}}{\pi}\,
    \sqrt{\bar{A} + \displaystyle\frac{\bar{B}}{a^{6}}}
\end{array}
\label{eq.8.1.3}
\end{equation}
After change of variable $a_{\rm new} = a_{\rm old}\, \sqrt{12}$ the non-stationary eq.~(\ref{eq.8.1.2}) transforms into our eq.~(\ref{eq.8.1.1}) simce the $V_{\rm eff}$ potential is independent on the $\tau$ variable (such a choice allows a correspondence between energy levels, convenient in comparative analysis).
The potential~(\ref{eq.8.1.2}) after such a transformation is shown in figs.~\ref{fig.model_Monerat.1}.
We shall analyze the behavior of the wave function.
% *******************************************************************************************************************

% *******************************************************************************************************************
\subsection{Tunneling of the packet through a barrier composed from arbitrary number of rectangular steps
\label{sec.7.3}}

Now let us come to another more difficult problem, namely that a packet penetrating through the radial barrier of arbitrary shape in a cosmological problem. In order to apply the idea of multiple internal refections for study the packet tunneling through the real barrier, we have to generalize the formalism of the multiple internal reflections presented above.
We shall assume that the total potential has successfully been approximated by finite number $N$ of rectangular steps:
\begin{equation}
  V(a) = \left\{
  \begin{array}{cll}
    V_{1},   & \mbox{at } a_{\rm min} < a \leq a_{1}      & \mbox{(region 1)}, \\
    V_{2},   & \mbox{at } a_{1} < a \leq a_{2}         & \mbox{(region 2)}, \\
%    V_{3},   & \mbox{at } a_{2} < a \leq a_{3}        & \mbox{(region 3)}, \\
    \ldots   & \ldots & \ldots \\
%    V_{N-1}, & \mbox{at } a_{N-2} < a \leq a_{N-1}     & \mbox{(region $N-1$)}, \\
    V_{N},   & \mbox{at } a_{N-1} < a \leq a_{\rm max} & \mbox{(region $N$)},
  \end{array} \right.
\label{eq.8.3.1}
\end{equation}
where $V_{i}$ are constants ($i = 1 \ldots N$).
Let us assume that the packet starts to propagate outside inside the region with some arbitrary number $M$ (for simplicity, we denote its left boundary $a_{M-1}$ as $a_{\rm start}$) from the left of the barrier. We are interested in solutions for energies above that of the barrier while the solution for tunneling could be obtained after by change $i\,\xi_{i} \to k_{i}$. A general solution of the wave function (up to its normalization) has the following form:
\begin{equation}
\varphi\,(a) = \left\{
\begin{array}{lll}
   \alpha_{1}\, e^{ik_{1}a} + \beta_{1}\, e^{-ik_{1}a}, \\
     \vspace{1mm}
     \quad \mbox{at } a_{\rm min} \leq a \leq a_{1} \quad \mbox{(region 1)}, \\
   \quad \ldots \\
   \alpha_{M-1}\, e^{ik_{M-1}a} + \beta_{M-1}\, e^{-ik_{M-1}a}, \\
     \vspace{3mm}
     \quad \mbox{at } a_{M-2} \leq a \leq a_{\rm M-1} \quad \mbox{(region $M-1$)}, \\

   e^{ik_{M}a} + A_{R}\,e^{-ik_{M}a}, \\
     \vspace{3mm}
     \quad \mbox{at } a_{\rm M-1} < a \leq a_{M} \quad \mbox{(region $M$)}, \\

   \alpha_{M+1}\, e^{ik_{M+1}a} + \beta_{M+1}\, e^{-ik_{M+1}a}, \\
     \vspace{1mm}
     \quad \mbox{at } a_{M} \leq a \leq a_{M+1} \quad \mbox{(region $M+1$)}, \\
%    \alpha_{3}\, e^{ik_{3}a} + \beta_{3}\, e^{-ik_{3}a}, & \mbox{at } a_{2} \leq a \leq a_{3} & \mbox{(region 3)}, \\
   \quad \ldots \\

   \alpha_{n-1}\, e^{ik_{N-1}a} + \beta_{N-1}\, e^{-ik_{N-1}a}, \\
     \vspace{3mm}
     \quad \mbox{at } a_{N-2} \leq a \leq a_{N-1} \quad \mbox{(region $N-1$)}, \\

   A_{T}\,e^{ik_{N}a},\; \mbox{at } a_{N-1} \leq a \leq a_{\rm max} \quad \mbox{(region $N$)},
\end{array} \right.
\label{eq.8.3.2}
\end{equation}
where $\alpha_{j}$ and $\beta_{j}$ are unknown amplitudes, $A_{T}$ and $A_{R}$ are unknown amplitudes of transmission and reflection, $k_{i} = \frac{1}{\hbar}\sqrt{2m(E-V_{i})}$ are complex wave numbers. We have fixed the normalization so that the modulus of the starting wave $e^{ik_{M}a}$ equals to one. We look for a solution of such a problem by the approach of the multiple internal reflections.

{\bf Let us consider the initial stage when the packet starts to propagate to the right in the region with number $M$.}
% To start with, we shall study its propagation inside the right part of the potential with barrier, starting from this region.
{\bf According to the method of the multiple internal reflections, propagation of the packet through the barrier is considered by steps of its propagation relatively to each boundary (see \citep{Maydanyuk.2002.JPS,Maydanyuk.2003.PhD-thesis,Maydanyuk.2006.FPL}, for details).
Each next step in such a consideration of propagation of the packet will be similar to the first $2N-1$ steps. From analysis of these steps recurrent relations are found for calculation of all unknown amplitudes $A_{T}^{(n)}$, $A_{R}^{(n)}$, $\alpha_{j}^{(n)}$ and $\beta_{j}^{(n)}$ for arbitrary step $n$ (for region with number $j$), summation of these amplitudes are calculated. }
We shall look for the unknown amplitudes, requiring the wave function and its derivative to be continuous at each boundary. We shall consider the coefficients $T_{1}^{\pm}$, $T_{2}^{\pm}$ \ldots and $R_{1}^{\pm}$, $R_{2}^{\pm}$ \ldots as additional factors to amplitudes $e^{\pm i\,k\,a}$. Here, the bottom index denotes the number of the region, upper (top) signs ``$+$'' and ``$-$'' denote directions of the wave to the right or to the left, correspondingly. To begin with, we calculate $T_{1}^{\pm}$, $T_{2}^{\pm}$ \ldots $T_{N-1}^{\pm}$ and $R_{1}^{\pm}$,
$R_{2}^{\pm}$ \ldots $R_{N-1}^{\pm}$:
\begin{equation}
\begin{array}{ll}
\vspace{2mm}
   T_{j}^{+} = \displaystyle\frac{2k_{j}}{k_{j}+k_{j+1}} \,e^{i(k_{j}-k_{j+1}) a_{j}}, &
   T_{j}^{-} = \displaystyle\frac{2k_{j+1}}{k_{j}+k_{j+1}} \,e^{i(k_{j}-k_{j+1}) a_{j}}, \\
   R_{j}^{+} = \displaystyle\frac{k_{j}-k_{j+1}}{k_{j}+k_{j+1}} \,e^{2ik_{j}a_{j}}, &
   R_{j}^{-} = \displaystyle\frac{k_{j+1}-k_{j}}{k_{j}+k_{j+1}} \,e^{-2ik_{j+1}a_{j}}.
\end{array}
\label{eq.8.3.3}
\end{equation}
Analyzing all possible ``paths'' of the propagations of all possible packets inside the barrier and internal well,
we obtain:
\begin{equation}
\begin{array}{lcl}
  \sum\limits_{n=1}^{+\infty} A_{\rm inc}^{(n)} & = &
    1 + \tilde{R}_{M}^{+}\,\tilde{R}_{M-1}^{-} +
    \tilde{R}_{M}^{+}\,\tilde{R}_{M-1}^{-} \cdot \tilde{R}_{M}^{+}\,\tilde{R}_{M-1}^{-} + ... = \\
\vspace{2mm}
  & = &
    1 + \sum\limits_{m=1}^{+\infty} \bigl(\tilde{R}_{M}^{+}\,\tilde{R}_{M-1}^{-}\bigr)^{m} =
    \displaystyle\frac{1}{1 - \tilde{R}_{M}^{+}\,\tilde{R}_{M-1}^{-}}, \\

  \sum\limits_{n=1}^{+\infty} A_{T}^{(n)} & = &
  \Bigl( \sum\limits_{n=1}^{+\infty} A_{\rm inc}^{(n)} \Bigr) \cdot
  \Bigl\{ \tilde{T}_{N-2}^{+}\,T_{N-1}^{+} + \\
    & + &
         \tilde{T}_{N-2}^{+}\cdot R_{N-1}^{+}\,\tilde{R}_{N-2}^{-}\cdot T_{N-1}^{+} + ... \Bigr\} = \\
\vspace{2mm}
  & = & \Bigl( \sum\limits_{n=1}^{+\infty} A_{\rm inc}^{(n)} \Bigr) \cdot \tilde{T}_{N-1}^{+}, \\

  \sum\limits_{n=1}^{+\infty} A_{R}^{(n)} & = &
    \tilde{R}_{M}^{+} + \tilde{R}_{M}^{+} \cdot \tilde{R}_{M-1}^{-}\,\tilde{R}_{M}^{+} + \\
  & + &
    \tilde{R}_{M}^{+} \cdot \tilde{R}_{M-1}^{-}\,\tilde{R}_{M}^{+} \cdot \tilde{R}_{M-1}^{-}\,\tilde{R}_{M}^{+} + ... =\\
  & = &
    \tilde{R}_{M}^{+} \cdot
      \Bigl( 1 + \sum\limits_{m=1}^{+\infty} \bigl(\tilde{R}_{M-1}^{-}\,\tilde{R}_{M}^{+}\bigr)^{m} \Bigr) = \\
  & = & \displaystyle\frac{\tilde{R}_{M}^{+}} {1 - \tilde{R}_{M-1}^{-}\,\tilde{R}_{M}^{+}} =
  \Bigl( \sum\limits_{n=1}^{+\infty} A_{\rm inc}^{(n)} \Bigr) \cdot \tilde{R}_{M}^{+},
\end{array}
\label{eq.8.3.4}
\end{equation}
where
\begin{equation}
\begin{array}{lcl}
   \vspace{1mm}
   \tilde{R}_{j-1}^{+} & = &
     R_{j-1}^{+} + T_{j-1}^{+} \tilde{R}_{j}^{+} T_{j-1}^{-}
     \Bigl(1 + \sum\limits_{m=1}^{+\infty} (\tilde{R}_{j}^{+}R_{j-1}^{-})^{m} \Bigr) = \\
     & = & R_{j-1}^{+} +
     \displaystyle\frac{T_{j-1}^{+} \tilde{R}_{j}^{+} T_{j-1}^{-}} {1 - \tilde{R}_{j}^{+} R_{j-1}^{-}}, \\

   \vspace{1mm}
   \tilde{R}_{j+1}^{-} & = &
     R_{j+1}^{-} + T_{j+1}^{-} \tilde{R}_{j}^{-} T_{j+1}^{+}
     \Bigl(1 + \sum\limits_{m=1}^{+\infty} (R_{j+1}^{+} \tilde{R}_{j}^{-})^{m} \Bigr) = \\
     & = & R_{j+1}^{-} +
     \displaystyle\frac{T_{j+1}^{-} \tilde{R}_{j}^{-} T_{j+1}^{+}} {1 - R_{j+1}^{+} \tilde{R}_{j}^{-}}, \\

   \tilde{T}_{j+1}^{+} & = &
     \tilde{T}_{j}^{+} T_{j+1}^{+}
     \Bigl(1 + \sum\limits_{m=1}^{+\infty} (R_{j+1}^{+} \tilde{R}_{j}^{-})^{m} \Bigr) =
     \displaystyle\frac{\tilde{T}_{j}^{+} T_{j+1}^{+}} {1 - R_{j+1}^{+} \tilde{R}_{j}^{-}}.
\end{array}
\label{eq.8.3.5}
\end{equation}
Choosing as starting points, the following:
\begin{equation}
\begin{array}{ccc}
  \tilde{R}_{N-1}^{+} = R_{N-1}^{+}, & \quad
  \tilde{R}_{M}^{-} = R_{M}^{-}, & \quad
  \tilde{T}_{M}^{+} = T_{M}^{+},
\end{array}
\label{eq.8.3.6}
\end{equation}
we calculate the coefficients $\tilde{R}_{N-2}^{+}$ \ldots $\tilde{R}_{M}^{+}$, $\tilde{R}_{M+1}^{-}$ \ldots
$\tilde{R}_{N-1}^{-}$ and $\tilde{T}_{M+1}^{+}$ \ldots $\tilde{T}_{N-1}^{+}$.

We shall consider propagation of all packets in the region with number $M$, to the left. Such packets are formed in result of all possible reflections from the right part of potential, starting from the boundary $a_{M}$. In the previous section to describe their reflection from the left boundary $R_{0}$ to the right one, we used coefficient $R_{0}^{-}$. Now since we want to pass from simple boundary $a_{M-1}$ to the left part of the potential well starting from this point up to $a_{\rm min}$, we generalize the coefficient $R_{M-1}^{-}$ to $\tilde{R}_{M-1}^{-}$.
{\bf The middle formula in (\ref{eq.8.3.5}) is applicable when we use eqs.~(\ref{eq.8.3.3}) for definition of $T_{i}^{\pm}$ and $R_{i}^{\pm}$. }
Finally, we determine coefficients $\alpha_{j}$ and $\beta_{j}$:
\begin{equation}
\begin{array}{lcl}
   \vspace{1mm}
   \sum\limits_{n=1}^{+\infty} \alpha_{j}^{(n)} & = &
     \tilde{T}_{j-1}^{+}
     \Bigl(1 + \sum\limits_{m=1}^{+\infty} (R_{j}^{+} \tilde{R}_{j-1}^{-})^{m} \Bigr) = \\
   \vspace{2mm}
     & = &
     \displaystyle\frac{\tilde{T}_{j-1}^{+}} {1 - R_{j}^{+} \tilde{R}_{j-1}^{-}} =
     \displaystyle\frac{\tilde{T}_{j}^{+}} {T_{j}^{+}}, \\

   \sum\limits_{n=1}^{+\infty} \beta_{j}^{(n)} & = &
     \tilde{T}_{j-1}^{+}
     \Bigl(1 + \sum\limits_{m=1}^{+\infty} (\tilde{R}_{j}^{+} \tilde{R}_{j-1}^{-})^{m} \Bigr)\, R_{j}^{+} = \\
     & = &
     \displaystyle\frac{\tilde{T}_{j-1}^{+}\, R_{j}^{+}} {1 - \tilde{R}_{j}^{+} \tilde{R}_{j-1}^{-}} =
     \displaystyle\frac{\tilde{T}_{j}^{+} R_{j}^{+}} {T_{j}^{+}},
\end{array}
\label{eq.8.3.7}
\end{equation}
the amplitudes of transmission and reflection:
\begin{equation}
\begin{array}{ll}
  A_{T} = \sum\limits_{n=1}^{+\infty} A_{T}^{(n)}, &
  A_{R} = \sum\limits_{n=1}^{+\infty} A_{R}^{(n)}, \\
  \alpha_{j} = \sum\limits_{n=1}^{+\infty} \alpha_{j}^{(n)} = \displaystyle\frac{\tilde{T}_{j}^{+}} {T_{j}^{+}}, &
  \beta_{j} = \sum\limits_{n=1}^{+\infty} \beta_{j}^{(n)} = \alpha_{j} \cdot R_{j}^{+}
\end{array}
\label{eq.8.3.8}
\end{equation}
and coefficients $T$ and $R$ describing penetration of the packet from the internal region outside
and its reflection from the barrier
\begin{equation}
\begin{array}{ll}
\vspace{1mm}
  T_{MIR} \equiv \displaystyle\frac{k_{N}}{k_{M}}\; \bigl|A_{T}\bigr|^{2} =
    \bigl|A_{\rm inc}\bigr|^{2} \cdot T_{\rm bar}, &
  T_{\rm bar} = \displaystyle\frac{k_{N}}{k_{M}}\; \bigl|\tilde{T}_{N-1}^{+} \bigr|^{2}, \\
  R_{MIR} \equiv \bigl|A_{R}\bigr|^{2} = \bigl|A_{\rm inc}\bigr|^{2} \cdot R_{\rm bar}, &
  R_{\rm bar} = \bigl|\tilde{R}_{M}^{+} \bigr|^{2}.
\end{array}
\label{eq.8.3.9}
\end{equation}
Choosing $a_{\rm min}=0$, we assume full propagation of the packet through such a boundary (with no possible reflection) and we have $R_{0}^{-} = -1$ (it could be interesting to analyze results with varying $R_{0}^{-}$).
{\bf We use the test:}
\begin{equation}
\begin{array}{ccc}
  \displaystyle\frac{k_{N}}{k_{M}}\; |A_{T}|^{2} + |A_{R}|^{2} = 1 & \mbox{ or }&
  T_{MIR} + R_{MIR} = 1.
\end{array}
\label{eq.8.3.10}
\end{equation}
Now if energy of the packet is located below then height of one step with number $m$, then the following change
\begin{equation}
  k_{m} \to i\,\xi_{m}
\label{eq.8.3.11}
\end{equation}
should be used for description of transition of this packet through such a barrier with its tunneling. In the case of a barrier consisting from two rectangular steps of arbitrary heights and widths we have already obtained coincidence between amplitudes calculated by method of MIR and the corresponding amplitudes found by standard approach of quantum mechanics up to first 15 digits. Even increasing the number of steps up to some thousands has the right accuracy to fulfill the property (\ref{eq.8.3.10}).

% (see Appendix~B in \citep{Maydanyuk.2011.JMP} where one can see details and algorithms of calculations of amplitudes for general quantum decay radial task).
In particular, we reconstruct completely the pictures of the probability and reflection presented in figs.~\ref{fig.model_Monerat.2} (a) and (b), figs.~\ref{fig.model_Monerat.3} (a) and (b), figs.~\ref{fig.model_Monerat.4}~(b), but using such a standard technique. So, the \emph{result concerning the oscillating dependence of the penetrability on the position of the starting point $a_{\rm start}$ in such figures is independent on the fully quantum method chosen for calculations}.

This is an important test which confirms reliability of the method MIR. So, we have obtained full coincidence between all amplitudes, calculated by method MIR and by standard approach of quantum mechanics. This is why we generalize the method MIR for description of tunneling of the packet through potential, consisting from arbitrary number of rectangular barriers and wells of arbitrary sizes.

% *******************************************************************************************************************

% *******************************************************************************************************************
\subsection{Results
\label{sec.7.4}}

\begin{figure*}
\resizebox{1\textwidth}{!}{%
\centerline{\includegraphics[width=70mm]{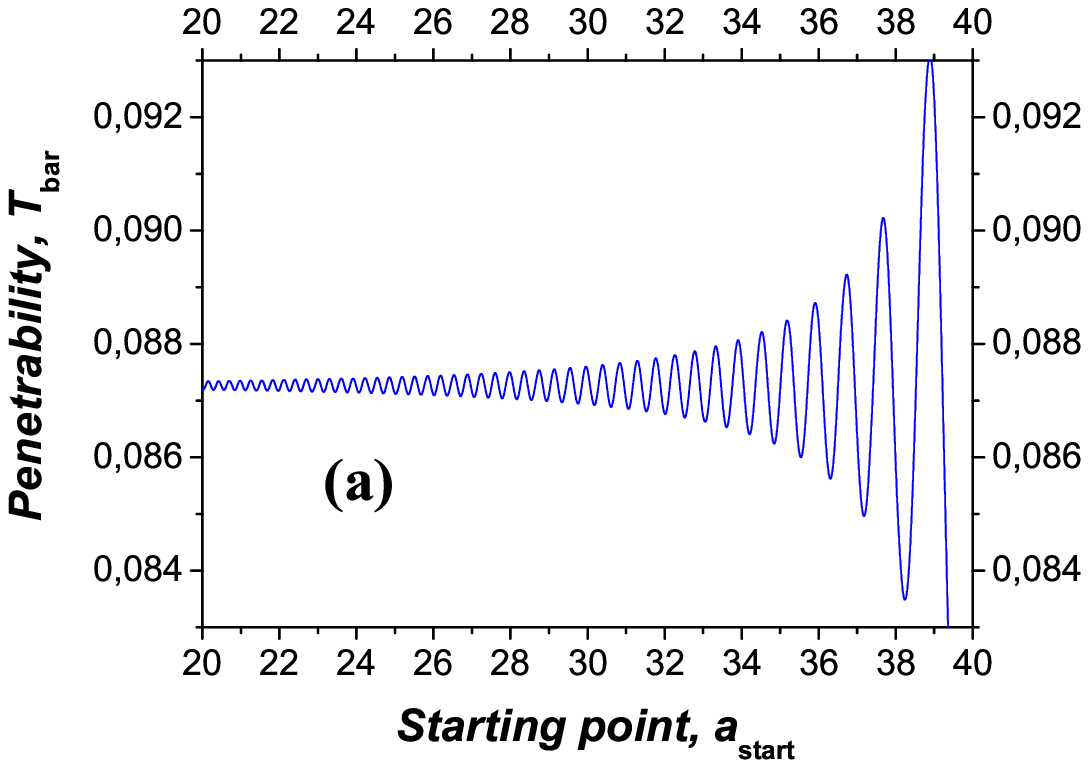}
\hspace{-7mm}\includegraphics[width=70mm]{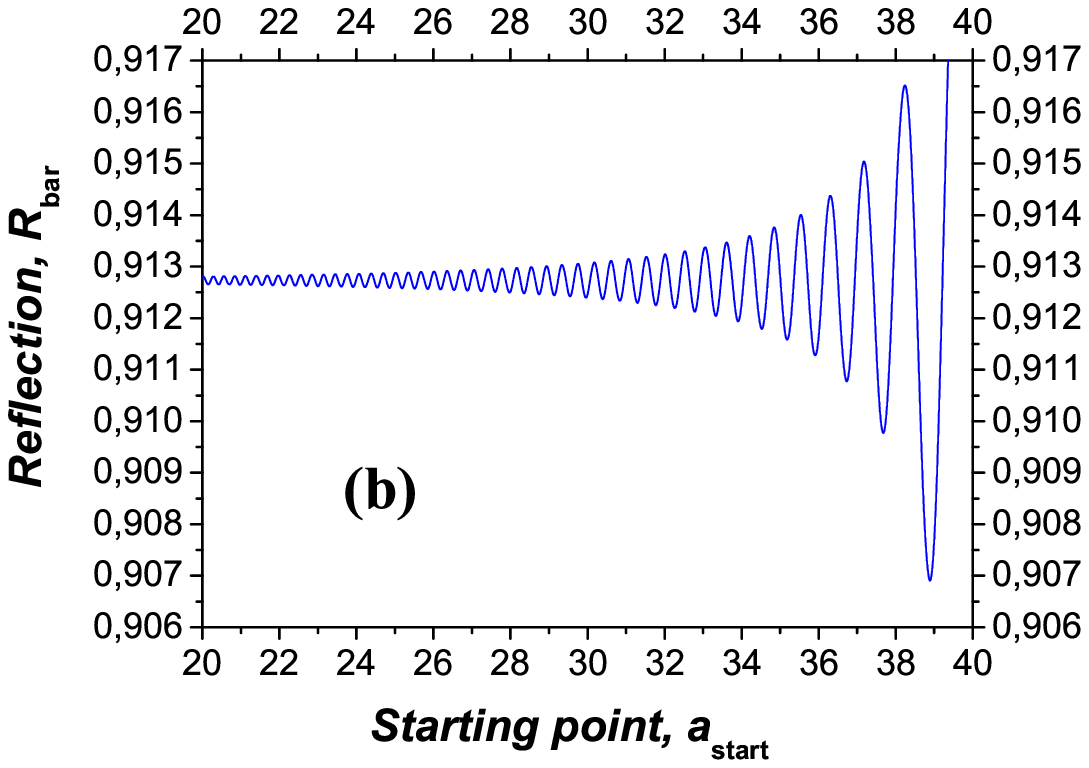}}}

\vspace{-7mm}
\resizebox{1\textwidth}{!}{%
\centerline{\includegraphics[width=70mm]{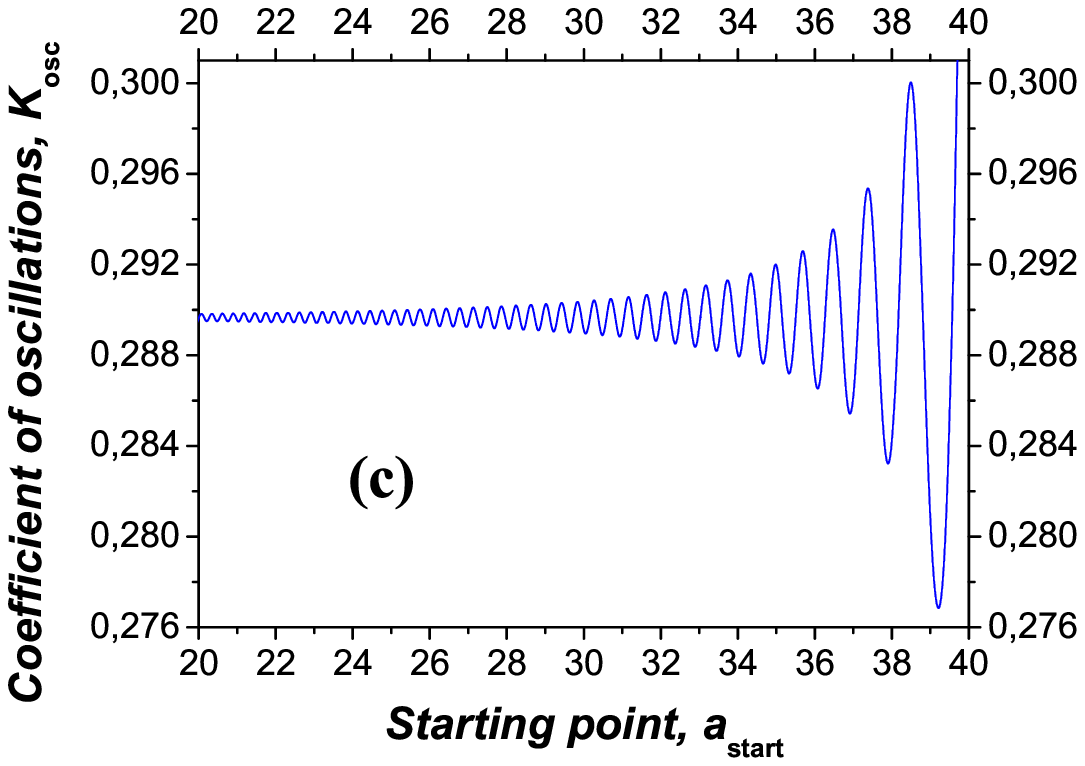}
\hspace{-7mm}\includegraphics[width=70mm]{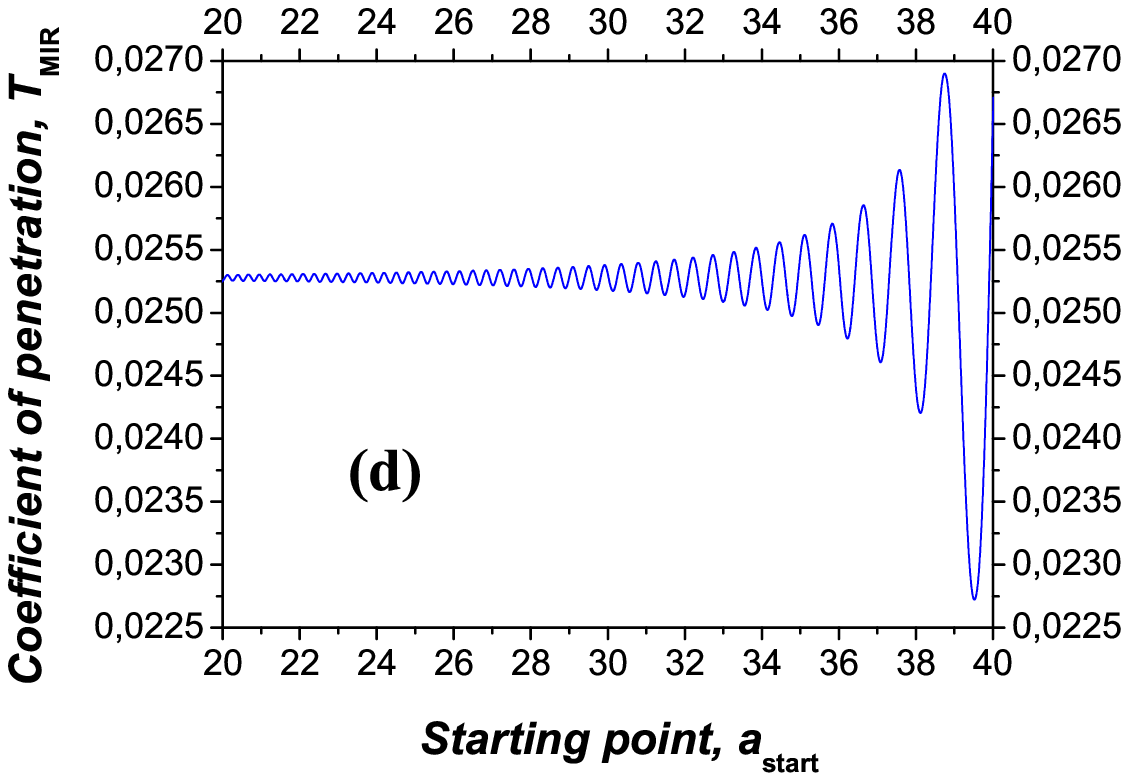}}
}

\vspace{3mm}
\caption{\small
Dependencies of the coefficients of the penetrability $T_{\rm bar}$ (a), reflection $R_{\rm bar}$ (b), coefficient of oscillations $K_{\rm osc}$ (c) and coefficient of penetration $T_{MIR}$ (d) in terms of the position of the starting point $a_{\rm start}$ for the energy $E=220$
($A=0.001, B=0.001$, $a_{\rm max} = 70$. The total number of intervals is 2000, for all presented cases the achieved accuracy is $|T_{\rm bar}+R_{\rm bar}-1| < 10^{-15}$).
These figures clearly demonstrate oscillating ({\it i.e.} not constant) behavior of all considered coefficients on $a_{\rm start}$.
\label{fig.model_Monerat.2}}
\end{figure*}

We have applied the above method to analyze the behavior of the packet tunneling through the barrier (\ref{eq.8.1.3}) (we used $a_{\rm new} \to \sqrt{12}\,a_{\rm old}$). The first interesting result is \emph{a visible change of the penetrability on the displacement of the starting point $a_{\rm min} \le a \le a_{1}$, where we put the packet}. Using the possibility of decreasing the width of intervals up to an enough small value (and choosing, for convenience, the width of each interval to be the same), we choose $a_{\rm min}$ as \emph{starting point} (and denote it as $a_{\rm start}$), from where the packet begins to propagate outside. We have analyzed how the position of such a point influences the penetrability. In fig.~\ref{fig.model_Monerat.2}~(a) one can see that the penetrability strongly changes in dependence of $a_{\rm start}$ for arbitrary values of energy of radiation $E_{\rm rad}$: it has oscillating behavior. Difference between its minimums and maximums is minimal at $a_{\rm start}$ in the center of the well (i.~e. its change tends to zero in the center of the well), this difference increases with increasing value of $a_{\rm start}$
and achieves the maximum close to the turning point. With this result, we may conclude that exists a \emph{dependence of penetrability on the starting point
$a_{\rm start}$ of the packet.}
%
% \begin{figure}[h]
The coefficients of reflection, oscillations and penetration on the position of the starting point $a_{\rm start}$ are presented in next figs.~\ref{fig.model_Monerat.2} (b), (c), (d) and have similar behavior.

Usually, in cosmological quantum models the penetrability is determined by the barrier shape. In the non-stationary approach one can find papers where the role of the initial condition is analyzed in calculations of rates, penetrability etc.\footnote{Such papers are very rare and questions about dynamics have not been studied deeply.} But, the stationary limit does not give us any choice on which to work. We conclude: (a) the penetrability should be connected with the initial condition (not only in non-stationary consideration, but also in the stationary one), which determines position (coordinate) of maximum of the packet which begins to propagate outside (at initial time moment $t=0$). (b) Even in the stationary consideration, the penetrability of the barrier should be determined in dependence on the initial condition.

% \begin{figure}[h]
\begin{figure*}
\resizebox{1\textwidth}{!}{%
\centerline{\includegraphics[width=70mm]{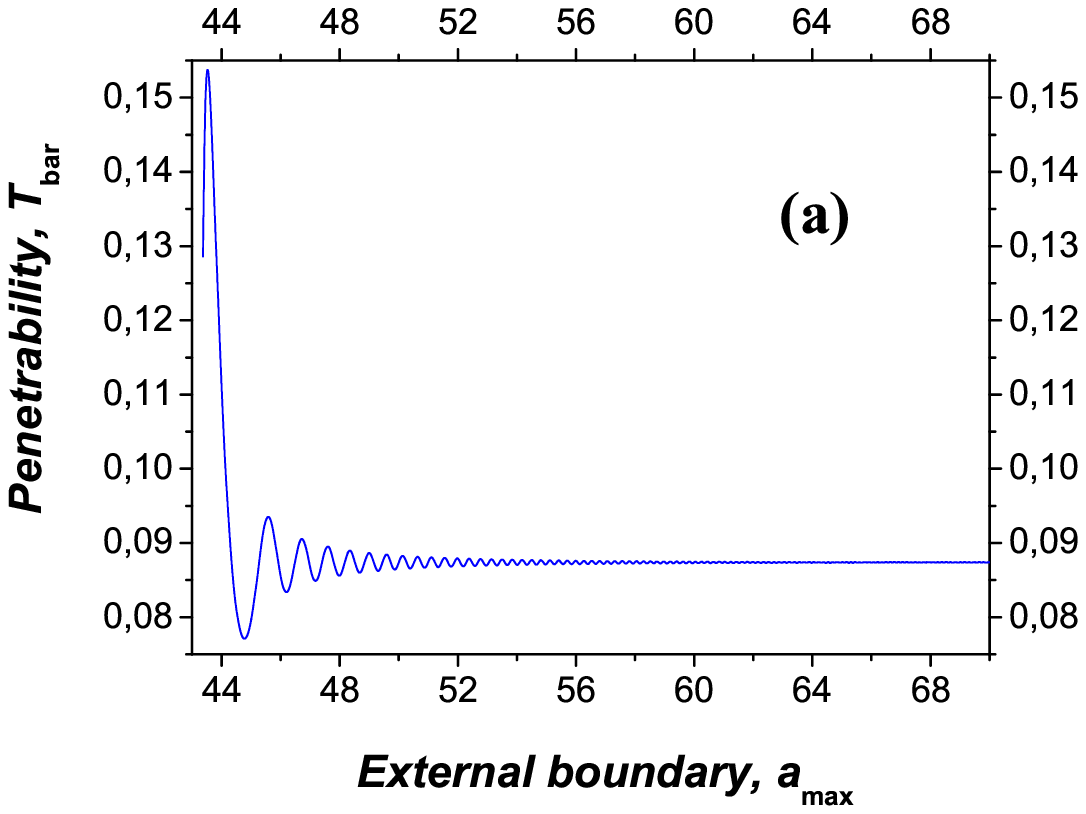}
\hspace{-7mm}\includegraphics[width=70mm]{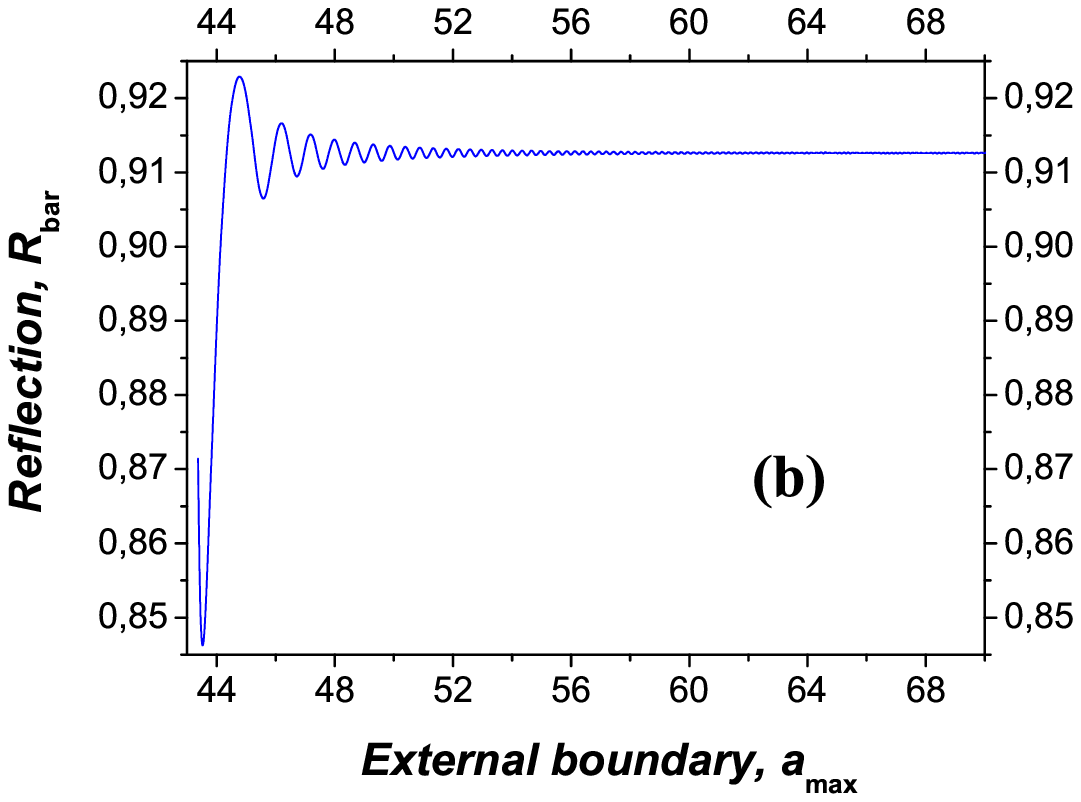}} }

\vspace{-7mm}
\resizebox{1\textwidth}{!}{%
\centerline{\includegraphics[width=70mm]{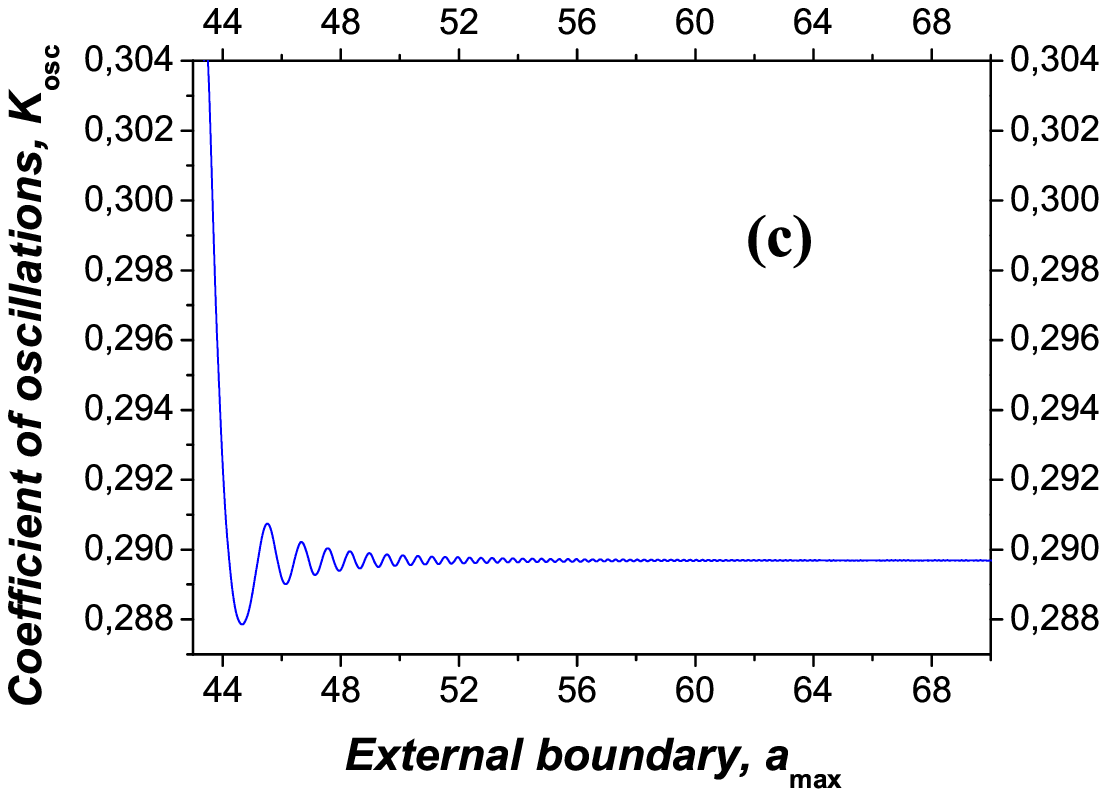}
\hspace{-7mm}\includegraphics[width=70mm]{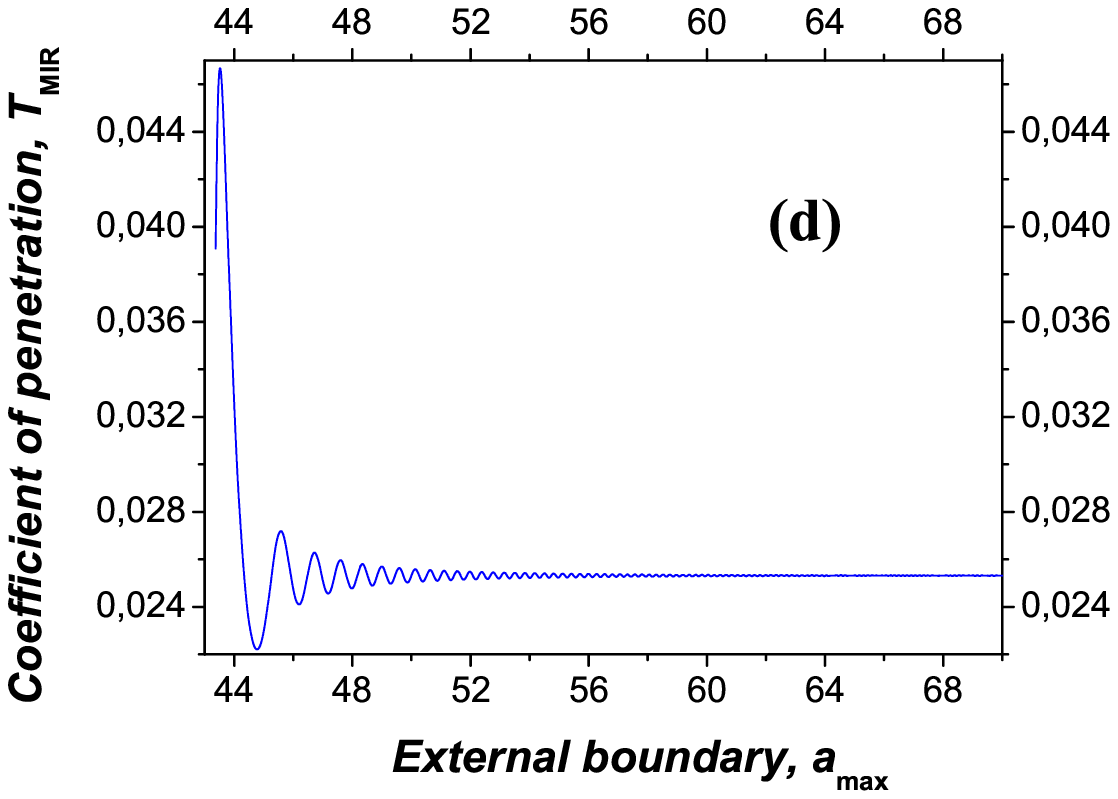}}
}
\vspace{-4mm}
\caption{\small
Dependencies of the coefficients of penetrability (a), reflection (b), oscillations (c) and penetration (d) on the position of the external region, $a_{\rm max}$ for the energy $E=223$ ($A=0.001, B=0.001$).
For all presented values we have achieved accuracy $|T_{\rm bar}+R_{\rm bar}-1| < 1 \cdot 10^{-15}$
(the maximum number of intervals is 2000).
\label{fig.model_Monerat.3}}
\end{figure*}

The first question is how much these results are reliable. In particular, how stable will such results be if we shift the external boundary outside?
% {\bf (while in the semiclassical approach we are restricted by two turning points only, then in the fully quantum approach the external tail of the barrier effects on the results inevitably and additionally)?
% Taking strong decreasing of the external tail of the barrier into account to minus infinity, one could even expect for this. }
The results of such calculations are presented in fig.~\ref{fig.model_Monerat.3}, where it is shown how the penetrability changes with $a_{\rm max}$
(for clearness sake, we have fixed the starting point: $a_{\rm start}=10$).
\begin{figure*}
\resizebox{1\textwidth}{!}{%
\centerline{\includegraphics[width=70mm]{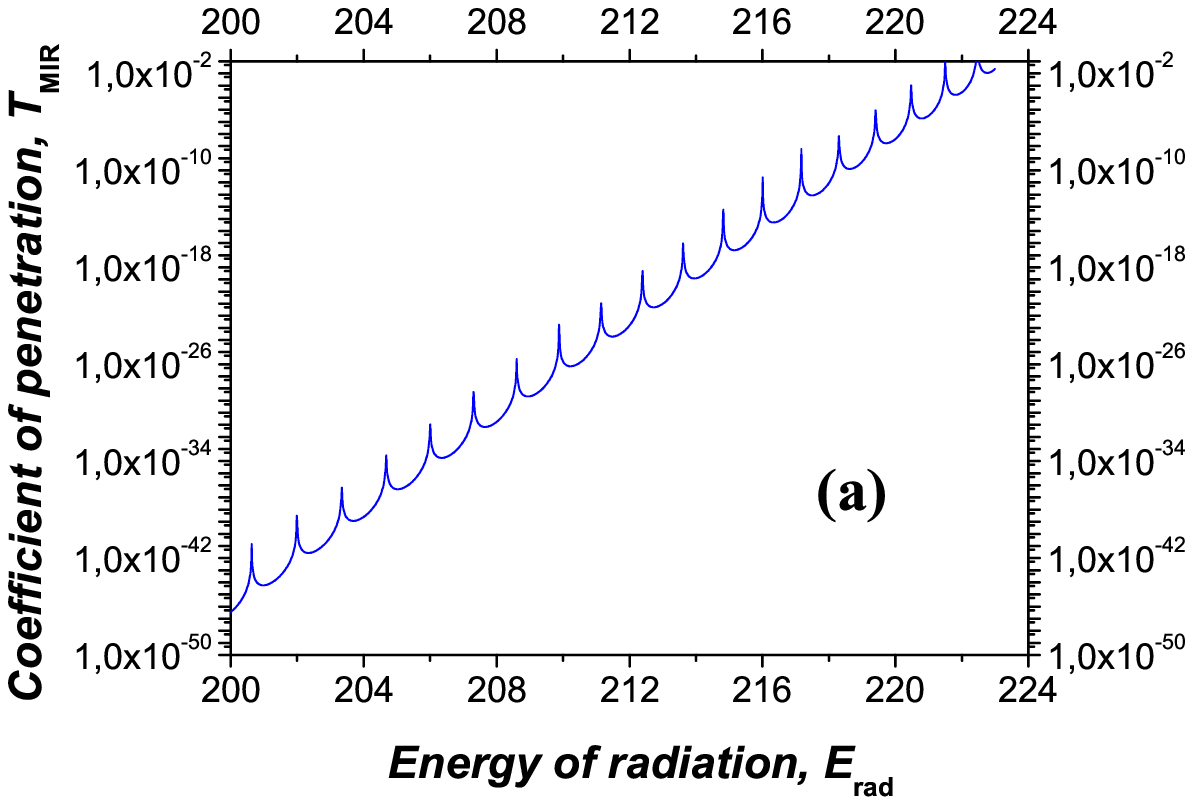}
\hspace{-4mm}\includegraphics[width=70mm]{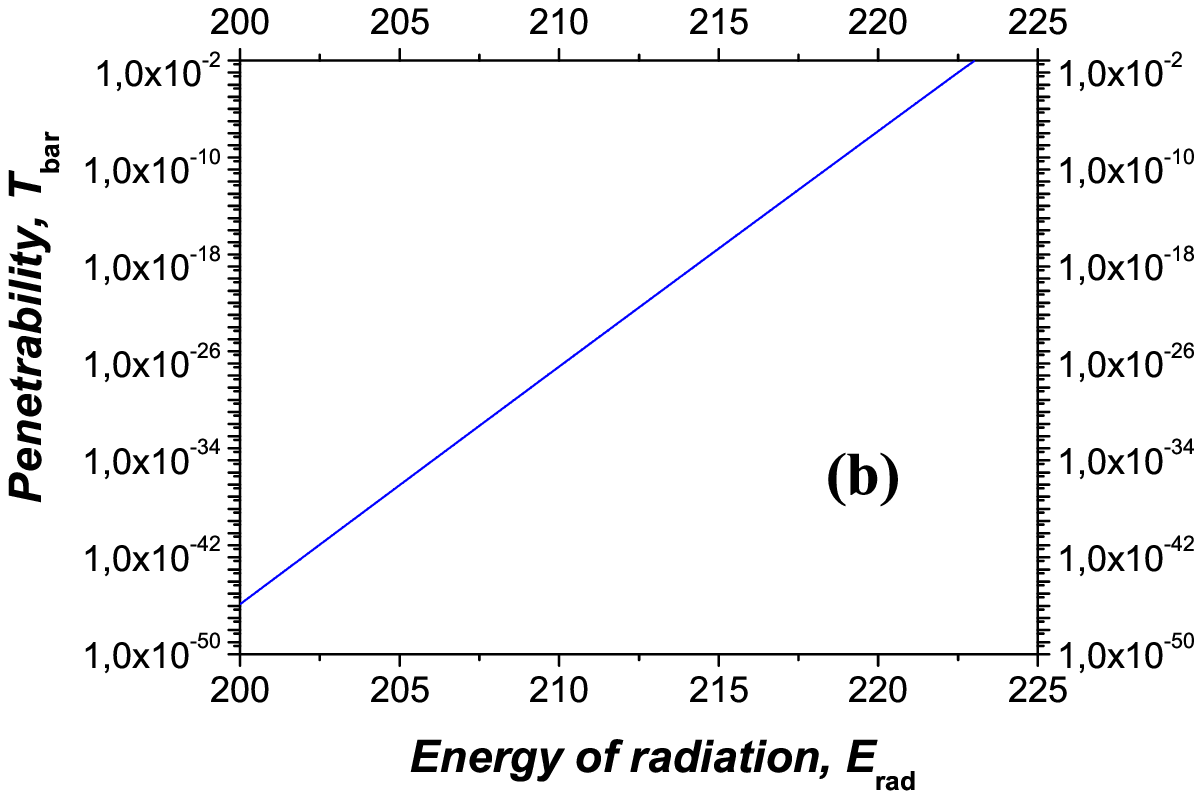}}}
\vspace{0mm}
\resizebox{1\textwidth}{!}{%
\centerline{\includegraphics[width=70mm]{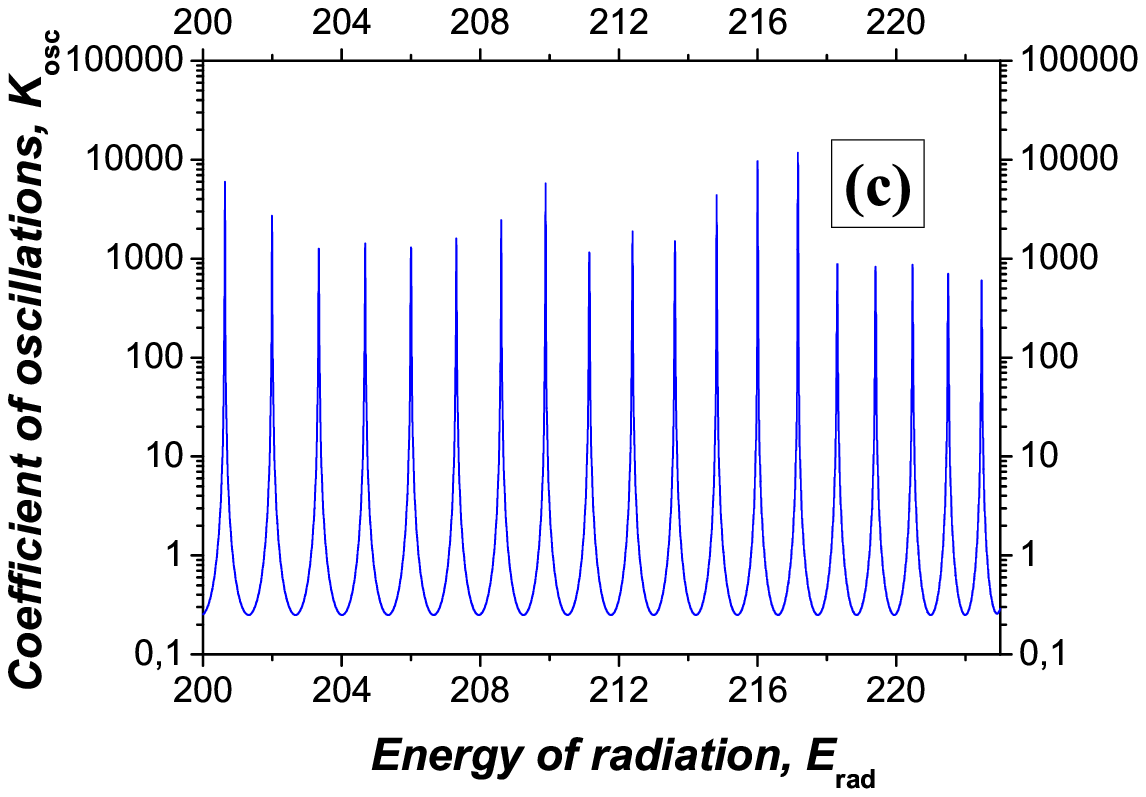}
\hspace{-4mm}\includegraphics[width=70mm]{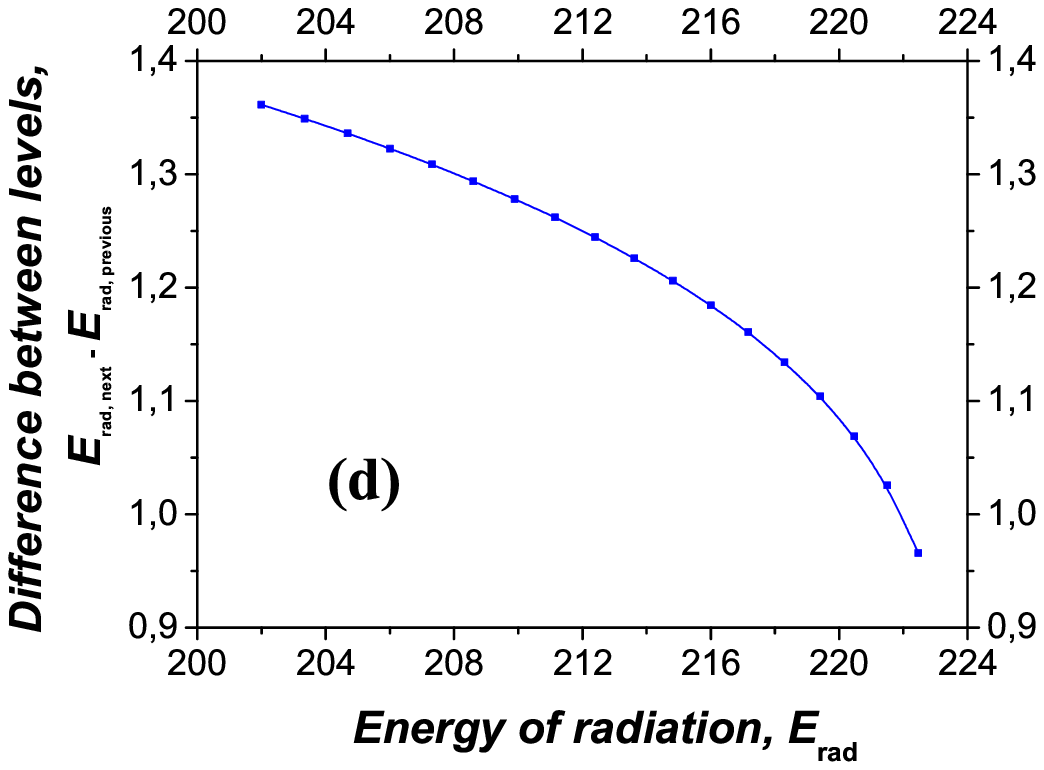}}}

\vspace{0mm}
\caption{\small
Dependencies of the coefficient of the coefficient of penetration $T_{MIR}$ (a), the coefficient of the penetrability $T_{\rm bar}$ (b), coefficient of oscillations $K_{\rm osc}$ (c) and difference $E_{\rm res,\, next} - E_{\rm res,\, previous}$ between two closest energy peaks (d) on the $E_{\rm rad}$ energy (we have choose: $A=0.001$ and $B=0.001$, $a_{\rm start}=10$, $a_{\rm max} = 70$, number of intervals inside the scale axis $a$ 1000, number of intervals of energy 100000). Inside the energy region $E_{\rm rad} = 200 - 223$ we observe 19 resonant peaks in the dependencies of coefficients $T_{MIR}$ and $K_{\rm osc}$ while the penetrability increases monotonously with increasing the $E_{\rm rad}$ energy.
\label{fig.model_Monerat.4}}
\end{figure*}
One can see that all calculations are well convergent, that confirms efficiency of the method of the multiple internal reflections.
% This points to that difficult attempts to properly determine the outgoing wave in the asymptotic limit properly ???? have no any practical sense.
On the basis of such results we choose $a_{\rm max}=70$ for further calculations. However, one can see that inclusion of the external region can change the coefficients of penetrability and penetration up to 2 times for the chosen energy level.

The second question is how strong this affects the calculations of the penetrability. If it was small than, the semiclassical approaches would have enough good approximation. From figs.~\ref{fig.model_Monerat.2} it follows that the penetrability is not strongly changed in dependence on shift of the starting point. However, such small variations are connected with relatively small height of the barrier and depth of the well, while they would be not small at another choice of parameters (the coefficient of oscillation and penetration turn out to change at some definite energies of radiation, see below). So, this effect is supposed to be larger at increasing height of the barrier and depth of the well, and also for near-barrier energies (i.~e. for energies comparable with the barrier height) and above-barrier energies of radiation).

We have analyzed how these characteristics change in dependence on the energy of radiation. We did not expect the results that we got (see figs.~\ref{fig.model_Monerat.4}).
%Result has turned out to be over-unexpected (see figs.~\ref{fig.model_Monerat.4}).
{\bf The coefficient of penetration has oscillations with peaks clearly shown.
These peaks are separated by similar distances and could be considered as resonances in energy scale.}
%, between which smooth and stable in calculations wells are observed with minimums. }
So, by using the fully quantum approach we observed for the first time clear pictures of resonances which could be connected with some early unknown quasi-stationary states.
%More detailed analysis shows the following.
At increasing energy of radiation the penetrability changes monotonously and determines a general tendency of change of the coefficient of penetration, while the coefficient of oscillations introduces the peaks. Now the reason of the presence of resonances has become clearer: oscillations of the packet inside the internal well produce them, while the possibility of the packet to penetrate through the barrier (described by the penetrability of the barrier) has no influence on them. In general, we observe 134 resonant levels inside energy range $E_{\rm rad}$ = 0--200, and else 19 levels inside $E_{\rm rad}$ = 200--223.

\begin{figure*}
\resizebox{1\textwidth}{!}{%
\centerline{\includegraphics[width=70mm]{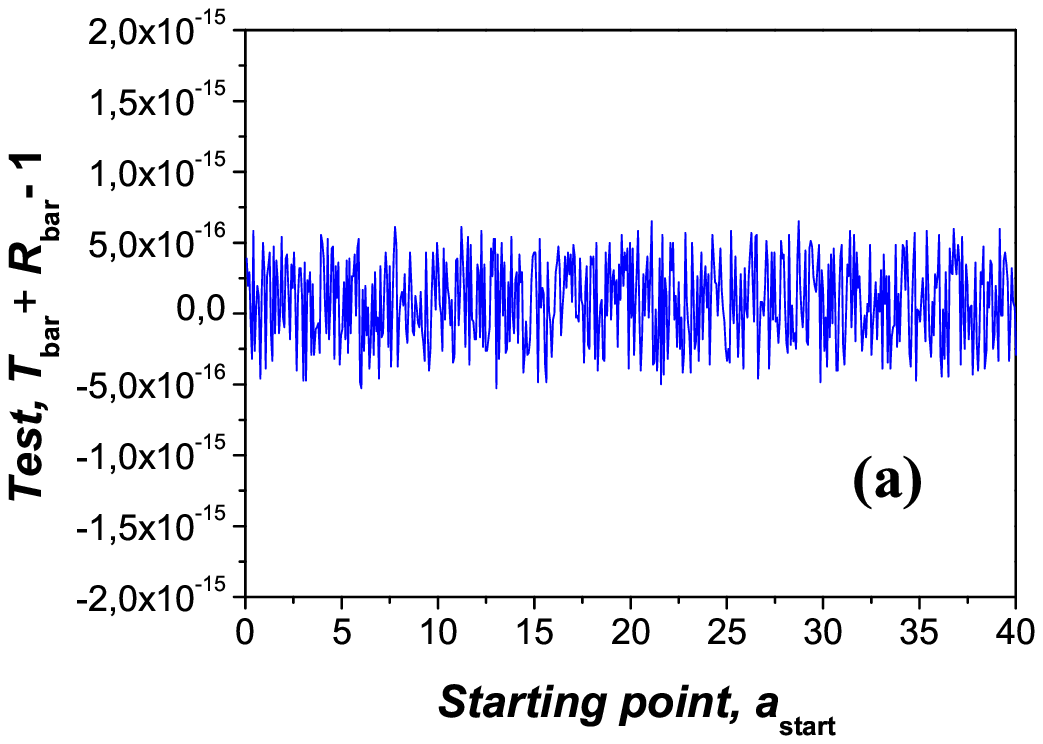}
\hspace{-7mm}\includegraphics[width=70mm]{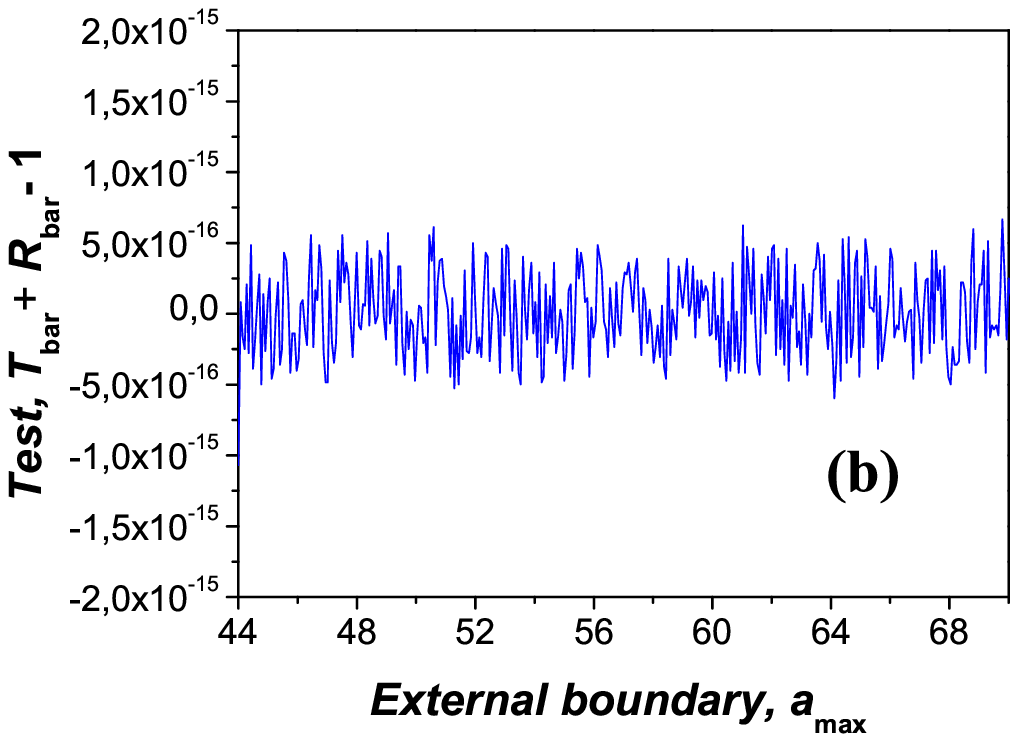}}}
% \vspace{-4mm}
\caption{\small
Accuracy of the obtained penetrability $T_{\rm bar}$ and reflection $R_{\rm bar}$ for the energy $E=220$ used in previous figs.~\ref{fig.model_Monerat.2} and \ref{fig.model_Monerat.3}.
As a test, we calculate $T_{\rm bar} + R_{\rm bar} - 1$ in dependence on the position of the starting point $a_{\rm start}$ (a) and the external boundary $a_{\rm max}$ (b) ($A=0.001, B=0.001$, total number of intervals is 2000).
\label{fig.model_Monerat.5}}
\end{figure*}

In the last fig.~\ref{fig.model_Monerat.5} one can see that we have achieved $|T_{\rm bar}+R_{\rm bar}-1| < 10^{-15}$ inside whole region of changes of $a_{\rm start}$ and $a_{\rm max}$ (such data were used in the previous figs.~\ref{fig.model_Monerat.2} and \ref{fig.model_Monerat.3}). This is the accuracy of the method of the multiple internal reflections in obtaining $T_{\rm bar}$ and $R_{\rm bar}$.

% *******************************************************************************************************************

% *******************************************************************************************************************
\subsection{The fully quantum penetrability versus semiclassical one in cosmology: a quick comparison
% Motivations: to implement fully quantum approach versus semiclassical in cosmology
\label{sec.introduction}}

Does the penetrability, determined according to the semiclassical theory by a shape of the barrier between two turning points, give exhaustive answers and the best estimations of rates of evolution of universe? If we look at figs.~\ref{fig.model_Monerat.2}~(a), we shall see that this is not the case. The penetrability is depended on the position (coordinate) of maximum of the packet which begins to propagate outside at time moment $t=0$. So, the penetrability should be a function of some parameters of the packet at its start.
For the first time, it has been demonstrated the difference between the fully quantum approach and the semiclassical
However, let us perform a general analysis.

% \vspace{2mm}
(1) {\bf If
%despite prevailing confidence in the semiclassical approach,
we wanted to check the semiclassical approach, we should miss some of the parameters. One can use test of $T + R = 1$ (where $T$ and $R$ are the penetrability through the barrier and reflection from it). But, note that the semiclassical approximation neglects the reflected waves in quantum mechanics (see~\citep{Landau.v3.1989}, eq.~(46.10), p.~205, p.~221--222). Therefore, we cannot use the test above for checking $T$ in the semiclassical theory.}

% \vspace{2mm}
(2) If we would like to determine the reflection coefficient, then we should find a more accurate semiclassical approximation (in order to take into account both decreasing and increasing components of the wave function in the tunneling region). In such a case, we shall face another problem, namely the presence of a non-zero interference between the incident and reflected waves. Now the relation $T + R = 1$ cannot be used as test, and one needs to take the third component $M$ of interference into account (see~\citep{Maydanyuk.2010.IJMPD}). If we improperly separate the exactly known full wave function in the incident and reflected waves\footnote{However, the semiclassical approaches have no apparatus for such an analysis.}, the interference component should increase without limit. In such a case, the penetrability and reflection can freely exceed unit and increase without limit. What is now the general meaning of the penetrability?

% \vspace{2mm}
(3) We shall give only some examples from quantum mechanics.
(i) If we consider two-dimensional penetration of the packet through the simplest rectangle barrier (with finite size), we shall see that the penetrability is directly dependent on direction of tunneling of the packet. So, the penetrability is not a single value but a function.
(ii) If we consider one-dimensional tunneling of the packet through the simplest rectangular barrier, we shall obtain ``interference picture'' of its amplitude in the transmitted region, which is dependent on time and space coordinates and is an exact analytical solution. Of course, the stationary part of such a result exactly coincides with well known stationary solutions \citep{Maydanyuk.2003.PhD-thesis}.
% From the previous arguments, one could think that the penetrability defined by the shape of the barrier between two turning points is nothing more than  {\bf the simplified understanding, while for accurate and deep analysis we need in stronger basis. }

% \vspace{2mm}
(4)
% Advance of the semiclassical approach is in simplicity of formula of the penetrability based on determination of the outgoing wave in the asymptotic region. }
A tunneling boundary condition \citep{Vilenkin.1995,Vilenkin.1994.PRD} seems to be natural and clear, where the wave function should represent an outgoing wave at large scale factor $a$. However, is such a wave free? In contrast to problems of quantum atomic and nuclear physics, in cosmology we deal with potentials, which modules increase with increasing the scale factor $a$ (their gradients increase, which have sense of force acting on the wave).
Therefore, in quantum cosmology we should define the boundary condition on the basis of the waves propagating inside strong fields (see~\citep{Maydanyuk.2010.IJMPD}).

% Therefore, ??? instead of the free wave in the asymptotic region (missing in problems of quantum cosmology), we should work with the waves propagating inside strong fields (see~\citep{Maydanyuk.2010.IJMPD}).

These points destroy the semiclassical basis of the cosmological models. Now the statement concerning reliability of the semiclassical approach become a question of `` faith'' (note that this is widespread \citep{Maydanyuk.2010.IJMPD,Maydanyuk.2011.EPJP}). The semiclassical approach could be compared with \emph{``black box''}, where deeper and more detailed information about the dynamics of the universe is hidden.

% {\bf In such a black box those missing elements are hidden, without which it is impossible to combine everything together in obtaining self-consistent quantum theory of the formation of the universe and its evolution in the first stage.}

% *******************************************************************************************************************

% *******************************************************************************************************************
\section{Conclusions and perspectives
\label{sec.conclusions}}

In this Chapter the closed Friedmann--Robertson--Walker model with quantization in the presence of a positive cosmological constant and radiation was studied. We have solved it numerically and have determined the tunneling probability for the birth of an asymptotically de-Sitter, inflationary Universe as a function of the radiation energy.
Note the following.
\begin{enumerate}
% \item
% A formalism for calculation of two linear independent partial solutions for the wave function of the Universe for the scale factor inside the region $0 \le a \le 100$ and the energy of radiation from zero up to the barrier height has been constructed.

\item
A fully quantum definition of the wave which propagates inside strong field and which interact minimally with them, has been formulated for the first time, and approach for its determination has been constructed.

\item
A new stationary approach for the determination of the incident, reflected and transmitted waves relatively to the barrier has been constructed. The tunneling boundary condition has been corrected.

\item
A quantum stationary method of determination of coefficients of penetrability and reflection relatively to the barrier with analysis of uniqueness of solution has been developed, where for the first time non-zero interference between the incident and reflected waves has been taken into account and for its estimation the coefficient of mixing has been introduced.

\item
In this chapter the a development of the method of multiple internal reflections is presented
% ???is further development of approach of multiple internal reflection???
(see Refs.~\citep{Maydanyuk.2000.UPJ,Maydanyuk.2002.JPS,Maydanyuk.2002.PAST,Maydanyuk.2003.PhD-thesis,%
Maydanyuk.2006.FPL,Maydanyuk.arXiv:0805.4165},
also Refs.~\citep{Fermor.1966.AJPIA,McVoy.1967.RMPHA,Anderson.1989.AJPIA}).
% According to such an approach, tunneling of the packet through the barrier is considered ???consequently by steps of its propagation relatively to each boundary of the barrier???.
When the barrier is composed from arbitrary number $n$ of rectangular potential steps, the exact analytical solutions for amplitudes of the wave function, the penetrability $T_{\rm bar}$ through the barrier and the reflection $R_{\rm bar}$ from it are found. At $n \to \infty$ these solutions can be considered as exact limits for potential with the barrier and well of arbitrary shapes.

% \item
% A criterion of estimation of accuracy of the determination of these coefficients has been proposed on the basis of check of eq.~(\ref{eq.4.2.3}).
\end{enumerate}
In such a quantum approach the penetrability of the barrier for the studied quantum cosmological model with parameters $A=36$, $B=12\,\Lambda$ ($\Lambda=0.01$) has been estimated with a comparison with results of other known methods.
Note the following.
\begin{enumerate}
% \item
% According to the calculations, inside whole region of energy of radiation the tunneling probability for the birth of an asymptotically deSitter, inflationary Universe is very close to its value, obtained in the semiclassical approach by eqs.~(\ref{eq.6.1}) and (\ref{eq.6.2}), but essentially differs on the results obtained before by the quantum non-stationary approach in Ref.~\citep{AcacioDeBarros.2007.PRD} (see Tabl.~1 and 2 in Appendix).

\item
The coefficient of reflection from the barrier in the internal region has been determined. According to calculations, this coefficient is different visibly from unity at the energy of radiation close enough to the barrier height.

\item
The modulus of the coefficient of mixing is very small (it is less $10^{-19}$). This points out that \emph{there is no interference between the found incident and reflected waves close to the internal turning point.}

\item
On the basis of the calculated coefficients we reconstruct a property (\ref{eq.4.2.3})
% with accuracy ???of the first 11--18 digits???
inside the whole studied range of energy of radiation (see Fig.~12).

% The new method for determination of probability of penetration of the packet from the internal well outside with its tunneling through one-dimensional barrier of arbitrary shape, used in problems of quantum cosmology, is presented.

% \item
% Accuracy of the method in determination of penetrability and reflection is $|T_{\rm bar}+R_{\rm bar}-1| < 10^{-15}$ (see figs.~\ref{fig.model_Monerat.5}). Author has not found other methods achieving such an accuracy.

\item
{\bf The probability of penetration of the packet from the internal well outside with its tunneling through the barrier of arbitrary shape is determined. We call such coefficient as \emph{coefficient of penetration}. This coefficient is separated on the penetrability and a new coefficient, which characterizes oscillating behavior of the packet inside the internal well and is called \emph{coefficient of oscillation}. The formula found, seems to be the fully quantum analogue of the semiclassical formula of $\Gamma$ width of decay in quasistationary state proposed in Ref.~\citep{Gurvitz.1987.PRL}. }Here, the coefficient of oscillations is the fully quantum analogue for the semiclassical $F$ factor of formation and the coefficient of penetration is analogue for the semiclassical $\Gamma$ width.

\item
The penetrability of the barrier visibly changes in dependence of the position of the starting point $R_{\rm start}$ inside the internal well, where the packet begins to propagate (see figs.~\ref{fig.model_Monerat.2}). We note the following peculiarities: the penetrability has oscillating behavior, difference between its minimums and maximums is minimal at $R_{\rm start}$ in the center of the well, with increasing $R_{\rm start}$ this difference increases achieving to maximum near the turning point. The coefficients of reflection, oscillations and penetration have similar behavior. We achieve coincidence (up to the first 15 digits) between the amplitudes of the wave function obtained by such a method, and the corresponding amplitudes obtained by the standard approach of quantum mechanics (see Appendix~B in \citep{Maydanyuk.2011.JMP} where solutions for amplitudes were calculated in general quantum decay problem). This confirms that this result does not depend on a choice of the fully quantum method applied for calculations. Such a peculiarity is shown in the fully quantum considerations and it is hidden after imposing the semiclassical restrictions.

\item
In the non-stationary and stationary considerations the penetrability of the barrier
should be connected with the initial condition.
% (which should define start of the packet ??????).
We suggest that a possible introduction of the initial condition into the known stationary semiclassical models can change the obtained results.

\item
If one takes into account the external tail of the barrier, the penetrability is visibly changed. For example, the penetrability is changed up to 2 times (see figs.~\ref{fig.model_Monerat.3}) for the barrier (\ref{eq.model.2.2}) with parameters $A=0.001$ and $B=0.001$ (see figs.~\ref{fig.model_Monerat.1}) at the energy of radiation $E_{\rm rad} = 223$. If one increases the external boundary $a_{\rm max}$, all amplitudes and coefficients are convergent. This confirms efficiency of the developed method.

\item
The coefficient of penetration has oscillating dependence on the energy of radiation. Here, peaks are clearly shown. They are localized at similar distances (see figs.~\ref{fig.model_Monerat.4}). So, for the first time we have obtained in the fully quantum approach a clear and stable picture of resonances, which indicate the presence of some early unknown quasistationary states. If the energy of radiation increases, the penetrability is monotonously changed. It describes a general tendency of behavior of the coefficient of penetration, while the coefficient of oscillations gives peaks. Now the reason of existence of resonances becomes clear: oscillations of the packet inside the internal well give rise to them. In particular, we establish 134 such resonant levels inside range $E_{\rm rad}$ = 0--223 for the barrier (\ref{eq.model.2.2}) with parameters $A=0.001$ and $B=0.001$.

\item
A dependence of the penetrability on the starting point has maxima and minima.
% On such a basis one can suppose that the most probable localization of maximum of the packet at start
% is in one point of such maxima ????.
This allows to predict some definite initial values of the scale factor, when the universe begins to expand. Such initial data is direct result of quantization of the cosmological model.

\item
The modulus of the wave function in the internal and external regions has minima and maxima which were clearly established
in~\citep{Maydanyuk.2008.EPJC,Maydanyuk.2010.IJMPD}.
{\bf This indicates, in terms of values of the scale factor, where the probable ``appearance'' of the universe is the maximal or minimal.}
So, the radius of the universe during its expansion changes not continuously, but consequently passes through definite discrete values connected with these maxima. It follows that space-time of universe on the first stage after quantization seems to be rather discrete than continuous. According to results~\citep{Maydanyuk.2008.EPJC,Maydanyuk.2010.IJMPD}, difference between maxima and minima is slowly smoothed with increasing of the scale factor $a$. In this way, we obtain the continuous structure of the space-time at latter times. The discontinuity of space-time is direct result of quantization of cosmological model. This new phenomenon is the most strongly shown on the first stage of expansion and disappears after imposition of the semiclassical approximations.
\end{enumerate}
% *******************************************************************************************************************

% *******************************************************************************************************************

\end{document}